\documentclass{ws-procs9x6}
\def\ga{\mathrel{\raise.3ex\hbox{$>$\kern-.75em\lower1ex\hbox{$\sim$}}}}
\def\la{\mathrel{\raise.3ex\hbox{$<$\kern-.75em\lower1ex\hbox{$\sim$}}}}

\def\he#1{\hbox{${}^{#1}$He}}
\def\li#1{\hbox{${}^{#1}$Li}}

\def\ohsq{\Omega_{\chi} h^2}
\def\m12{m_{1\!/2}}
\def\mst{m_{\tilde\tau_1}}

\def\st{{\widetilde \tau}_{\scriptscriptstyle\rm 1}}

\def\gev{{\rm \, Ge\kern-0.125em V}}
\def\iso#1#2{\mbox{${}^{#2}{\rm #1}$}}

\def\he#1{\iso{He}{#1}}
\def\li#1{\iso{Li}{#1}}
\def\be#1{\iso{Be}{#1}}

\def\PL{{\it Phys. Lett.} }
\def\PR{{\it Phys. Rev.} }
\def\PRL{{\it Phys. Rev. Lett.} }
\def\NP{{\it Nucl. Phys.} }
\usepackage{epsf}
\usepackage{epsfig}
\begin{document}

\title{TASI Lectures on AstroParticle Physics}

\author{Keith A. Olive\footnote{ \uppercase{T}his work was supported in part by 
\uppercase{DOE} grant \uppercase{DE-FG02-94ER40823} at \uppercase{M}innesota.}}

\address{William I. Fine Theoretical Physics
Institute, School of Physics and Astronomy, \\
University of Minnesota, Minneapolis, MN 55455 USA \\ 
E-mail: olive@umn.edu}

 
\maketitle

\abstracts{
\vskip -2.5in
\rightline{astro-ph/0503065}
\rightline{UMN--TH--2346/05}
\rightline{FTPI--MINN--05/05}
\rightline{March 2005}
\vskip 1.9in
Selected topics in Astroparticle Physics including the CMB,
dark matter, BBN, and the variations of fundamental couplings are discussed.}

\section{Introduction}

The background for all of the topics to be discussed in these lectures
is the Big bang model.
The observed homogeneity and isotropy enable us to
describe the overall geometry and evolution of the Universe in terms
of two cosmological parameters accounting for the spatial curvature and
the overall expansion (or contraction) of the Universe. These two quantities
appear in the most general expression for a space-time metric which has a
(3D) maximally symmetric subspace of a 4D space-time, known as the
Robertson-Walker metric:
\begin{equation}
ds^2 = dt^2 - R^2(t) \left[ \frac{d r ^2}{1 - k r ^2} + r^2 (d \theta^2 + \sin^2\theta d \phi^2 ) \right] 
\label{frw}
\end{equation}
where $R(t)$ is the cosmological scale factor and $k$ is the curvature constant.
By rescaling the radial coordinate, we can choose  $k$
to take only the discrete values $+1$,
$-1$, or 0 corresponding to closed, open, or spatially flat geometries.

The cosmological equations of motion are derived from 
Einstein's equations
\begin{equation}
 {\mathcal R}_{\mu\nu} - {1 \over 2} g_{\mu\nu} {\mathcal R} 
= 8 \pi G_{\rm N} T_{\mu\nu} + \Lambda g_{\mu\nu}
\end{equation}
where $\Lambda$ is the cosmological constant.         
It is common to assume that the matter content of
the Universe is a perfect fluid, for which
\begin{equation}
        T_{\mu\nu} = - p g_{\mu\nu} +
        \left (
        p + \rho
        \right )u_\mu u_\nu
\end{equation}
where $g_{\mu\nu}$ is the space-time metric described by (\ref{frw}), $p$ is the
isotropic pressure, $\rho$ is the energy density and $u  = (1,0,0,0)$ is the
velocity vector for the isotropic fluid in co-moving coordinates.
With the perfect fluid source, Einstein's equations lead to the
Friedmann-Lema\^\i tre equations
\begin{equation}
H^2 = {{\dot R}^2 \over R^2} = {8 \pi G_{\rm N} \rho \over 3} - {k \over R^2} +
{\Lambda \over 3}
\label{fried}
\end{equation}
and 
\begin{equation}
        \frac{\ddot R}{R}
        =
        \frac{\Lambda}{3}
        -
        \frac{4 \pi G_{\rm N}}{3}
        \;\;( \rho + 3 p)
        \label{fried2}
\end{equation}
where $H(t)$ is the Hubble parameter.
Energy conservation  via
$T^{\mu\nu}_{~;\mu} = 0$, leads to a third useful equation 
[which can also be derived from Eqs. (\ref{fried}) and (\ref{fried2})] 
\begin{equation}
        \dot \rho
        =
        -3H \left( \rho + p \right)
\label{econ}
\end{equation}

The Friedmann equation can be rewritten as
\begin{equation}
	(\Omega - 1)H^2  = {k \over R^2}	
\label{o-1}	
\end{equation}
so that $k = 0, +1, -1$ corresponds to $\Omega = 1, \Omega > 1$
 and $\Omega < 1$.  However, the value of $\Omega$ appearing in Eq.
(\ref{o-1}) represents the sum $\Omega = \Omega_m + \Omega_\Lambda$
of contributions from the matter density ($\Omega_m$) and the
cosmological constant $(\Omega_\Lambda = \Lambda/3H^2)$.

\section{The CMB}

There has been a great deal of progress in the last several years
concerning the determination of both $\Omega_m$ and $\Omega_\Lambda$.
Cosmic Microwave Background (CMB) anisotropy experiments have been able
to determine the curvature (i.e. the sum of $\Omega_m$ and
$\Omega_\Lambda$) to within  a few percent, while observations of type Ia
supernovae at high redshift provide information on a (nearly) orthogonal
combination of the two density parameters.

The CMB is of course deeply rooted in the development and verification of
the big bang model and big bang nucleosynthesis (BBN)\cite{ah}. 
Indeed, it was the formulation of BBN that
led to the prediction of the microwave background.  The argument is rather
simple. BBN requires temperatures greater than 100 keV, which according
to the standard model time-temperature relation,
$t_{\rm s} T^2_{\rm MeV} = 2.4/\sqrt{N}$, where
$N$ is the number of relativistic degrees of freedom at temperature
$T$, and corresponds to timescales less than about 200 s. The typical
cross section for the first link in the nucleosynthetic chain is
\begin{equation} 
\sigma v (p + n \rightarrow D + \gamma) \simeq 5 \times 10^{-20} 
{\rm cm}^3/{\rm s}
\end{equation}
This implies that it was necessary to achieve a density
\begin{equation}
n \sim {1 \over \sigma v t} \sim 10^{17} {\rm cm}^{-3}
\end{equation}
for nucleosynthesis to begin.
The density in baryons today is known approximately from the density of
visible matter to be ${n_B}_o \sim 10^{-7}$ cm$^{-3}$ and since
we know that that the density $n$ scales as $R^{-3} \sim T^3$, 
the temperature today must be
\begin{equation}
T_o = ({n_B}_o/n)^{1/3} T_{\rm BBN} \sim 10 {\rm K}
\end{equation}
thus linking two of the most important tests of the big bang theory.

An enormous amount of cosmological information is encoded in the
angular expansion of the CMB temperature
\begin{equation}
T(\theta,\phi) = \sum_{\ell m} a_{\ell m} Y_{\ell m}(\theta, \phi) .
\end{equation}
The monopole term characterizes the mean background temperature
of $T_\gamma = 2.725 \pm 0.001$ K as determined by COBE\cite{cobe},
whereas the dipole term can be associated with the Doppler shift
produced by our peculiar motion with respect to the CMB.
In contrast, the higher order multipoles, are directly related to
energy density perturbations in the early Universe. When compared
with theoretical models, the higher order anisotropies can be used
to constrain several key cosmological parameters.
In the context of simple adiabatic cold dark matter (CDM) models,
there are nine of these: the cold dark matter density, $\Omega_\chi h^2$;
the baryon density, $\Omega_B h^2$; the curvature - characterized by $\Omega_{\rm total}$;
the hubble parameter, $h$; the optical depth, $\tau$; the spectral indices of scalar and 
tensor perturbations, $n_s$ and $n_t$; the ratio of tensor to scalar perturbations, $r$;
and the overall amplitude of fluctuations, $Q$.

\begin{figure}[htbp]
\hspace{0.5truecm}
\centering
\epsfxsize=10cm   
\epsfbox{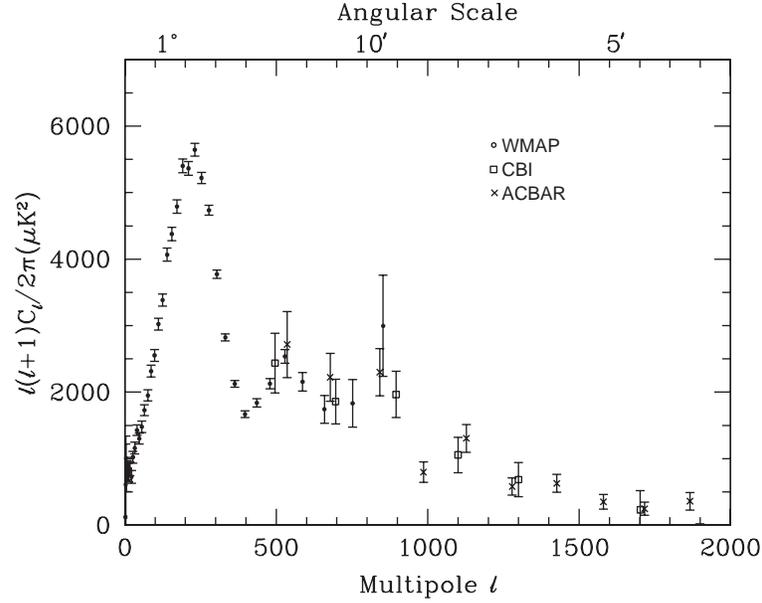}
\caption{The power in the microwave background anisotropy spectrum
as measured by WMAP\protect\cite{wmap}, CBI\protect\cite{cbi}, and ACBAR\protect\cite{acbar}.
Taken from \protect\cite{scsm}. }
\label{freq1}
\end{figure}

Microwave background anisotropy measurements have made tremendous
advances in the last few years. The power spectrum\cite{max,boom,dasi1,cbi,vsa,arch,wmap,acbar}
has been measured relatively accurately out to multipole moments
corresponding to $\ell \sim 2000$. A compilation of recent data is shown in Fig. 
\ref{freq1} \cite{scsm}, where the power in at each ${\ell}$ is given by $(2{\ell} + 1) C_\ell/(4 \pi)$, 
and $C_\ell = <|a_{\ell m}|^2>$.

As indicated above, the details of this spectrum enable
one to make accurate predictions of a large number of fundamental
cosmological parameters. The results of the WMAP data 
(with other information concerning the power spectrum) is shown in Table 1.
For details see ref. \cite{wmap}.

\begin{table}
\centering
\begin{tabular}{l|ccc}
& WMAP alone & WMAPext $+$ 2dFGRS & WMAPext $+$ 2dFGRS $+$ Lyman $\alpha$\\
& power-law & power-law & running \\
\hline
$\Omega_{{\rm m}} h^2$ & $0.14 \pm 0.02$ & $0.134 \pm 0.006$& $0.135^{+0.008}_{-0.009}$\\
$\Omega_{{\rm B}} h^2$ & $0.024 \pm 0.001$ & $0.023 \pm 0.001$&$0.0224 \pm 0.0009$\\
$h$ & $0.72 \pm 0.05$ & $0.73 \pm 0.03$&$0.71^{+0.04}_{-0.03}$\\
$n_s$ & $0.99 \pm 0.04$ & $0.97 \pm 0.03$&$0.93 \pm 0.03$\\
$\tau$ & $0.166^{+0.076}_{-0.071}$ & $0.148^{+0.073}_{-0.071}$&$0.17 \pm 0.06$\\
\end{tabular}
\end{table}

\begin{figure}
\begin{center}
\epsfxsize=10cm   
\epsfbox{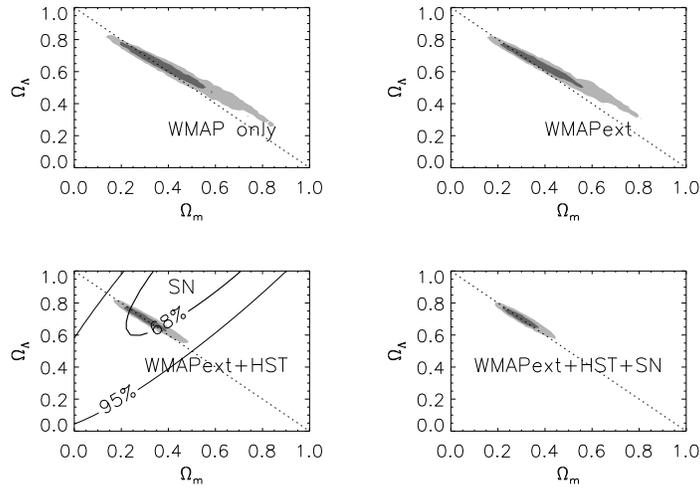}
\end{center}
\caption{\footnotesize Two-dimensional confidence regions
in the ($\Omega_{\rm M},\Omega_{\Lambda}$) plane\protect\cite{wmap}.}
\label{lm}
\end{figure}

Of particular interest to us here is the CMB determination of the total
density, $\Omega_{\rm tot}$, as well as the matter density $\Omega_m$. 
There is strong evidence that the Universe is flat or
very close to it. The best constraint on $\Omega_{\rm total}$ is $1.02 \pm 0.02$. 
Furthermore, the matter density is significantly larger than the baryon
density implying the existence of cold dark matter  and the baryon density,
as we will see below, is consistent with the BBN production of D/H and its
abundance in quasar absorption systems. 
The apparent discrepancy between the CMB value
of $\Omega_{\rm tot}$ and $\Omega_m$, though not conclusive on its own,
is a sign that a contribution from the vacuum energy density or
cosmological constant, is also required.  The preferred region in the
$\Omega_m - \Omega_\Lambda$ plane is shown in Fig. \ref{lm} under four 
different assumptions\cite{wmap}.

The presence or absence of a cosmological constant is a long standing
problem in cosmology.   We know that 
the cosmological term is at most a factor of a few times larger than the
current mass density.  Thus from Eq. (\ref{fried}), we see that the
dimensionless combination, $G_N \Lambda \la 10^{-121}$. Nevertheless,
even a small non-zero value for $\Lambda$ could greatly affect the
future history of the Universe: allowing open Universes to recollapse (if
$\Lambda < 0 $), or closed Universes to expand forever (if $\Lambda > 0$
and sufficiently large).

When the SN 1a results\cite{sn1} are included (see the last panel of Fig. \ref{lm}) 
we are led to a seemingly conclusive
picture.  The Universe is nearly flat with $\Omega_{\rm tot} \simeq 1$.
However the density in matter makes up only 23\% of this total, with
the remainder in a cosmological constant or some other form of dark
energy.

\section{Dark Matter}

\subsection{Observational Evidence}

Direct observational
evidence for dark matter is found from a variety of sources. On the scale
of galactic halos, the  observed flatness of the rotation curves of
spiral galaxies is a clear indicator for dark matter. There is also
evidence for dark matter in elliptical galaxies, as well as clusters of
galaxies coming from the X-ray observations of these objects. Also, direct
evidence has been obtained through the study of gravitational lenses. 

\begin{figure}[t]
\hskip 2cm
\epsfxsize=6cm   
\epsfbox{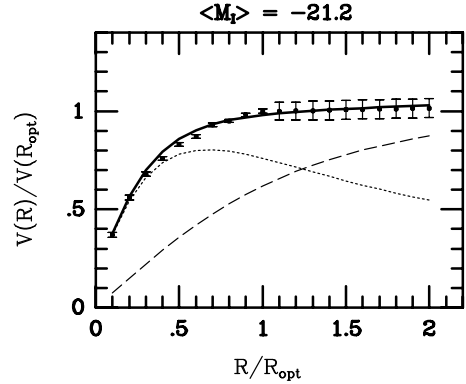}
\caption{Synthetic rotation curve\protect\cite{pss} for galaxies with
$\langle M
\rangle = -21.2$.  The dotted curve shows the disk contribution,
whereas the dashed curve shows the halo contribution.  
\label{rot}}
\end{figure}

For example,  assuming that galaxies are in virial equilibrium,
 one expects that one can relate 
the mass at a given distance $r$, from the center of 
a galaxy to its rotational velocity by
\begin{equation}
	M(r) \propto v^2 r/G_N 	
\end{equation}
The rotational velocity, $v$, is measured\cite{fg,rft}
 by observing 21 cm 
emission lines in HI regions (neutral hydrogen) beyond the point 
where most of the light in the galaxy ceases.  A subset of a
compilation\cite{pss}  of nearly 1000
 rotation curves of spiral galaxies is shown in Fig. \ref{rot}. The
subset shown is restricted to a narrow range in brightness, but is
characteristic for a wide range of spiral galaxies. Shown is the
rotational velocity as a function of
$r$ in units of the optical radius.  If the bulk of the mass is 
associated with light, then beyond the point where most of the light 
stops, $M$ would be  constant and $v^2  \propto 1/r$.  This is not the
case, as the rotation
 curves appear to be flat, i.e., $v \sim$ constant outside the
 core of the galaxy. This implies that $M \propto r$ beyond the point
 where the light stops.  This is one of the strongest pieces of 
evidence for the existence of dark matter on galactic scales. Velocity measurements indicate
dark matter in elliptical galaxies as well\cite{sag}.
For a more complete discussion see \cite{otasi3}.

\subsection{Theory}

Theoretically, there is no lack of support for the dark matter hypothesis.
The standard big bang model including inflation almost requires 
$\Omega_{\rm tot} = 1~$\cite{infl}. This can be seen from the following
simple solution to the curvature problem.
 The 
unfortunate fact that at present we do not even know whether $\Omega$ is
larger or smaller than one, indicates that we do not know the sign of
the curvature term further implying that it is subdominant in Eq.
(\ref{fried})
\begin{equation}
	  { k \over R^2 } < {8 \pi G \over 3} \rho
\end{equation}
In an adiabatically expanding Universe, $R \sim T^{-1}$   where $T$ is the
temperature of the thermal photon background.  Therefore the quantity
\begin{equation}
	\hat{k} = { k \over R^2 T^2} <  {8 \pi G \over 3 T_o^2} < 2 \times 10^{-58}
\label{khat}
\end{equation}
is dimensionless and constant in the standard model.  This is known as
the curvature problem and can be resolved by a period of inflation.
Before inflation, let us write $R = R_i$, $T = T_i$  and $R \sim T^{-1}$.  During
inflation, $R \sim T^{-1} \sim e^{Ht}$, where $H$ is constant.  After inflation, $R =
R_f  \gg R_i$  but $T = T_f  = T_R  \la T_i$  where $T_R$  is the temperature
 to which the
Universe reheats.  Thus $R \not\sim T$   and $\hat{k} \rightarrow 0$
 is not constant.  But from
Eqs. (\ref{o-1}) and (\ref{khat}) if $\hat{k} \rightarrow 0$ then
 $\Omega \rightarrow 1$, and since typical
inflationary models contain much more expansion than is necessary, $\Omega$
becomes exponentially close to one.

The existence of non-baryonic dark matter can be immediately
inferred from the determination of the cosmological 
parameters through the microwave background anisotropy as described above.
If $\Omega_m h^2 \simeq 0.13$ and $\Omega_B h^2 \simeq 0.02$, then the 
difference must be dark matter which contributes to the total density
$\Omega_{\rm DM} h^2 \simeq 0.11$.  In addition, because the amplitude
of fluctuations is relatively small, dark matter is necessary to have sufficient time 
to grow primordial perturbations into galaxies (for a more complete discussion see \cite{otasi3}).

\subsection{Candidates}

\subsubsection{Baryons}

Accepting the dark matter hypothesis, the first choice for a
 candidate should be something we know to exist, baryons.  
Though baryonic dark matter can not be the whole story if $\Omega_m >
0.1$,  the identity of the
 dark matter in galactic halos, which appear to contribute at the 
level of $\Omega \sim 0.05$,  remains an important question needing to be
resolved.  A baryon density of this magnitude is not excluded by 
nucleosynthesis. 
 Indeed we know some of the baryons are dark since $\Omega \la 0.01$ 
in the disk of the galaxy.

It is interesting to note that until recently, there seemed to be some
difficulty in reconciling the baryon budget of the Universe.
By counting the visible contribution to $\Omega$ in stellar populations
and  the X-ray producing hot gas, Persic and Salucci\cite{ps} found only
$\Omega_{\rm vis} \simeq 0.003$. A subsequent accounting by Fukugita,
Hogan and Peebles\cite{fhp} found slightly more ($\Omega \sim 0.02$) by
including the contribution from plasmas in groups and clusters.
At high redshift on the other hand, all of the baryons can be accounted
for. The observed opacity of the Ly $\alpha$ forest in QSO absorption
spectra requires a large baryon density consistent with the
determinations by the CMB and BBN\cite{hae}.

In galactic halos, however, it is quite difficult to hide large
amounts  of baryonic matter. Sites for halo baryons that 
have been discussed
include Hydrogen (frozen, cold or hot gas), low mass stars/Jupiters,
remnants of massive stars such as white dwarfs, neutron stars or black
holes.  In almost every case, a serious theoretical or observational
problem is encountered\cite{hio12}.

\subsubsection{Neutrinos}

	Light neutrinos ($m \le 30 eV$) are 
a long-time standard when it comes to
 non-baryonic dark matter\cite{ss}.    Light neutrinos
 are, however, ruled out as a dominant form of dark matter because they 
 produce too much large scale structure\cite{nu3}.
  Because the smallest non-linear structures have mass scale 
 $M_J \approx 3 \times 10^{18} M_\odot / m_\nu^2 ({\rm eV})$ and 
the typical galactic mass scale is $\simeq 10^{12} M_\odot$, galaxies must 
fragment out of the larger pancake-like objects.  The problem with
such a scenario is that galaxies form late\cite{nu2,nu4} 
 ($z \le 1$) whereas
 quasars and galaxies are seen out to redshifts $z \ga 6$.

The neutrino decoupling scale of ${\mathcal O}(1)$ MeV has an important
consequence  on the final relic density of massive neutrinos. Neutrinos
more massive  than 1 MeV will begin to annihilate prior to decoupling,
and while in  equilibrium, their number density will become exponentially
suppressed.  Lighter neutrinos decouple as radiation on the other hand,
and hence do  not experience the suppression due to annihilation.
Therefore, the  calculations of the number density of light ($m_\nu \la
1$ MeV) and  heavy ($m_\nu \ga 1$ MeV) neutrinos differ substantially.

\begin{figure}[t]
\hspace{0.5truecm}
\epsfxsize=8cm   
\epsfbox{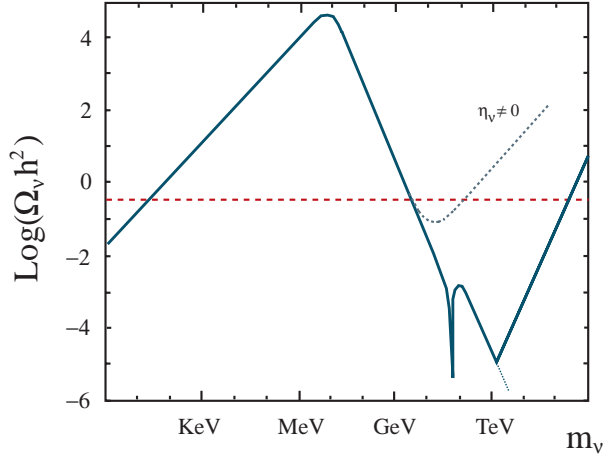}
\caption{{Summary plot\protect\cite{kko} of the relic density of Dirac
neutrinos  (solid) including a possible neutrino asymmetry of $\eta_\nu =
5\times 10^{-11}$ (dotted).}}
\label{nu}
\end{figure}

The energy of density of light neutrinos with $m_\nu \la 1$ MeV can be 
expressed at late times as
$  \rho_\nu  = m_\nu Y_\nu n_\gamma  $
where $Y_\nu = n_\nu/n_\gamma$ is the number density of $\nu$'s relative to 
the density of photons, which today is 411 photons per cm$^3$. It is 
easy to show that in an adiabatically expanding universe $Y_\nu = 
3/11$. This suppression is a result of the $e^+ e^-$ annihilation 
which occurs after neutrino decoupling and heats the photon bath 
relative to the neutrinos.
Imposing the constraint $\Omega_\nu h^2 \la 0.13$, translates into a 
 strong constraint (upper bound) on Majorana
neutrino masses\cite{cows}:
\begin{equation}
m_{\rm tot} =   \sum_\nu  m_\nu   \la 12 {\rm eV}.
\label{ml1}
\end{equation}
where the sum runs over neutrino mass eigenstates. The limit for Dirac
neutrinos depends on the interactions of the right-handed states.  
The limit (\ref{ml1}) and the
corresponding  initial rise in
$\Omega_\nu h^2$ as a function of $m_\nu$ is  displayed in the
Figure~\ref{nu}.

Combining the rapidly improving data on key cosmological 
parameters with the better statistics from large redshift surveys 
has made it possible to go a step forward along this path. It is 
now possible to set stringent limits on the light neutrino mass
density $\Omega_\nu h^2$, and hence on the neutrino mass based on
the power spectrum of the Ly 
$\alpha$ forest\cite{strong}, $m_{\rm tot} < 5.5$ eV, and the limit 
is even stronger if the total matter density, $\Omega_m$ is less 
than 0.5. Adding additional observation constraints from the CMB and
galaxy clusters drops this limit\cite{wang} to 4.2 eV. This limit has
recently been improved by the 2dF Galaxy redshift\cite{2dF} 
survey by comparing the derived power spectrum of fluctuations
with structure formation models. 
 Focussing on the the 
presently favoured $\Lambda$CDM model, the neutrino mass bound becomes
$m_{\rm tot} < 1.8 $ eV for $\Omega_m < 0.5$.  When even more constraints
such as HST Key project data,  supernovae type Ia data, and BBN are
included\cite{Lewis} the limit can be pushed to $m_{\rm tot} < 0.9 $ eV.
With WMAP data, an upper limit of  $m_{\rm tot} < 0.7 $ eV has been derived\cite{wmap}.

The calculation of the relic density for neutrinos more massive than
$\sim 1$ MeV, is substantially more involved. The relic density is now
determined by the freeze-out of neutrino annihilations which occur at
$T \la m_\nu$, after annihilations have begun to seriously reduce their
number density\cite{lw}.  For particles
which annihilate through approximate weak scale interactions, annihilations freeze out
when $T \sim m_\chi /20$.  

\begin{figure}
\begin{center}
\hspace{0.5truecm}
\epsfxsize=8cm   
\epsfbox{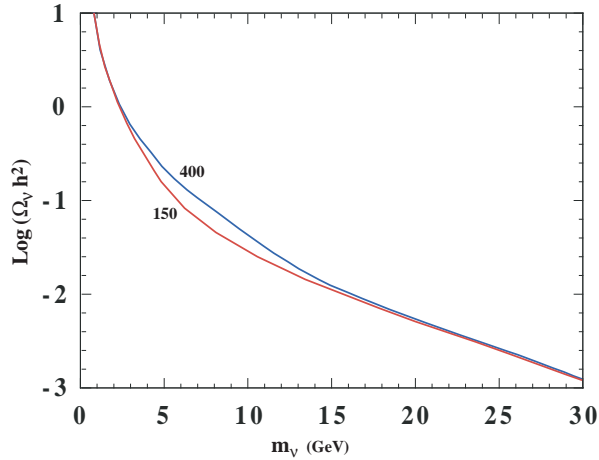}
\caption{The relic density of heavy Dirac neutrinos due to 
annihilations\protect\cite{wso}. The curves are labeled by
the assumed quark-hadron phase transition temperature in MeV.}
\label{swo}       
\end{center}
\end{figure}

Roughly, the solution to the Boltzmann equation, which
tracks the neutrino abundance, goes as $Y_\nu \sim f \sim (m
\langle
\sigma v
\rangle_{ann} )^{-1}$ and hence $\Omega_\nu h^2 \sim {\langle \sigma v
\rangle_{ann}}^{-1}$, so that parametrically $\Omega_\nu h^2  \sim
1/{m_\nu^2}$. As a result, the constraint on $\Omega$ now leads to
a {\em lower}  bound\cite{lw,ko,wso} on the neutrino mass, of about
$m_\nu \ga 3-7$  GeV, depending on whether it is a Dirac or Majorana
neutrino.  This bound and the corresponding downward trend $\Omega_\nu
h^2 \sim 1/m^2_\nu$ can again be seen in Figure~\ref{nu}. The result of a
more detailed calculation is shown in
Figure~\ref{swo}~\cite{wso} for the case of a Dirac neutrino. 
The two curves show the slight sensitivity on the temperature scale
associated with the quark-hadron transition. The result for a Majorana
mass neutrino is qualitatively similar. 
Indeed, any particle with roughly weak scale cross-sections will tend to
give an interesting  value of  $\Omega h^2 \sim 1$.

The deep drop in $\Omega_\nu h^2$, visible in Figure~\ref{nu}
at around $m_\nu = M_Z/2$, is due to a very strong annihilation 
cross section at $Z$-boson pole. For yet higher neutrino masses the 
$Z$-annihilation channel cross section drops as $\sim 1/m_\nu^2$, 
leading to a brief period of an increasing trend in $\Omega_\nu h^2$. 
However, for $m_\nu \ga m_W$ the cross section regains its parametric 
form $\langle \sigma v \rangle_{ann} \sim m_\nu^2$ due to the opening up
of a new annihilation channel to $W$-boson pairs\cite{Enqvist:1988we}, 
and the density drops again as $\Omega_\nu h^2 \sim 1/m^2_\nu$. 
The tree level $W$-channel cross section breaks the unitarity at 
around ${\mathcal O}({\rm few})$ TeV~\cite{Enqvist:yz} however, and the
full  cross section must be bound by the unitarity
limit\cite{Griest:1989wd}. This behaves again as $1/m_\nu^2$, whereby
$\Omega_\nu h^2$  has to start increasing again, until it becomes too
large again at 200-400 TeV~\cite{Griest:1989wd,Enqvist:yz}.

If neutrinos are Dirac particles, and 
have a nonzero asymmetry  the relic density could 
be governed by the asymmetry rather than by the annihilation cross section. 
Indeed, it is easy to see that the neutrino mass density corresponding 
to the asymmetry $\eta_\nu \equiv (n_\nu - n_{\bar \nu})/n_\gamma$ is 
given by\cite{ho} $
  \rho = m_\nu \eta_\nu n_\gamma$, which implies
\begin{equation}
  \Omega_\nu h^2 \simeq 0.004 \,\eta_{\nu 10}\, (m_\nu/{\rm GeV}).
\end{equation}
where $\eta_{\nu 10}\equiv 10^{10}\eta_\nu$.
The behaviour of the energy density of neutrinos
with an asymmetry is shown by the dotted line in the Figure~\ref{nu}.
In the  figure, we have assumed an asymmetry of
$\eta_\nu \sim 5 \times  10^{-11}$ for neutrinos with standard weak 
interaction strength.

Based on the leptonic and invisible width of the $Z$ boson, 
experiments at LEP have determined that the number of neutrinos is 
$N_\nu = 2.994 \pm 0.012$~\cite{RPP}. Conversely, any new physics 
must fit within these brackets, and thus LEP excludes additional 
neutrinos (with standard weak interactions) with masses $m_\nu 
\la 45$ GeV.  Combined with the limits displayed in Figures~\ref{nu} 
and \ref{swo}, we see that the mass density of ordinary heavy
neutrinos is bound to be very small, $\Omega_\nu {h}^2 < 0.001$ for
masses  $m_\nu > 45$ GeV up to $m_\nu \sim {\mathcal O}(100)$ TeV. 
Lab constraints for 
Dirac neutrinos are available\cite{dir}, excluding neutrinos 
with masses between
10 GeV and 4.7 TeV. This is significant, since it precludes the possibility 
of neutrino dark matter based on an asymmetry between $\nu$ and ${\bar \nu}$ ~\cite{ho}. 

\subsubsection{Axions}

Due to space limitations, the discussion of this candidate will be very
brief.
Axions are
pseudo-Goldstone bosons which arise in solving the strong CP
problem\cite{ax1,ax2} via a global U(1) Peccei-Quinn symmetry.  The
invisible axion\cite{ax2} is associated with the flat direction of the
spontaneously broken PQ symmetry.  Because the PQ symmetry is also
explicitly broken (the CP violating $\theta F {\widetilde F}$ coupling is
not PQ invariant) the axion picks up a small mass similar to pion picking
up a mass when chiral symmetry is broken.  We can expect that $m_a  \sim
m_\pi f_\pi /f_a$  where $f_a$, the axion decay constant, is the vacuum
expectation value of the PQ current and can be taken to be quite large. 
If we write the axion field as $a = f_a \theta$, near the minimum, the
potential produced by QCD instanton effects looks like $V \sim m_a^2
\theta^2 f_a^2$.  The axion equations of motion lead to
a relatively stable oscillating solution.  The energy density stored in
the oscillations exceeds the critical density\cite{axden}  unless $f_a 
\la 10^{12}$ GeV.

	Axions may also be emitted stars and supernova\cite{raff}.
In supernovae, axions are produced via nucleon-nucleon bremsstrahlung with
a coupling $g_AN \propto m_N/f_a$.  As was noted above the cosmological
density limit requires $f_a \la 10^{12}$  GeV.  Axion emission from red
giants imply\cite{dss}   $ f_a  \ga 10^{10}$ GeV (though this limit
depends on an adjustable axion-electron coupling), the supernova limit
requires\cite{sn}   $ f_a \ga 2 \times 10^{11}$   GeV for naive quark
model couplings of the axion to nucleons.   
Thus only a narrow window exists for the axion as a viable dark matter
candidate.

\section{Supersymmetric Dark Matter}

For the remaining discussion of dark matter, I will restrict my attention to
supersymmetry and in particular, the minimal supersymmetric standard model (MSSM)
with R-parity conservation. R-parity is necessary if one wants to forbid all
new baryon and lepton number violating interactions at the weak scale.
 If R-parity, which
distinguishes between ``normal" matter and the  supersymmetric
partners and can be defined in terms of baryon, lepton and spin
as $R = (-1)^{3B + L + 2S}$, is unbroken, there is at least one 
supersymmetric particle (the lightest supersymmetric particle or LSP)
which must be stable.  Thus, the minimal model contains the fewest number of new
particles and interactions necessary to make a consistent theory.

There are very strong constraints, however, forbidding the existence of stable or
long lived particles which are not color and electrically neutral\cite{EHNOS}. Strong
and electromagnetically interacting LSPs would become bound with normal matter
forming anomalously heavy isotopes. Indeed, there are very strong upper limits on the
abundances, relative to hydrogen, of nuclear isotopes\cite{isotopes},
$n/n_H \la 10^{-15}~~{\rm to}~~10^{-29}
$
for 1 GeV $\la m \la$ 1 TeV. A strongly interacting stable relic is expected
to have an abundance $n/n_H \la 10^{-10}$
with a higher abundance for charged particles.

There are relatively few supersymmetric candidates which are not colored and
are electrically neutral.  The sneutrino\cite{snu} is one possibility,
but in the MSSM, it has been excluded as a dark matter candidate by
direct\cite{dir} and indirect\cite{indir} searches.  In fact, one can set
an accelerator based limit on the sneutrino mass from neutrino counting, 
$m_{\tilde\nu}\ga$ 44.7 GeV~\cite{EFOS}. In this case, the direct relic
searches in
underground low-background experiments require  
$m_{\tilde\nu}\ga$ 20 TeV~\cite{dir}. Another possibility is the
gravitino which is probably the most difficult to exclude. 
I will concentrate on the remaining possibility in the MSSM, namely the
neutralinos but will return to the case of gravitino dark matter as well.

\subsection{Parameters}

The most general version of the MSSM, despite its minimality in particles and
interactions contains well over a hundred new parameters. The study of such a model
would be untenable were it not for some (well motivated) assumptions.
These have to do with the parameters associated with supersymmetry breaking.
It is often assumed that, at some unification scale, all of the gaugino masses
receive a common mass, $m_{1/2}$. The gaugino masses at the weak scale are
determined by running a set of renormalization group equations.
Similarly, one often assumes that all scalars receive a common mass, $m_0$,
at the GUT scale. These too are run down to the weak scale. 
The remaining
supersymmetry breaking parameters are the trilinear mass terms, $A_0$,
which I will also assume are unified at the GUT scale,  and the bilinear
mass term
$B$. There are, in addition, two physical CP violating phases which will
not be considered here.
Finally, there is the Higgs mixing mass
parameter, $\mu$, and since there are two Higgs doublets in the MSSM, there are 
two vacuum expectation values. One combination of these is related to the $Z$ mass,
and therefore is not a free parameter, while the other combination, the ratio of the
two vevs, $\tan \beta$, is free. 

The natural boundary conditions at the GUT scale for the MSSM would
include
$\mu$ and $B$ in addition to
$m_{1/2}$,
$m_0$, and $A_0$. In this case, upon running the RGEs down to a low energy
scale and minimizing the Higgs potential, one would predict the values of $M_Z$, 
$\tan \beta$ (in addition to all of the sparticle masses).
Since $M_Z$ is known, it is more useful to analyze supersymmetric models
where $M_Z$ is input rather than output.  It is also common to treat
$\tan \beta$ as an input parameter. This can be done at the expense of 
shifting $\mu$ (up to a sign) and $B$ from inputs to outputs. 
This model is often referred
to as the constrained MSSM or CMSSM. Once these parameters are set, the
entire spectrum of sparticle masses at the weak scale can be calculated. 

\begin{figure}
\begin{center}
\epsfxsize=8cm   
\epsfbox{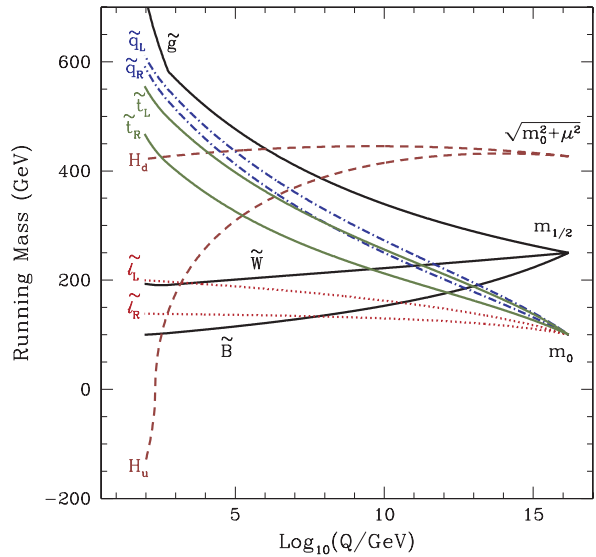}
\caption{RG evolution of the mass parameters in the CMSSM.}
\label{running}       
\end{center}
\end{figure}

In Fig. \ref{running}, an example of the running of the mass parameters
in the CMSSM is shown.  Here, we have chosen $m_{1/2} = 250$ GeV, $m_0 = 100$
GeV, $\tan \beta = 3$, $A_0 = 0$, and $\mu < 0$.
Indeed, it is rather amazing that from so few input parameters, all of the
masses of the supersymmetric particles can be determined. 
The characteristic features that one sees in the figure, are for example, that
the colored sparticles are typically the heaviest in the spectrum.  This is
due to the large positive correction to the masses due to $\alpha_3$ in the
RGE's.  Also, one finds that the ${\widetilde B}$ (the partner of the
$U(1)_Y$ gauge boson), is typically the lightest sparticle.  But most
importantly, notice that one of the Higgs mass$^2$, goes negative
triggering electroweak symmetry breaking\cite{rewsb}. 
(The negative sign in the figure refers to the sign of the
mass$^2$, even though it is the mass of the sparticles which are
depicted.)

\subsection{Neutralinos}

 There are four neutralinos, each of which is a  
linear combination of the $R=-1$ neutral fermions\cite{EHNOS}: the wino
$\tilde W^3$, the partner of the
 3rd component of the $SU(2)_L$ gauge boson;
 the bino, $\tilde B$;
 and the two neutral Higgsinos,  $\tilde H_1$ and $\tilde H_2$.
Assuming gaugino mass universality at the  GUT scale, the identity and
mass of the LSP are determined by the gaugino mass $m_{1/2}$, 
$\mu$, and  $\tan \beta$. In general,
neutralinos can  be expressed as a linear combination
\begin{equation}
	\chi = \alpha \tilde B + \beta \tilde W^3 + \gamma \tilde H_1 +
\delta
\tilde H_2
\end{equation}
The solution for the coefficients $\alpha, \beta, \gamma$ and $\delta$
for neutralinos that make up the LSP 
can be found by diagonalizing the mass matrix
\begin{equation}
      ({\tilde W}^3, {\tilde B}, {{\tilde H}^0}_1,{{\tilde H}^0}_2 )
  \left( \begin{array}{cccc}
M_2 & 0 & {-g_2 v_1 \over \sqrt{2}} &  {g_2 v_2 \over \sqrt{2}} \\
0 & M_1 & {g_1 v_1 \over \sqrt{2}} & {-g_1 v_2 \over \sqrt{2}} \\
{-g_2 v_1 \over \sqrt{2}} & {g_1 v_1 \over \sqrt{2}} & 0 & -\mu \\
{g_2 v_2 \over \sqrt{2}} & {-g_1 v_2 \over \sqrt{2}} & -\mu & 0 
\end{array} \right) \left( \begin{array}{c} {\tilde W}^3 \\
{\tilde B} \\ {{\tilde H}^0}_1 \\ {{\tilde H}^0}_2 \end{array} \right)
\end{equation}
where $M_1 (M_2)$ is a soft supersymmetry breaking
 term giving mass to the U(1) (SU(2))  gaugino(s).
  In a unified
 theory $M_1 = M_2 = m_{1/2}$ at the unification scale (at the weak scale,
$ M_1
\simeq {5 \over 3}  {\alpha_1 \over \alpha_2}  M_2	$).   As one can see, 
the coefficients
$\alpha, \beta, \gamma,$ and $\delta$ depend only on
$m_{1/2}$, $\mu$, and $\tan \beta$. 
In the CMSSM, the solutions for $\mu$ generally lead to a lightest neutralino
which is very nearly a pure $\tilde B$.

\subsection{The Relic Density}

The relic abundance of LSP's is 
determined by solving
the Boltzmann
 equation for the LSP number density in an expanding Universe.
 The technique\cite{wso} used is similar to that for computing
 the relic abundance of massive neutrinos\cite{lw} with the appropriate substitution of  the
cross section.
The relic density depends on additional parameters in the MSSM beyond $m_{1/2},
\mu$, and $\tan \beta$. These include the sfermion masses, $m_{\tilde f}$ and the
Higgs pseudo-scalar mass, $m_A$, derived from $m_0$ (and $m_{1/2}$). To
determine the relic density it is necessary to obtain the general
annihilation cross-section for neutralinos.  In much of the parameter
space of interest, the LSP is a bino and the annihilation proceeds mainly
through sfermion exchange. Because of the p-wave suppression associated
with Majorana fermions, the s-wave part of the annihilation cross-section
is suppressed by the outgoing fermion masses.  This means that it is
necessary to expand the cross-section to include p-wave corrections which
can be expressed as a term proportional to the temperature if neutralinos
are in equilibrium. Unless the neutralino mass happens to lie near near a
pole, such as $m_\chi \simeq$
$m_Z/2$ or $m_h/2$, in which case there are large contributions to the
annihilation through direct $s$-channel resonance exchange, the dominant
contribution to
the $\tilde{B} \tilde{B}$ annihilation cross section comes from crossed
$t$-channel sfermion exchange.

Annihilations in the early
Universe continue until the annihilation rate
$\Gamma
\simeq \sigma v n_\chi$ drops below the expansion rate. 
The final neutralino relic density expressed as a fraction of the critical
energy density  can be written as\cite{EHNOS}
\begin{equation}
\Omega_\chi h^2 \simeq 1.9 \times 10^{-11} \left({T_\chi \over
T_\gamma}\right)^3 N_f^{1/2} \left({{\rm GeV} \over ax_f + {1\over 2} b
x_f^2}\right)
\label{relic}
\end{equation} 
where $(T_\chi/T_\gamma)^3$ accounts for the subsequent reheating of the
photon temperature with respect to $\chi$, due to the annihilations of
particles with mass $m < x_f m_\chi$~\cite{oss} and $x_f = T_f/m_\chi$ is
proportional to the freeze-out temperature. The coefficients $a$ and $b$
are related to the partial wave expansion of the cross-section, $\sigma v = a + b x +
\dots $. Eq. (\ref{relic} ) results in a very good approximation to the relic density
expect near s-channel annihilation poles,  thresholds and in regions where the LSP is
nearly degenerate with the next lightest supersymmetric particle\cite{gs}.

\subsection{The CMSSM after WMAP}

\begin{figure}[h]
\includegraphics[height=2.2in]{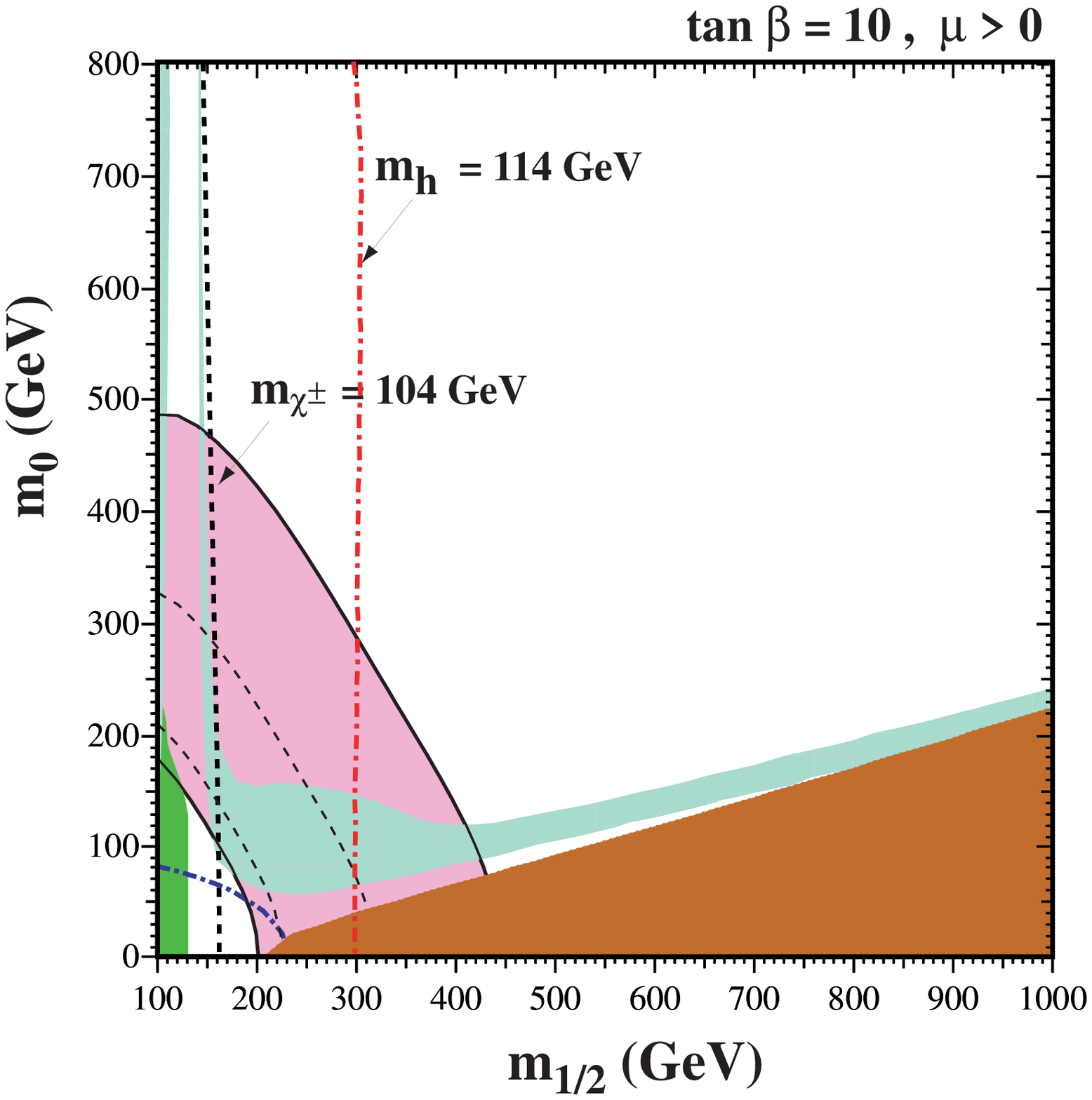}
\includegraphics[height=2.2in]{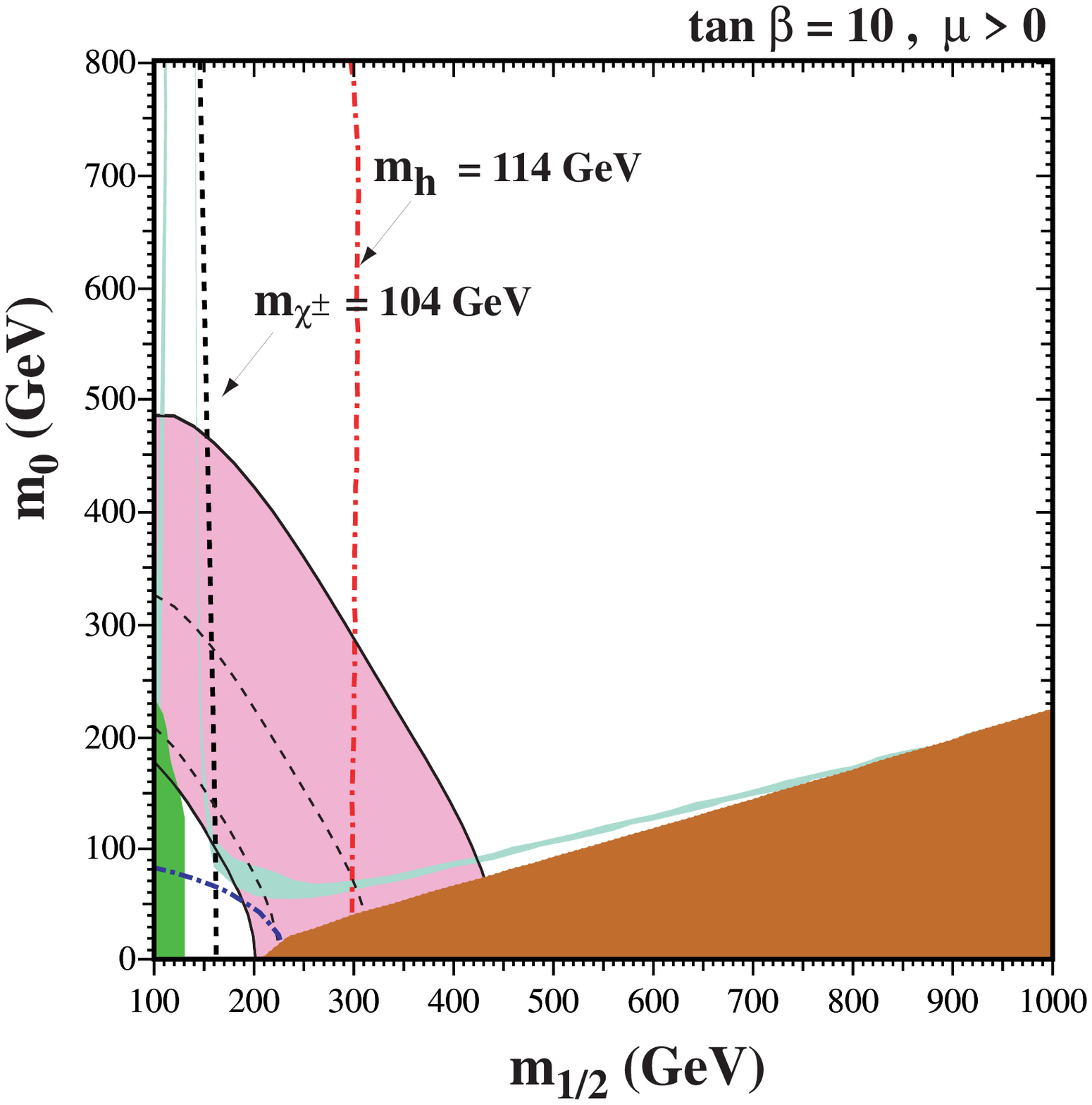}
\caption{\label{fig:UHM}
{\it The $(m_{1/2}, m_0)$ planes for  (a) $\tan \beta = 10$ and  $\mu > 0$, 
assuming $A_0 = 0, m_t = 175$~GeV and
$m_b(m_b)^{\overline {MS}}_{SM} = 4.25$~GeV. The near-vertical (red)
dot-dashed lines are the contours $m_h = 114$~GeV, and the near-vertical (black) dashed
line is the contour $m_{\chi^\pm} = 104$~GeV. Also
shown by the dot-dashed curve in the lower left is the corner
excluded by the LEP bound of $m_{\tilde e} > 99$ GeV. The medium (dark
green) shaded region is excluded by $b \to s
\gamma$, and the light (turquoise) shaded area is the cosmologically
preferred regions with \protect\mbox{$0.1\leq\ohsq\leq 0.3$}. In the dark
(brick red) shaded region, the LSP is the charged ${\tilde \tau}_1$. The
region allowed by the E821 measurement of $a_\mu$ at the 2-$\sigma$
level, is shaded (pink) and bounded by solid black lines, with dashed
lines indicating the 1-$\sigma$ ranges. In (b), the relic density is restricted to the
range $0.094 < \ohsq < 0.129$. }}
\end{figure}

For a given value of $\tan \beta$, $A_0$,  and $sgn(\mu)$, the resulting regions of 
acceptable relic density and which satisfy the phenomenological constraints
can be displayed on the  $m_{1/2} - m_0$ plane.
In Fig. \ref{fig:UHM}a,  the light
shaded region corresponds to that portion of the CMSSM plane
with $\tan \beta = 10$, $A_0 = 0$, and $\mu > 0$ such that the computed
relic density yields \mbox{$0.1<\ohsq<0.3$}.
At relatively low values of 
$m_{1/2}$ and $m_0$, there is a large  `bulk' region  which tapers off
as $\m12$ is increased.  At higher values of $m_0$,  annihilation cross sections
are too small to maintain an acceptable relic density and $\ohsq > 0.3$.
Although sfermion masses are also enhanced at large $\m12$ (due to RGE running),
co-annihilation processes between the LSP and the next lightest sparticle 
(in this case the $\st$) enhance the annihilation cross section and reduce the
relic density.  This occurs when the LSP and NLSP are nearly degenerate in mass.
The dark shaded region has $m_{\st}< m_\chi$
and is excluded.   Neglecting coannihilations, one would find an upper
bound of $\sim450\gev$ on $\m12$, corresponding to an upper bound of
roughly $200\gev$ on $m_{\tilde B}$.  
The effect of coannihilations is
to create an allowed band about 25-50 $\gev$ wide in $m_0$ for $\m12 \la
1400\gev$, which tracks above the $\mst=m_\chi$ contour\cite{efo}.

Also shown in Fig. \ref{fig:UHM}a are
the relevant phenomenological constraints.  
These include the limit on the chargino mass: $m_{\chi^\pm} > 104$~GeV~\cite{LEPsusy}, 
on the selectron mass: $m_{\tilde e} > 99$~GeV~ \cite{LEPSUSYWG_0101} 
and on the Higgs mass: $m_h >
114$~GeV~\cite{LEPHiggs}. The former two constrain $m_{1/2}$ and $m_0$ directly
via the sparticle masses, and the latter indirectly via the sensitivity of
radiative corrections to the Higgs mass to the sparticle masses,
principally $m_{\tilde t, \tilde b}$. 
{\tt FeynHiggs}~\cite{FeynHiggs} is used for the calculation of $m_h$. 
The Higgs limit  imposes important constraints
principally on $m_{1/2}$ particularly at low $\tan \beta$.
Another constraint is the requirement that
the branching ratio for $b \rightarrow
s \gamma$ is consistent with the experimental measurements\cite{bsgex}. 
These measurements agree with the Standard Model, and
therefore provide bounds on MSSM particles\cite{gam,bsgth},  such as the chargino and
charged Higgs masses, in particular. Typically, the $b\rightarrow s\gamma$
constraint is more important for $\mu < 0$, but it is also relevant for
$\mu > 0$,  particularly when $\tan\beta$ is large. The constraint imposed by
measurements of $b\rightarrow s\gamma$ also excludes small
values of $m_{1/2}$. 
Finally, there are
regions of the $(m_{1/2}, m_0)$ plane that are favoured by
the BNL measurement\cite{newBNL} of $g_\mu - 2$ at the 2-$\sigma$ level, corresponding to 
a deviation  from the Standard Model 
calculation\cite{Davier} using $e^+ e^-$ data.  One should be  however 
aware that this constraint is still under active discussion.

The preferred range of the relic LSP density  has been altered
significantly by the recent improved determination of the allowable range
of the cold dark matter density obtained by combining WMAP and other
cosmological data:  $0.094 < \Omega_{CDM} < 0.129$ at the 2-$\sigma$
level\cite{wmap}.  In the second panel of Fig.
\ref{fig:UHM}, we see the effect of imposing the WMAP range on the
neutralino density\cite{eoss,Baer,morewmap}.
We see immediately that (i) the cosmological regions are
generally much narrower, and (ii) the `bulk' regions at small $m_{1/2}$
and $m_0$ have almost disappeared, in particular when the laboratory
constraints are imposed. Looking more closely at the coannihilation
regions, we see that (iii) they are significantly truncated as well as
becoming much narrower, since the reduced upper bound on $\Omega_\chi h^2$
moves the tip where $m_\chi = m_{\tilde \tau}$ to smaller $m_{1/2}$
so that the upper limit is now $m_{1/2} \la 950$ GeV or $m_\chi \la 400$ GeV. 

\begin{figure}[h]
\includegraphics[height=2.2in]{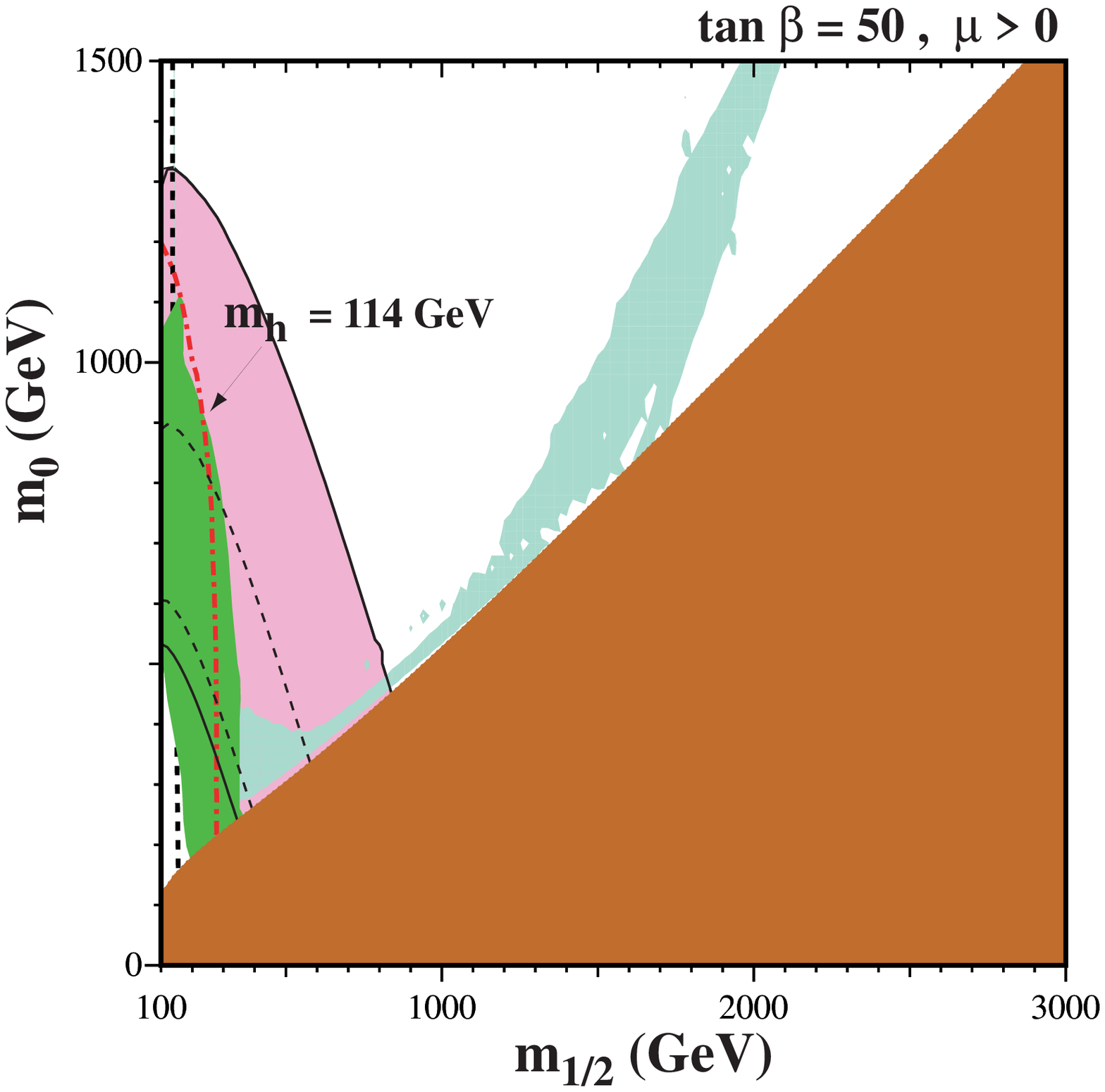}
\includegraphics[height=2.2in]{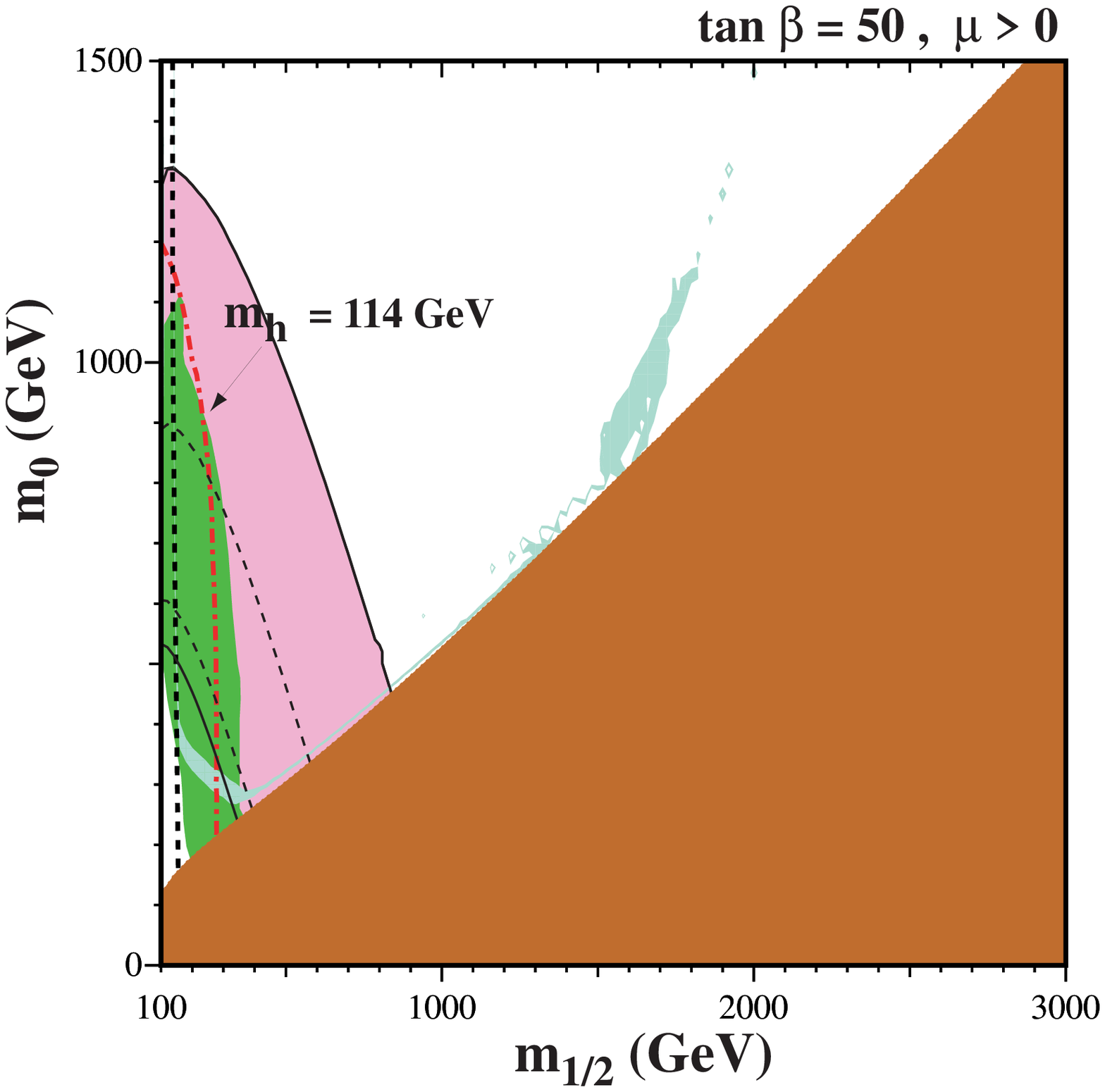}
\caption{\label{fig:UHM50}
{\it As in Fig. \protect{\ref{fig:UHM}} for $\tan \beta = 50$.}}
\end{figure}

Another
mechanism for extending the allowed CMSSM region to large
$m_\chi$ is rapid annihilation via a direct-channel pole when $m_\chi
\sim {1\over 2} m_{A}$~\cite{funnel,EFGOSi}. Since the heavy scalar and
pseudoscalar Higgs masses decrease as  
$\tan \beta$ increases, eventually  $ 2 m_\chi \simeq  m_A$ yielding a
`funnel' extending to large
$m_{1/2}$ and
$m_0$ at large
$\tan\beta$, as seen in the high $\tan \beta$ strips of Fig.~\ref{fig:UHM50}.
As one can see, the impact of the Higgs mass constraint is reduced (relative to 
the case with $\tan \beta = 10$) while that of $b \to s \gamma$ is enhanced.

\begin{figure}
\begin{center}
\includegraphics[height=2.6in]{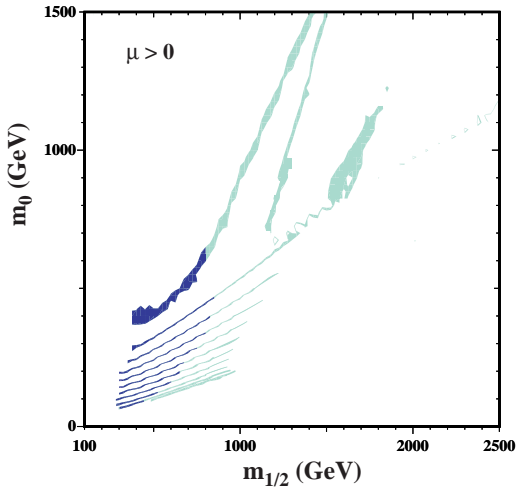}
\end{center}
\caption{\label{fig:strips}\it
The strips display the regions of the $(m_{1/2}, m_0)$ plane that are
compatible with $0.094 < \Omega_\chi h^2 < 0.129$ and the laboratory
constraints for $\mu > 0$ and $\tan \beta = 5, 10, 15, 20, 25, 30,
35, 40, 45, 50, 55$. The parts of the strips compatible with $g_\mu - 2$ 
at the 2-$\sigma$ level have darker shading.
}
\end{figure}

Shown in Fig.~\ref{fig:strips} are the WMAP lines\cite{eoss} of the $(m_{1/2}, m_0)$
plane allowed by the new cosmological constraint $0.094 < \Omega_\chi h^2
< 0.129$ and the laboratory constraints listed above, for $\mu > 0$ and
values of $\tan \beta$ from 5 to 55, in steps $\Delta ( \tan \beta ) = 5$.
We notice immediately that the strips are considerably narrower than the
spacing between them, though any intermediate point in the $(m_{1/2},
m_0)$ plane would be compatible with some intermediate value of $\tan
\beta$. The right (left) ends of the strips correspond to the maximal
(minimal) allowed values of $m_{1/2}$ and hence $m_\chi$. 
The lower bounds on $m_{1/2}$ are due to the Higgs 
mass constraint for $\tan \beta \le 23$, but are determined by the $b \to 
s \gamma$ constraint for higher values of $\tan \beta$.

\begin{figure}
\begin{center}
\includegraphics[height=2.6in]{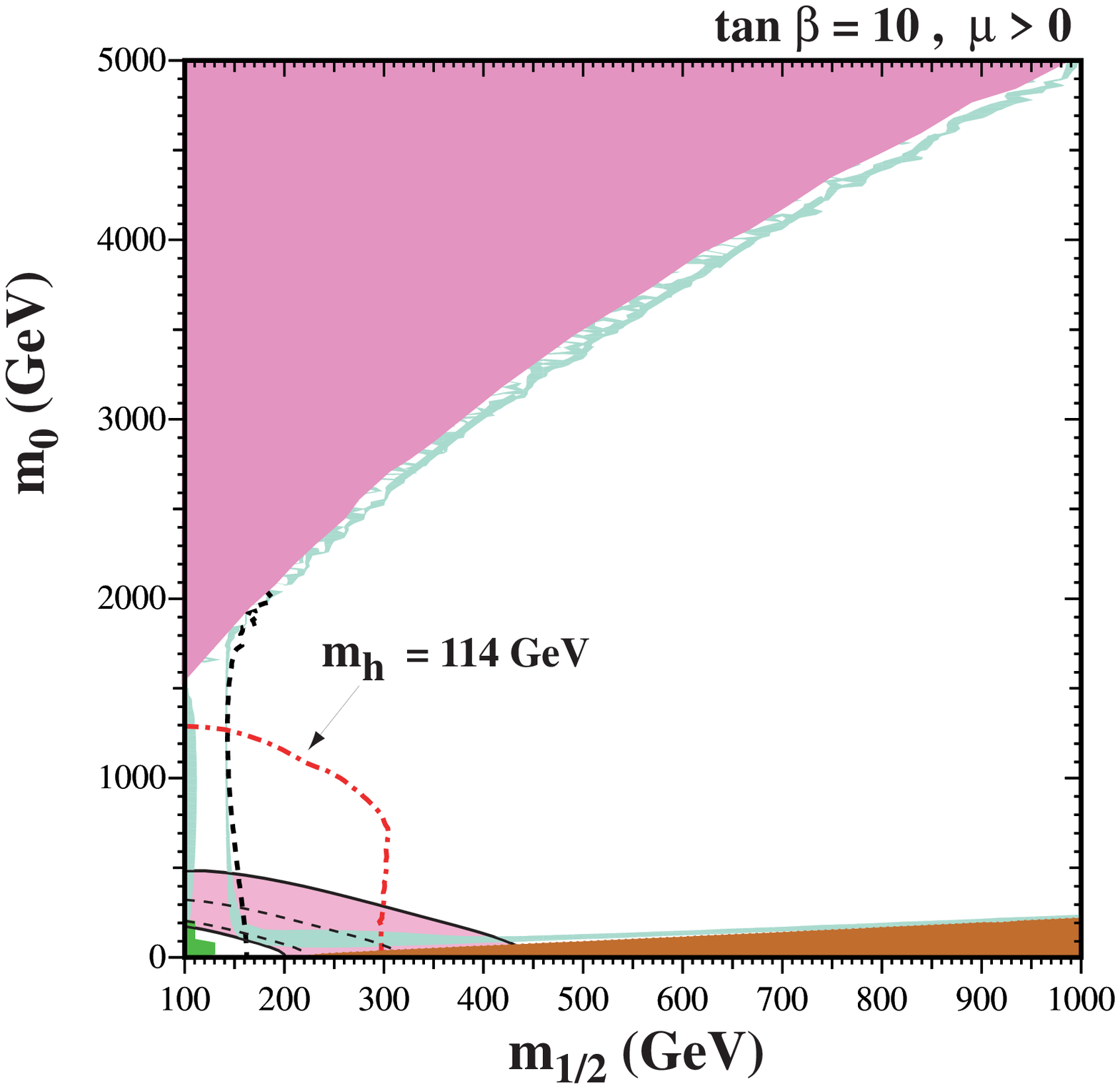}
\end{center}
\caption{\label{figfp}\it
As in Fig. \protect\ref{fig:UHM}a, where the range in $m_0$ is extended to
5 TeV.  In the shaded region at very high $m_0$, there are no solutions 
for $\mu$ which respect the low energy electroweak symmetry breaking conditions.
}
\end{figure}

\begin{figure}
\begin{center}
\includegraphics[height=2.6in]{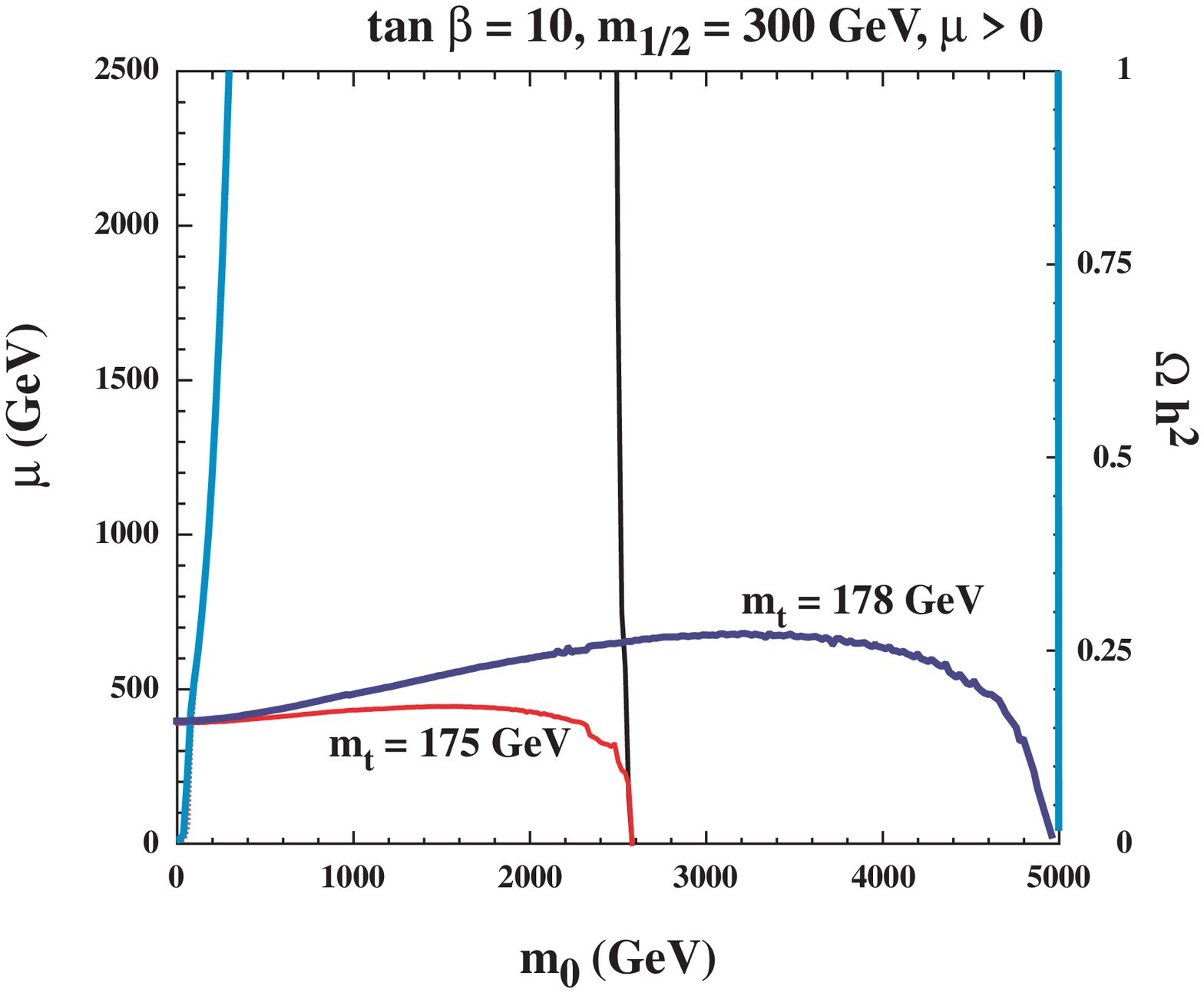}
\end{center}
\caption{\label{fignofp}\it
The value of $\mu$ as a function of $m_0$ for fixed $m_{1/2} = 300$ GeV
and $\tan \beta = 10$ for two choices of $m_t$ as indicated. The scale on the right
gives the value of $\Omega h^2$.  The curves corresponding to this is scale
rise sharply at low $m_0$ to values much larger than 1.  For $m_t = 175$ GeV and 
$m_0 \approx 2500$ GeV, the value of $\Omega h^2$ drops to acceptable values 
when $\mu$ becomes small.  When the $m_t = 178$ GeV,  
$\Omega h^2$ drops at $m_0 \approx 5000$ GeV.}
\end{figure}

Finally, there is one additional region of acceptable relic density known as the
focus-point region\cite{fp}, which is found
at very high values of $m_0$. An example showing this region is found in Fig. \ref{figfp},
plotted for $\tan \beta = 10$, $\mu > 0$, and $m_t = 175$ TeV.
As $m_0$ is increased, the solution for $\mu$ at low energies as determined
by the electroweak symmetry breaking conditions eventually begins to drop. 
When $\mu \la m_{1/2}$, the composition of the LSP gains a strong Higgsino
component and as such the relic density begins to drop precipitously.  These effects
are both shown in  Fig. \ref{fignofp} where the value of $\mu$ and $\Omega h^2$ 
are plotted as a function of $m_0$ for fixed $m_{1/2} = 300$ GeV and $\tan \beta = 10$. 
As $m_0$ is increased further, there are no longer any solutions for $\mu$.  This 
occurs in the shaded region in the upper left corner of Fig. \ref{figfp}. 

Fig. \ref{fignofp} also exemplifies the degree of fine tuning associated with the
focus-point region.  While the position of the focus-point region in the $m_0, m_{1/2}$
plane is not overly sensitive to supersymmetric parameters, it is highly sensitive to 
the top quark Yukawa coupling which contributes to the evolution of $\mu$~\cite{rs,ftuning}. 
As one can see in the figure, a change in $m_t$ of 3 GeV produces a shift
of about 2.5 TeV in $m_0$.  Note that the position of the focus-point region
is also highly sensitive to the value of $A_0/m_0$. In Fig. \ref{fignofp}, $A_0 = 0$
was chosen.  For $A_0/m_0 = 0.5$, the focus point shifts from 2.5 to 4.5 TeV
and moves to larger $m_0$ as $A_0/m_0$ is increased.

\subsection{A Likelihood analysis of the CMSSM}

In displaying acceptable regions of cosmological density
in the $m_0, m_{1/2}$ plane,  it has been assumed that the input
parameters are known with perfect accuracy so that the relic density can
be calculated precisely.  While all of the beyond the standard model
parameters are completely unknown and therefore carry no
formal uncertainties, standard model parameters such as 
the top and bottom Yukawa couplings are known but do 
carry significant uncertainties.  

The optimal way to combine the various constraints (both phenomenological and cosmological)
is via a likelihood
analysis. When performing such an analysis,
in addition to the formal experimental errors, it is also essential to
take into account theoretical errors, which introduce systematic
uncertainties that are frequently non-negligible. 
Recently, we have preformed an extensive likelihood analysis of the CMSSM\cite{eoss4}.

The interpretation of the combined Higgs likelihood, ${ L}_{exp}$,  
in the $(m_{1/2},
m_0)$ plane depends on uncertainties in the theoretical calculation of
$m_h$. These include the experimental error in $m_t$ and (particularly at
large $\tan \beta$) $m_b$, and theoretical uncertainties associated with
higher-order corrections to $m_h$. Our default assumptions are that $m_t =
175 \pm 5$~GeV for the pole mass, and $m_b = 4.25 \pm 0.25$~GeV for the
running $\overline {MS}$ mass evaluated at $m_b$ itself.
The theoretical uncertainty in $m_h$,  $\sigma_{th}$,  is dominated by
the experimental uncertainties in $m_{t,b}$, which are 
treated as uncorrelated Gaussian errors:
\begin{equation}
{\sigma_{th}}^2 =  \left( \frac{\partial m_h}{\partial m_t} \right)^2 \Delta 
m_t^2 + \left( \frac{\partial m_h}{\partial m_b} \right)^2 \Delta m_b^2 \,.
\label{eq:sigmath}
\end{equation}
Typically, we find that $(\partial m_h/\partial m_t) \sim 0.5$, so that 
$\sigma_{th}$ is roughly 2-3 GeV.

The
combined experimental likelihood, ${\mathcal L}_{exp}$, from 
direct searches at LEP~2 and a global
electroweak fit is then convolved with a theoretical likelihood
(taken as a Gaussian) with uncertainty given by $\sigma_{th}$ from
(\ref{eq:sigmath}) above. Thus, we define the
total Higgs likelihood function, ${\mathcal L}_h$, as
\begin{equation}
{\mathcal L}_h(m_h) = { {\mathcal N} \over {\sqrt{2 \pi}\, \sigma_{th}  }}
 \int d m^{\prime}_h \,\, {\mathcal L}_{exp}(m^{\prime}_h)
 \,\, e^{-(m^{\prime}_h-m_h)^2/2 \sigma_{th}^2 }\, ,
\label{eq:higlik}
\end{equation}
where ${\mathcal N}$ is a factor that normalizes  the experimental likelihood
distribution. 
In addition to the Higgs likelihood function, we  have included the likelihood 
function based on $b \to s \gamma$. 
While the likelihood function based on the measurements of the 
anomalous magnetic moment of the muon was considered in~\cite{eoss4}, 
it will not be discussed here. 

Finally, in calculating the likelihood of the CDM density, we take into
account the contribution of the uncertainties in $m_{t,b}$. 
We will see that the theoretical uncertainty plays a
very significant role in this analysis. The likelihood for $\Omega h^2$ is therefore,
\begin{equation}
{\mathcal L}_{\Omega h^2}=\frac{1}{\sqrt{2 \pi} \sigma}
e^{-({\Omega h^2}^{th}-{\Omega h^2}^{exp})^2/2 \sigma^2} \,,
\label{eq:likamu}
\end{equation} 
where $\sigma^2=\sigma_{exp}^2+\sigma_{th}^2$, with
$\sigma_{exp}$ taken from the WMAP\cite{wmap}  result and $\sigma_{th}^2$
from (\ref{eq:sigmath}), replacing $m_h$ by $\Omega h^2$.

The total likelihood function is computed by combining all the components
described above: 
\begin{equation} {\mathcal L}_{tot} = {\mathcal L}_h \times {\mathcal
L}_{bs\gamma} \times {\mathcal L}_{\Omega_\chi h^2} (\times {\mathcal L}_{a_\mu})
\end{equation}
The likelihood function in the CMSSM can be considered  a function of
two variables, ${\mathcal L}_{tot}(m_{1/2},m_0)$, where $m_{1/2}$ and $m_0$
are the unified GUT-scale gaugino and scalar masses respectively.

\begin{figure}[h]
\includegraphics[height=2.3in]{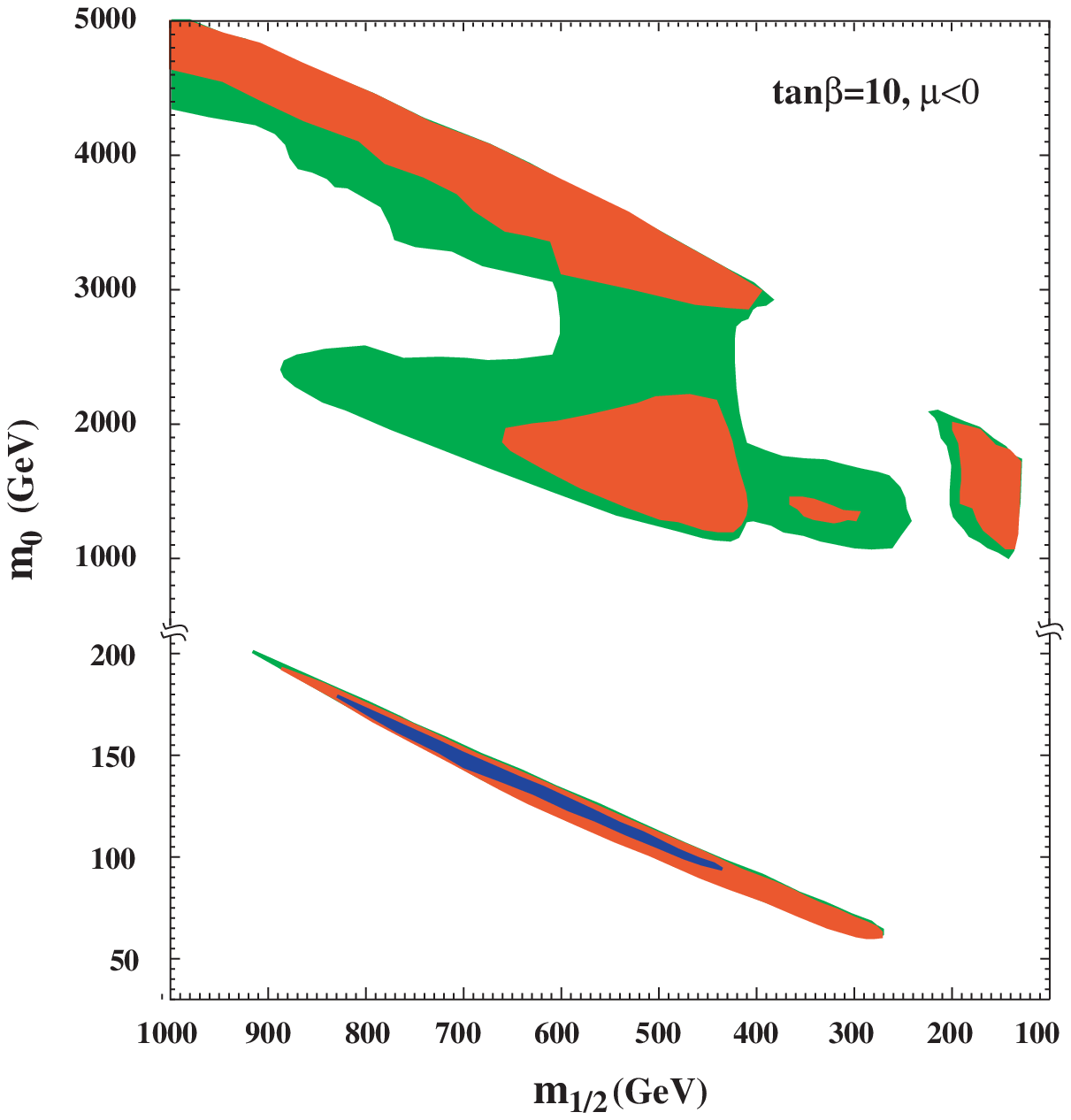}
\hspace {-.17in}
\includegraphics[height=2.25in]{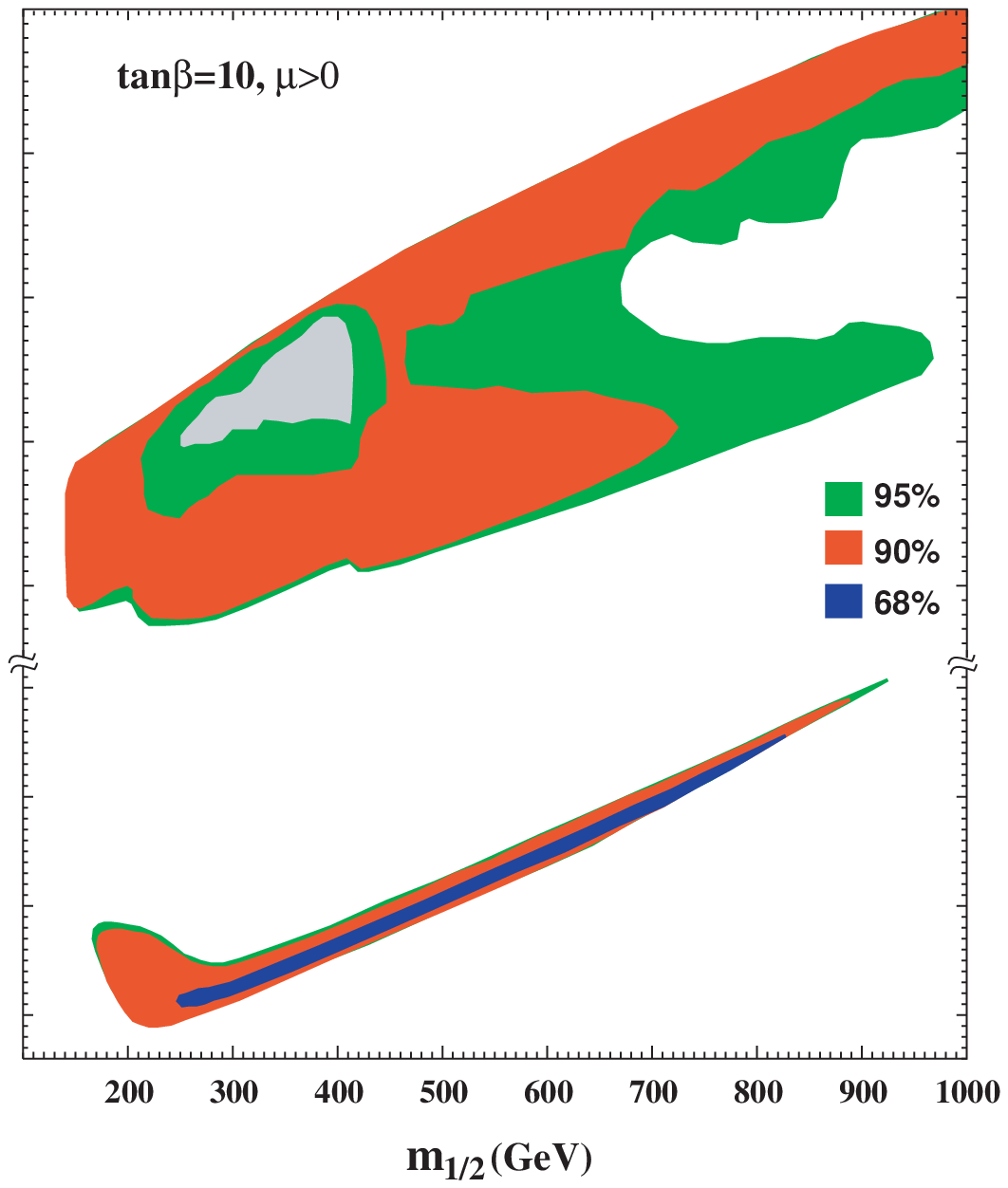}
\caption{\label{fig:contours}
{\it Contours of the likelihood at the 68\%, 90\% and 95\% levels for $\tan 
\beta = 10$, $A_0 = 0$ and $\mu > 0$ (left panel) or $\mu < 0$ (right 
panel), calculated 
using information of $m_h$, $b \to s \gamma$ and $\Omega_{CDM} h^2$ and 
the current uncertainties in $m_t$ and $m_b$. }}
\end{figure}

Using a fully normalized likelihood function ${\mathcal L}_{tot}$ obtained
by combining both signs of $\mu$ for each value of $\tan \beta$, we can
determine the regions in the $(m_{1/2}, m_0)$ planes which correspond to
specific CLs as shown in
Fig.~\ref{fig:contours}.  The darkest (blue), intermediate (red) and lightest
(green) shaded regions are, respectively, those where the likelihood is
above 68\%, above 90\%, and  above 95\%.

\begin{figure}[h]
\includegraphics[height=2.3in]{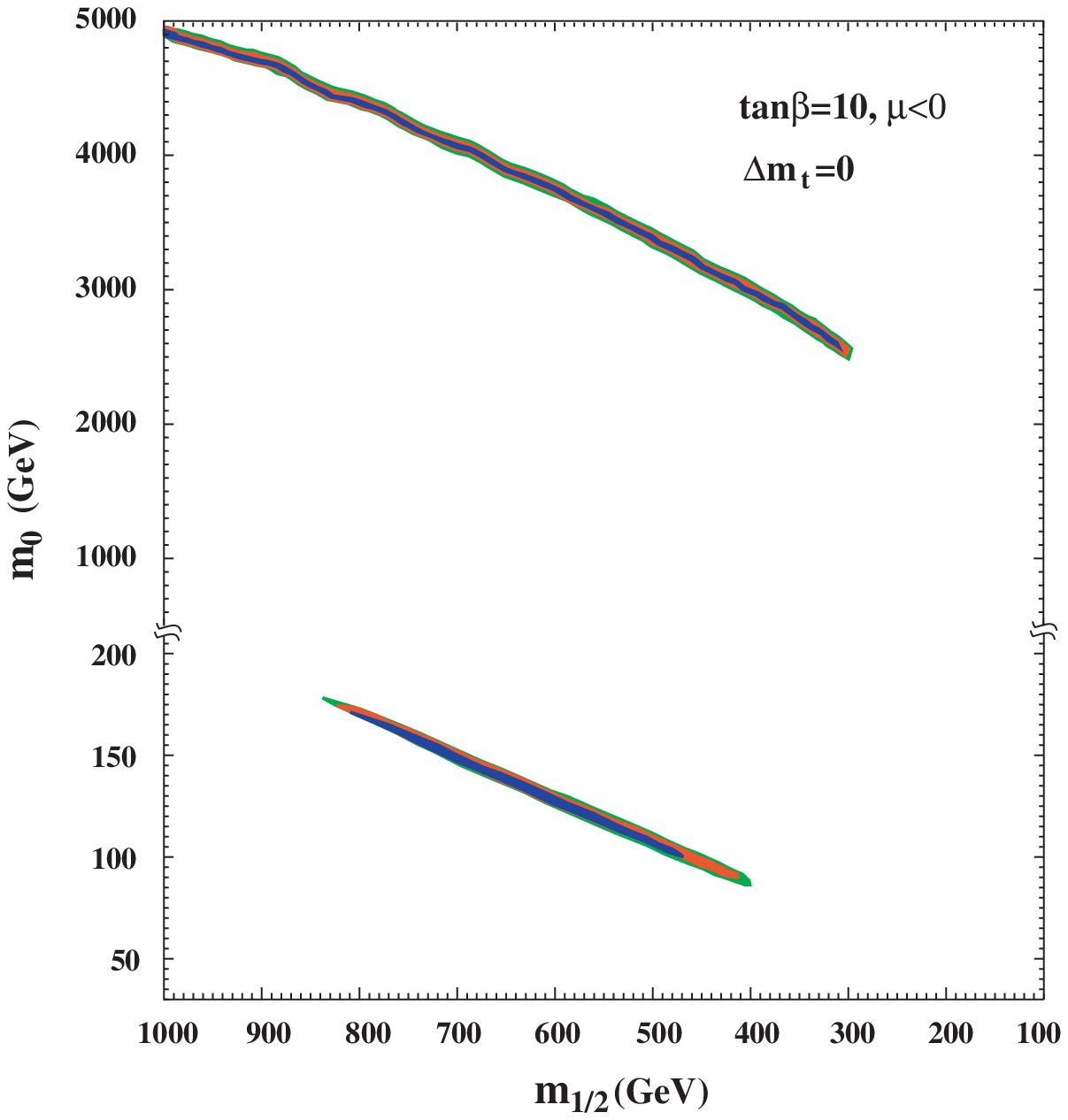}
\hspace {-.17in}
\includegraphics[height=2.25in]{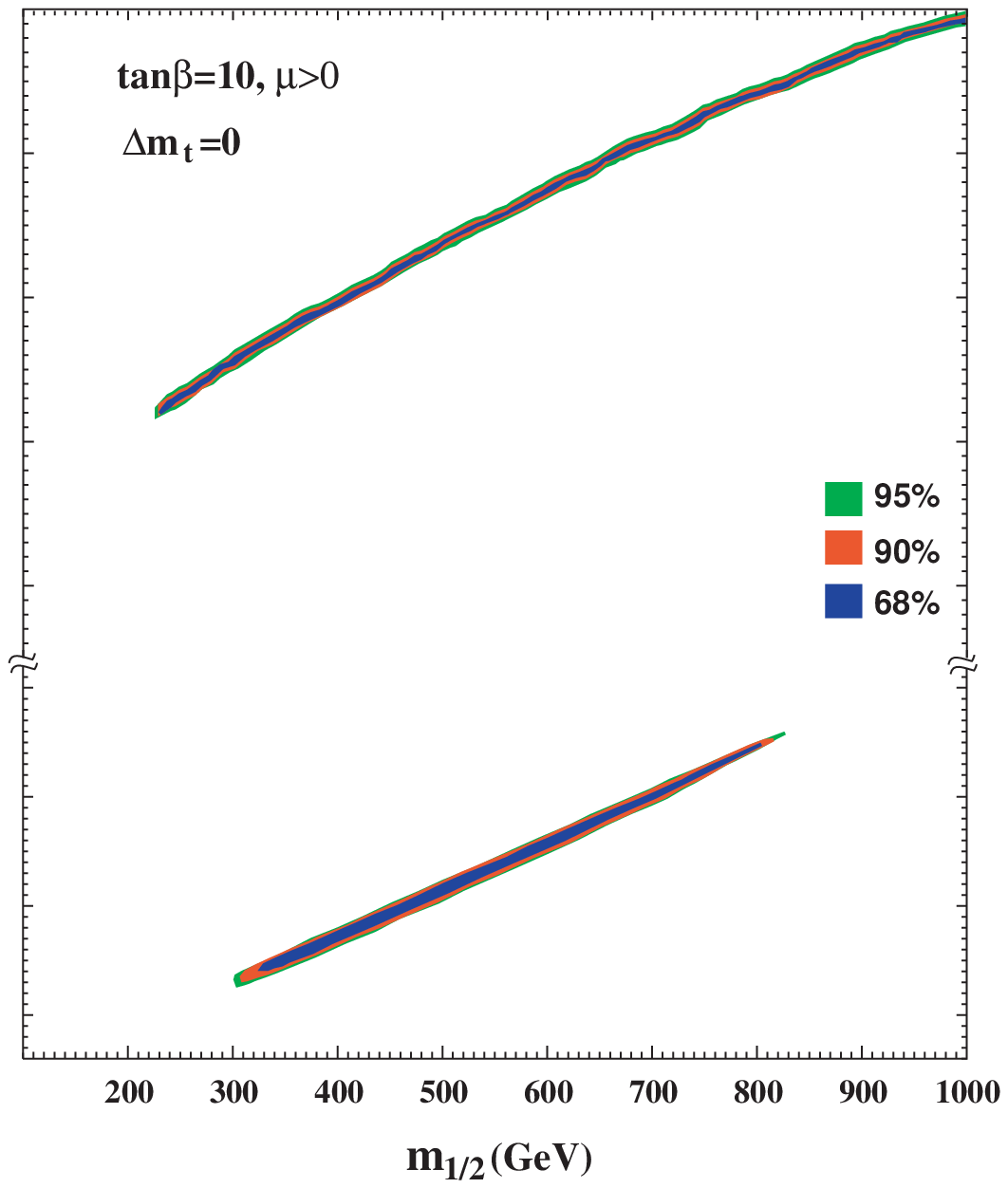}
\caption{\label{fig:contourswithoutmt}
{\it As in Fig.~\protect\ref{fig:contours} but assuming zero uncertainty in 
$m_t$.}}
\end{figure}

The bulk region is more apparent in the right panel of
Fig.~\ref{fig:contours} for $\mu > 0$ than it would be if the experimental
error in $m_t$ and the theoretical error in $m_h$ were neglected. 
Fig.~\ref{fig:contourswithoutmt} complements the previous figures by 
showing the likelihood functions as they would appear if there were no 
uncertainty in $m_t$, keeping the other inputs the same. 
We see that, in this case, both the 
coannihilation and focus-point strips rise above the 68\% CL.

\subsection{Beyond the CMSSM}

\begin{figure}[hbtp]
\includegraphics[height=2.2in]{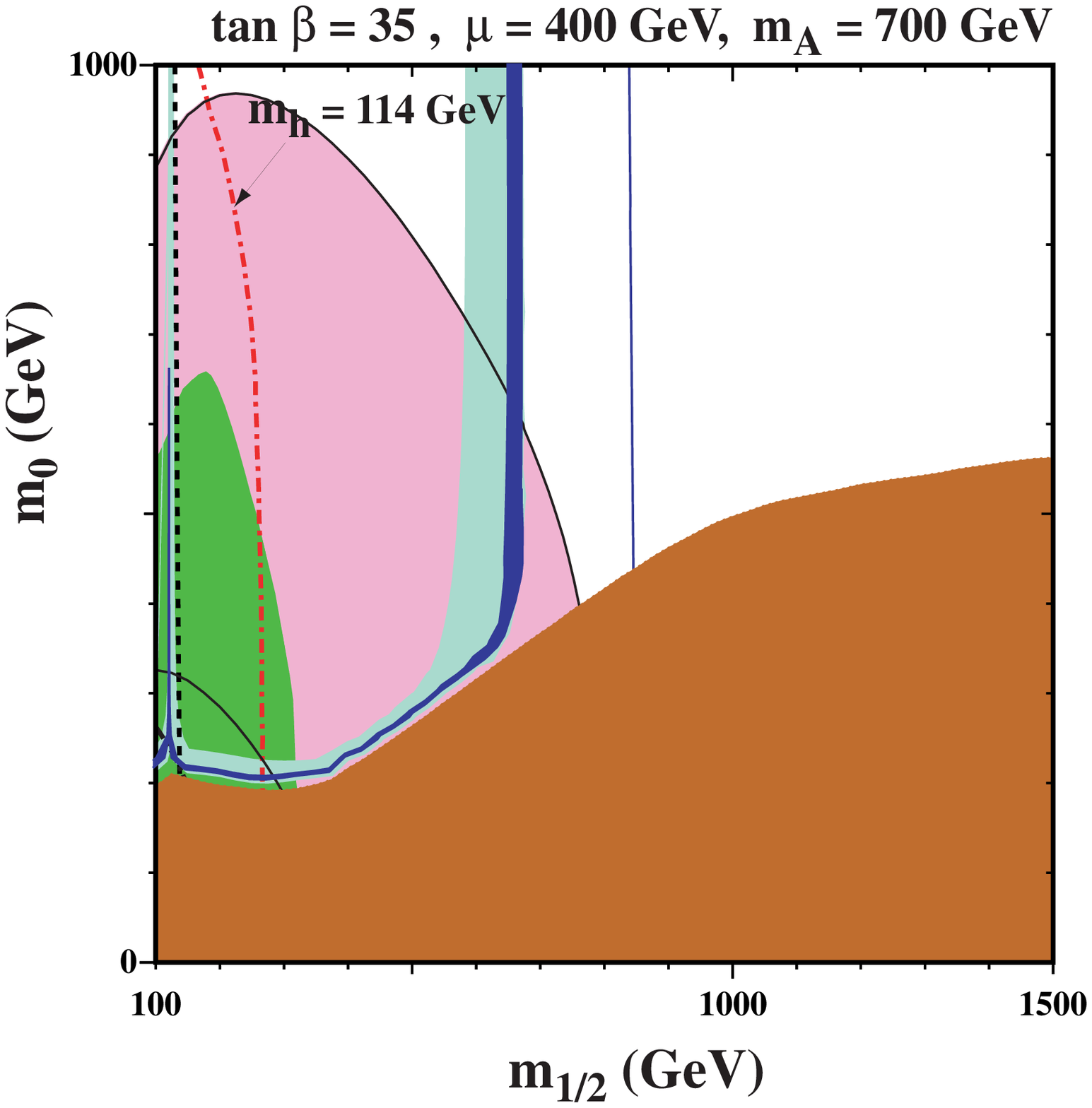}
\includegraphics[height=2.2in]{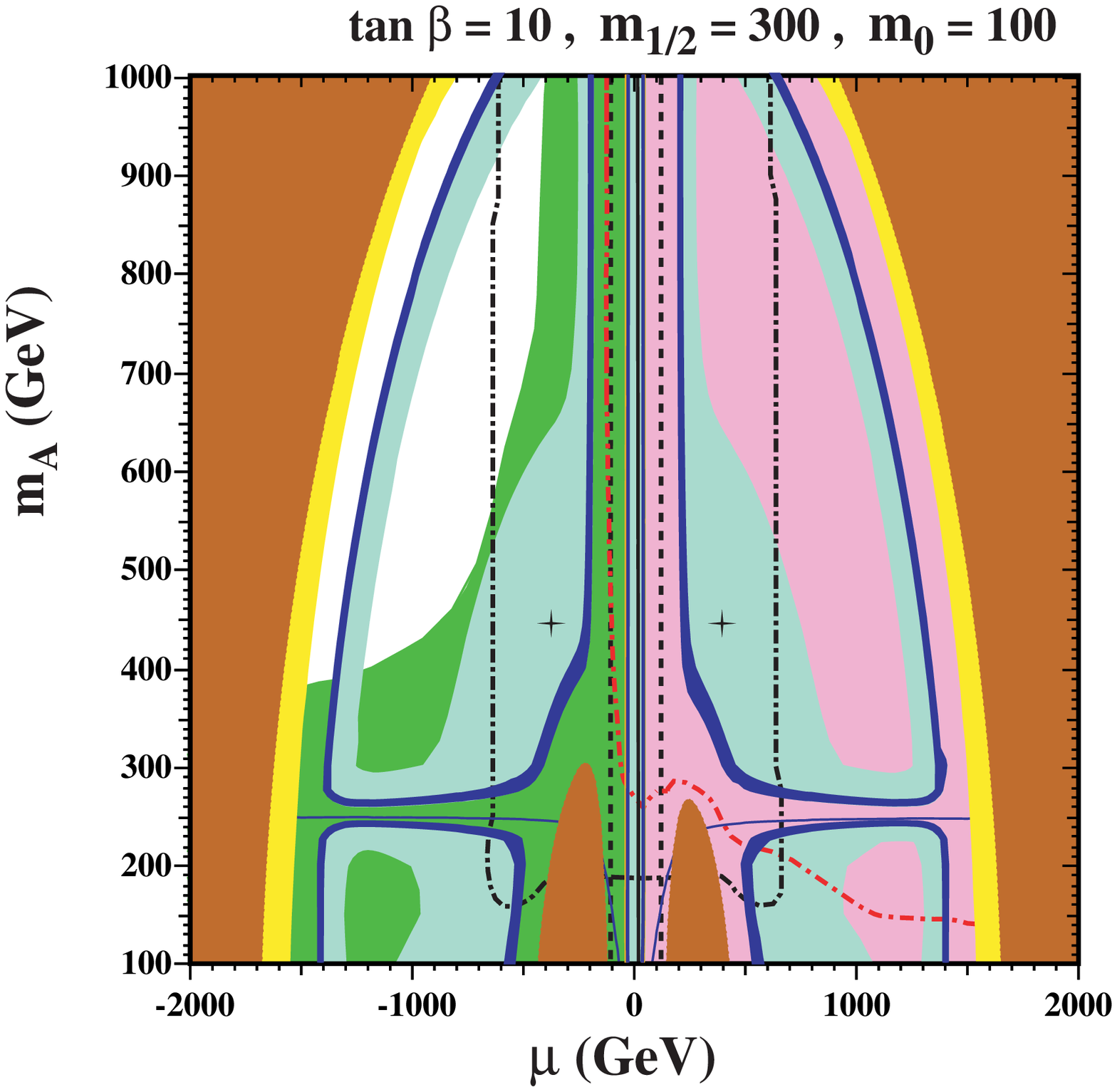}
\caption{\it   a) The NUHM $(m_{1/2}, m_0)$ plane for $\tan \beta = 35$, (a) $\mu = 400$~GeV 
and $m_{A} = 700$~GeV  b)the NUHM $(\mu, m_A)$ plane for $\tan \beta = 10$, $m_0
= 100$~GeV and $m_{1/2} = 300$~GeV,  with $A_0 = 0$.
The (red)
dot-dashed lines are the contours $m_h = 114$~GeV, and the near-vertical
(black) dashed lines are the contours $m_{\chi^\pm} = 103.5$~GeV. The
dark (black) dot-dashed lines indicate the GUT stability constraint. Only
the areas inside these curves (small $\mu$) are allowed by this
constraint. The light (turquoise) shaded areas are the cosmologically
preferred regions with
\protect\mbox{$0.1\leq\ohsq\leq 0.3$}. 
The darker (blue) portion of this region corresponds to the 
WMAP densities.  The dark (brick red) shaded
regions is excluded because the stau is the LSP, 
and the lighter (yellow) shaded regions is excluded because
the LSP is a sneutrino. The medium
(green) shaded region is excluded by $b \to s \gamma$.
The regions allowed by the $g-2$ constraint 
are shaded (pink) and bounded by solid black lines. The solid (blue)
curves correspond to $m_\chi = m_A/2$. }
	\label{muma}
\end{figure}

The results of  the CMSSM described in the previous sections are based heavily
on the assumptions of universality of the supersymmetry breaking parameters.
One of the simplest generalizations of this model relaxes the assumption of
universality of the Higgs soft masses and is known as the NUHM\cite{eos3}
In this case, the 			
input parameters include $ \mu$ and $m_A,$ in addition to the standard CMSSM inputs.
In order to switch $\mu$ and $m_A$ from outputs to inputs,
the two soft Higgs masses,  $m_1,
m_2$ can no longer be set equal to $m_0$ and instead are 
calculated from the electroweak symmetry breaking conditions. The NUHM parameter space was recently analyzed\cite{eos3} and a sample of the results
are shown in Fig. \ref{muma}.

In the left panel of Fig. \ref{muma}, we see a $m_{1/2},m_0$ plane
with a relative low value of $\mu$.  In this case, an allowed region is found
when the LSP contains a non-negligible Higgsino component
which moderates the relic density independent of $m_0$.  To the right of this region,
the relic density is too small.  In the right panel, we see an example of the $m_A,\mu$
plane.  The crosses correspond to CMSSM points.  In this single pane, we
see examples of acceptable cosmological regions corresponding to the 
bulk region, co-annihilation region and s-channel annihilation through the Higgs
pseudo scalar.

Rather than relax the CMSSM, it is in fact possible to further constrain the model.
While the CMSSM models described above are certainly
mSUGRA inspired, minimal supergravity models
can be argued to be still more predictive.  
In the simplest version of the theory\cite{pol} where supersymmetry is
broken in a hidden sector, the
universal trilinear soft
supersymmetry-breaking terms  are $A = (3 -
\sqrt{3}) m_{0}$ and bilinear
soft supersymmetry-breaking term is $B = (2 - \sqrt{3}) m_{0}$, i.e., a
special case of a general relation  between $B$ and
$A$, $B_0 = A_0 - m_0$. 

\begin{figure}
\includegraphics[height=2.3in]{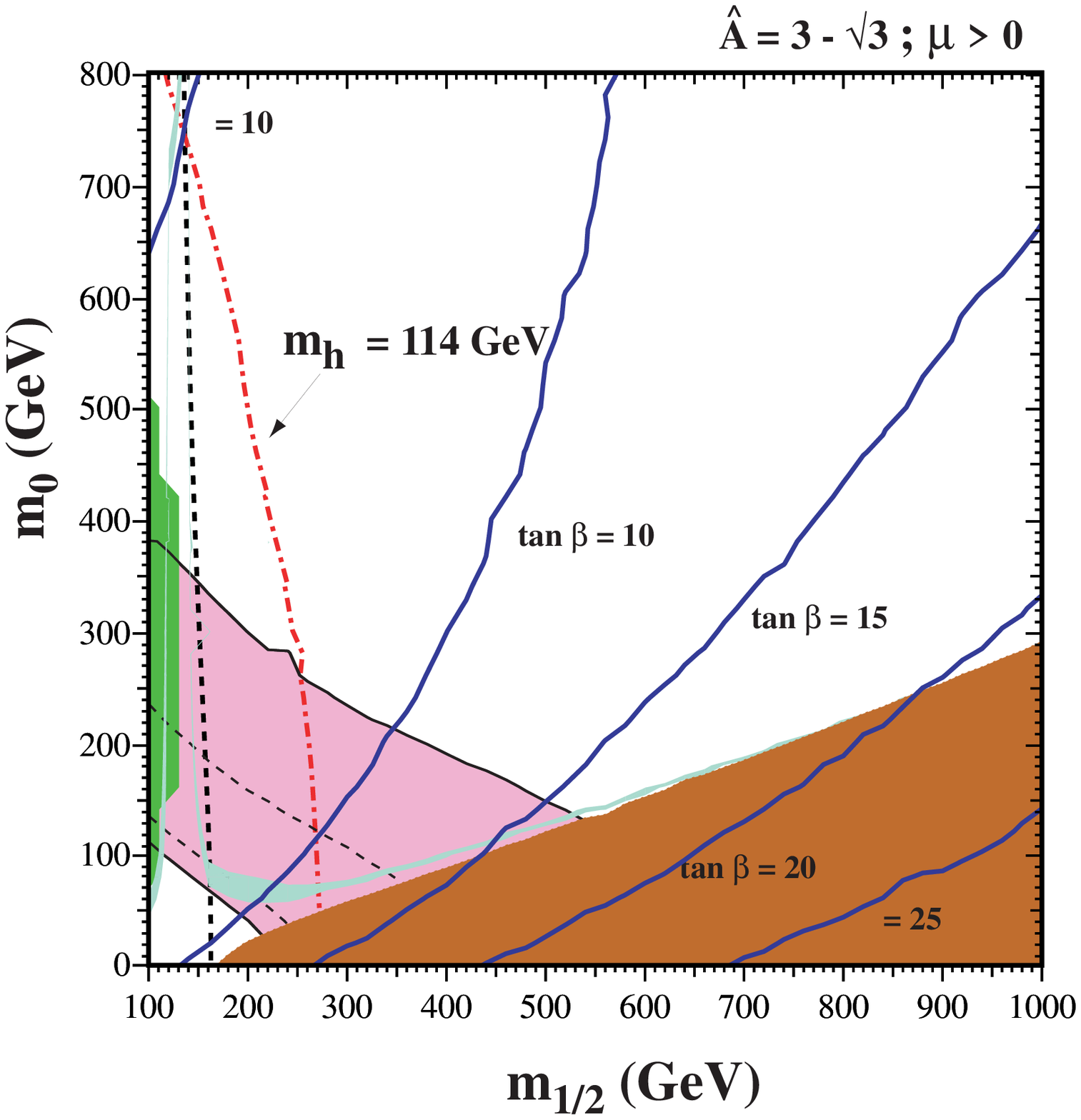}
\includegraphics[height=2.3in]{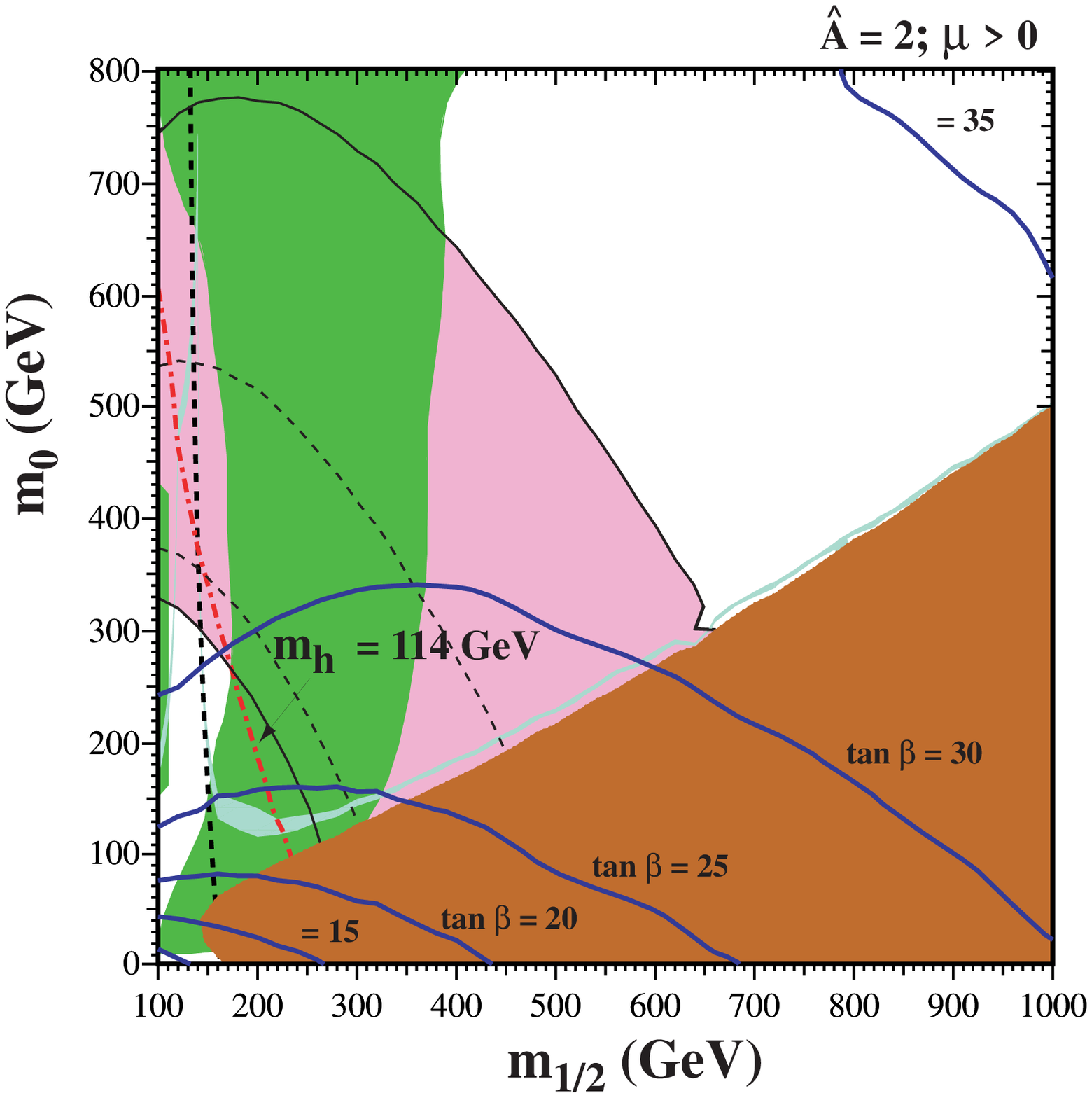}
\caption{\it
Examples of $(m_{1/2}, m_0)$ planes with contours of $\tan \beta$ 
superposed, for $\mu > 0$ and (a)  the simplest Polonyi model with ${\hat A} = 3 - 
\sqrt{3}, {\hat B} = {\hat A} -1$ and (b) ${\hat A} = 2.0, {\hat B} =
{\hat A} -1$. In each panel, we show the regions excluded by 
the LEP lower limits on MSSM particles, those ruled out by $b
\to s \gamma$ decay (medium green shading), and those 
excluded 
because the LSP would be charged (dark red shading). The region favoured 
by the WMAP range  has light turquoise shading. The region 
suggested by $g_\mu - 2$ is medium (pink) shaded.}
\label{fig:Polonyi}
\end{figure}

Given a relation between $B_0$ and $A_0$, we can no longer use the
standard CMSSM boundary conditions, in which $m_{1/2}$, $m_0$, $A_0$,
$\tan \beta$, and $sgn(\mu)$ are input at the GUT scale with $\mu$ and $B$ 
determined by the electroweak symmetry breaking condition.
Now, one is forced to input $B_0$ and instead $\tan \beta$ is 
calculated from the minimization of the Higgs potential\cite{eoss2}.
In Fig.~\ref{fig:Polonyi}, the contours of $\tan \beta$ (solid
blue lines) in the $(m_{1/2}, m_0)$ planes for two values of ${\hat
A}  = A_0/m_0$, ${\hat B} = B_0/m_0 = {\hat A} - 1$ and the sign of $\mu$ are displayed\cite{eoss2}.

In panel (a) of Fig.~\ref{fig:Polonyi}, we
see that the Higgs constraint combined with the relic density requires $\tan \beta
\ga 11$, whilst the relic density also enforces $\tan \beta \la 
20$. For a given point in the $m_{1/2} - m_0$ plane, the calculated value of $\tan \beta$
increases as ${\hat A}$ increases.
This is seen in panel (b) of
Fig.~\ref{fig:Polonyi}, when ${\hat A} = 2.0$, close to its maximal value
for $\mu > 0$, the $\tan \beta$ contours turn over towards smaller
$m_{1/2}$, and only relatively large values $25 \la \tan \beta \la 35$ are
allowed by the $b \to s \gamma$ and $\Omega_{CDM} h^2$ constraints,
respectively.  For any given value of ${\hat A}$, there is only a relatively narrow
range allowed for $\tan \beta$.

\subsection{Detectability}

\begin{figure}[h]
\centering
\includegraphics[height=3.2in]{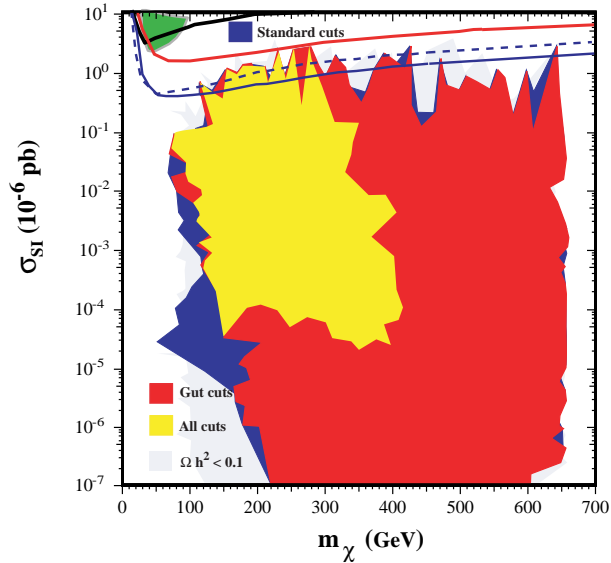}
\caption{\label{fig:Andyall}
{\it 
Ranges of the spin-independent 
cross section in the NUHM.
The ranges allowed by the  
cuts on $\ohsq$, $m_h$ and $b \to s \gamma$ have dark shading, those still 
allowed by the GUT stability cut have medium shading, and those still 
allowed after applying all the cuts including $g_\mu - 2$ have light 
shading.   The pale shaded region
corresponds to the extra area of points with low relic 
densities, whose cross sections have been rescaled appropriately. 
Also shown are the limits from the CDMS\protect\cite{cdms} and Edelweiss\protect\cite{edel}
experiments  as well as the recent CDMSII result \protect\cite{cdms2}
on the neutralino-proton elastic scattering cross section as a function
of the neutralino mass. The CDMSII limit is stronger than the 
Edelweiss limit which is stronger than the previous CDMS limit at higher $m_\chi$. 
The result reported by DAMA \protect\cite{dama} is found in the upper left.}}
\end{figure}

Direct detection techniques
rely on an ample neutralino-nucleon scattering cross-section.
In Fig.~\ref{fig:Andyall}, we display the allowed ranges of 
the spin-independent  cross sections in the NUHM when we
sample randomly $\tan \beta$ as well as the other NUHM parameters\cite{efloso}. 
The raggedness of the boundaries of the shaded regions
reflects the finite sample size. The dark shaded regions includes
all sample points after the constraints discussed above (including the relic
density constraint) have been applied.  
In a random sample, one often hits points which are
are perfectly acceptable at low energy scales but
when the parameters are run to high energies approaching the GUT scale,
one or several of the sparticles mass squared runs negative.  
This has been referred to as the GUT constraint here.
The medium shaded region
embodies those points after the GUT constraint has been applied.
After incorporating all
the cuts, including that motivated by $g_\mu - 2$, we find that the light shaded
region where the
scalar cross section has the range $10^{-6}$~pb $\ga
\sigma_{SI} \ga 10^{-10}$~pb, with
somewhat larger (smaller) values being possible in exceptional cases.

The results from this analysis\cite{efloso} for the scattering cross section in the NUHM
(which by definition includes all CMSSM results)
are compared with the previous CDMS\cite{cdms} and Edelweiss\cite{edel}
bounds as well as the recent CDMSII results\cite{cdms2} in Fig.~\ref{fig:Andyall}. 
 While previous experimental
sensitivities were not strong enough to probe predictions of the NUHM, the current
CDMSII bound has begun to exclude realistic models and it is expected that 
these bounds improve by a factor of about 20.
See ref.~\cite{eoss8} for updated direct detection calculations in  
the MSSM.

\section{Big Bang Nucleosynthesis}

The standard model\cite{wssok} of big bang nucleosynthesis (BBN)
is based on the relatively simple idea of including an extended nuclear
network into a homogeneous and isotropic cosmology.  Apart from the
input nuclear cross sections, the theory contains only a single parameter,
namely the baryon-to-photon ratio,
$\eta$. Other factors, such as the uncertainties in reaction rates, and
the neutron mean-life can be treated by standard statistical and Monte
Carlo techniques\cite{kr,skm,nb,cfo1,cvcr01}.  The theory then allows one to make
predictions (with well-defined uncertainties) of the abundances of the
light elements, D,
\he3, \he4, and \li7.

\subsection{Theory}
Conditions for the synthesis of the light elements were attained in the
early Universe at temperatures  $T \ga $ 1 MeV.  In the early Universe,
the energy density was dominated by radiation with
\begin{equation}
\rho = {\pi^2 \over 30} ( 2 + {7 \over 2} + {7 \over 4}N_\nu) T^4
\label{rho}
\end{equation}
from the contributions of photons, electrons and positrons, and $N_\nu$
neutrino flavors (at higher temperatures, other particle degrees of
freedom should be included as well). At these temperatures, weak
interaction rates were in equilibrium. In particular, the processes
\begin{eqnarray}
n + e^+ & \leftrightarrow  & p + {\bar \nu_e} \nonumber \\
n + \nu_e & \leftrightarrow  & p + e^- \nonumber \\
n  & \leftrightarrow  & p + e^- + {\bar \nu_e} 
\label{beta}
\end{eqnarray}
fix the ratio of
number densities of neutrons to protons. At $T \gg 1$ MeV, $(n/p) \simeq
1$. 

The weak interactions do not remain in equilibrium at lower temperatures.
Freeze-out occurs when the weak interaction rate, $\Gamma_{wk} \sim G_F^2
T^5$ falls below the expansion rate which is given by the Hubble
parameter, $H \sim \sqrt{G_N \rho} \sim T^2/M_P$, where $M_P =
1/\sqrt{G_N} \simeq 1.2
\times 10^{19}$ GeV. The $\beta$-interactions in eq.
(\ref{beta}) freeze-out at about 0.8 MeV.
 As the temperature falls
and approaches the point where the weak interaction rates are no longer
fast enough to maintain equilibrium, the neutron to proton ratio is given
approximately by the Boltzmann factor,
$(n/p)
\simeq e^{-\Delta m/T} \sim 1/6$, where $\Delta m$ is the neutron-proton
mass difference. After freeze-out, free neutron decays drop the ratio
slightly to about 1/7 before nucleosynthesis begins. 

The nucleosynthesis chain begins with the formation of deuterium
by the process, $p+n \rightarrow$ D $+~\gamma$.
However, because of the large number of photons relative to nucleons,
$\eta^{-1} = n_\gamma/n_B \sim 10^{10}$, deuterium production is delayed
past the point where the temperature has fallen below the deuterium
binding energy, $E_B = 2.2$ MeV (the average photon energy in a blackbody
is ${\bar E}_\gamma \simeq 2.7 T$).  This is because there are many
photons in the exponential tail of the photon energy distribution with
energies $E > E_B$ despite the fact that the temperature or ${\bar
E}_\gamma$ is less than $E_B$.  The degree to which deuterium production
is delayed can be found by comparing the qualitative expressions for the
deuterium production and destruction rates,
\begin{eqnarray}
\Gamma_p & \approx & n_B \sigma v \\ \nonumber
\Gamma_d & \approx & n_\gamma \sigma v e^{-E_B/T}
\end{eqnarray}
When the quantity $\eta^{-1}
{\rm exp}(-E_B/T)
\sim 1$, the rate for  deuterium destruction (D $+ ~\gamma
\rightarrow p + n$) finally falls below the deuterium
production rate and the nuclear chain begins at a temperature
$T \sim 0.1 MeV$.

The dominant product of big bang nucleosynthesis is \he4 and its
abundance is very sensitive to the
$(n/p)$ ratio
\begin{equation}
Y_p = {2(n/p) \over \left[ 1 + (n/p) \right]} \approx 0.25
\label{ynp}
\end{equation}
i.e., an
abundance of close to 25\% by mass. Lesser amounts of the other light
elements are produced: D and \he3 at the level of about $10^{-5}$ by
number,  and \li7 at the level of $10^{-10}$ by number. 

\begin{figure}[t]
 \includegraphics[height=.65\textheight, angle=270]{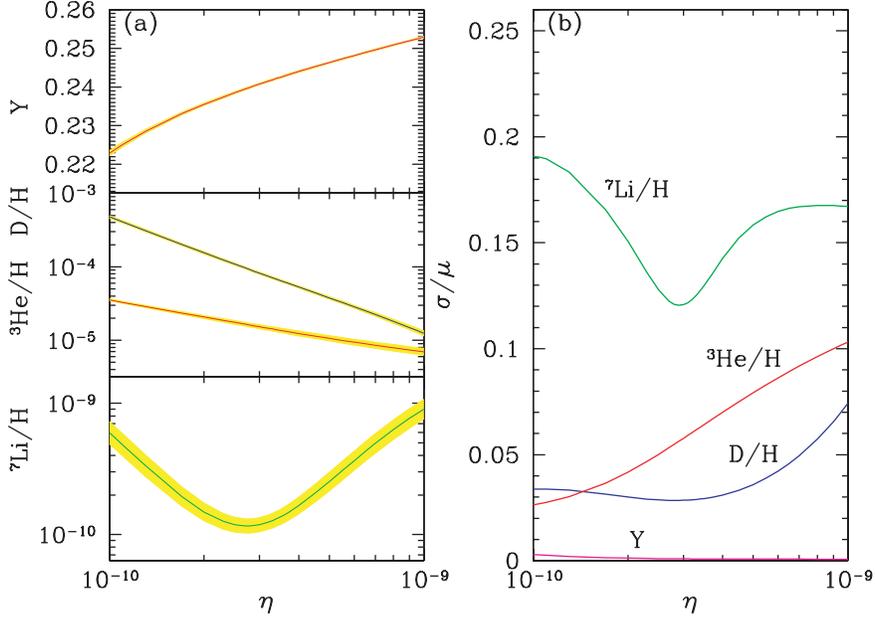}
\caption{{The light element abundances from big bang
nucleosynthesis as a function of $\eta_{10}$.}}
\label{cfo1}
\end{figure}

Recently the
input nuclear data have been carefully reassessed\cite{nb,cfo1,cvcr01,nacre,coc},
leading to improved precision in the abundance predictions. 
The NACRE collaboration presented an updated nuclear
compilation~\cite{nacre}.  
For example, notable improvements include a reduction in the uncertainty
in the rate for \he3$(n,p)$T from 10\%~\cite{skm} to 3.5\% and
for T$(\alpha, \gamma)$\li7 from $\sim 23 - 30\%$~\cite{skm} to $\sim 4\%$.
Since then, new data and techniques have become available, motivating
new compilations.  Within the last year, several new BBN compilations
have been presented\cite{cyburt,coc2,cuoco}.

The resulting elemental abundances predicted by standard BBN 
are shown in Fig. \ref{cfo1}
 as a function of $\eta$~\cite{cfo1}. 
 The left plot shows the
abundance of \he4 by mass, $Y$, and the abundances of the other three
isotopes by number.  The curves indicate the central predictions from
BBN, while the bands correspond to the uncertainty in the predicted
abundances.
This theoretical
uncertainty is shown explicitly in the right panel as a function of
$\eta$.

In the standard model with $N_\nu = 3$, the only free parameter is the
density of baryons 
which sets the rates of the strong reactions.
Thus, 
any abundance measurement determines $\eta$, while additional measurements
overconstrain the theory and thereby provide a consistency check.
BBN has thus historically been the premier means of determining
the cosmic baryon density.
With the increased precision of microwave background anisotropy
measurements, it is now possible to use the the CMB to independently
determine the baryon density.
The WMAP value for $\Omega_B h^2 = 0.0224$ translates into
\begin{equation}
\label{eq:eta-cmb}
\eta_{10}  = 6.14 \pm 0.25
\end{equation}
With $\eta$ fixed by the CMB, precision comparisons to the observations can now be 
attempted\cite{cfo2}.

\subsection{Light Element Observations and Comparison with Theory} 

BBN theory predicts the universal abundances of D, \he3, \he4, and
\li7, which are essentially determined by $t\sim180$~s. Abundances are
however observed at much later epochs, after stellar nucleosynthesis
has commenced. The ejected remains of this stellar processing
can alter the light element abundances from their primordial values,
and produce heavy elements such as C, N, O, and Fe
(``metals''). Thus one seeks astrophysical sites with low metal
abundances, in order to measure light element abundances which are
closer to primordial.
For all of the light elements, 
systematic errors are an important and often dominant
limitation to the precision of derived primordial abundances.

\subsubsection{D/H}

In recent years, high-resolution spectra have revealed the presence of
D in high-redshift, low-metallicity quasar absorption systems (QAS),
via its isotope-shifted Lyman-$\alpha$ absorption. These are the first measurements of light
element abundances at cosmological distances. It is believed that
there are no astrophysical sources of deuterium\cite{els}, so any
measurement of D/H provides a lower limit to primordial D/H and thus
an upper limit on $\eta$; for example, the local interstellar value of
D/H=$(1.5\pm0.1)\times10^{-5}$~cite{lin} requires that
$\eta_{10}\le9$. In fact, local interstellar D may have been
depleted by a factor of 2 or more due to stellar processing; however,
for the high-redshift systems, conventional models of galactic
nucleosynthesis (chemical evolution) do not predict significant D/H
depletion\cite{fie}.

\begin{figure}[ht]
\begin{center}
 \includegraphics[height=.48\textheight, angle=0]{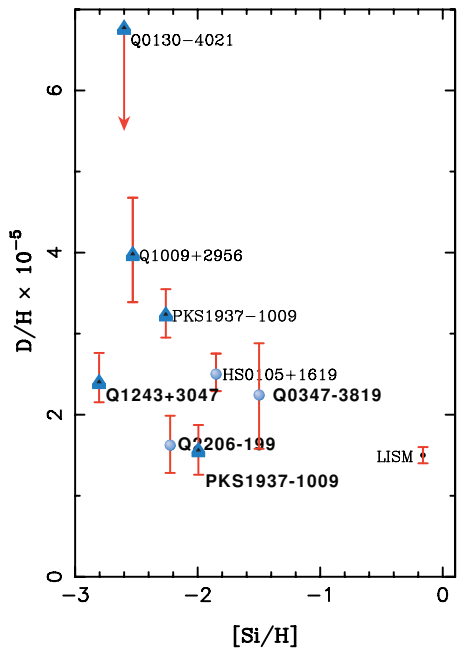}
\end{center}
\caption{{D/H abundances shown as a function of [Si/H].}}
\label{D}
\end{figure}

The five most precise observations of deuterium\cite{bt,omeara,kirkman,pet} 
in QAS give
D/H = $(2.78 \pm 0.29) \times 10^{-5}$, 
where the error is
statistical only. 
These are shown in Fig. \ref{D} along with some other recent measurements\cite{ketal,dodo,cr}.
Inspection of the data shown in the figure clearly indicates the need for
concern over systematic errors. We thus conservatively bracket the observed values 
with a range  D/H = $2 - 5 \times10^{-5}$ which corresponds to a range
in $\eta_{10}$ of 4 -- 8 which easily brackets the CMB determined value.

\vskip .2in

Using the WMAP value for the baryon density (\ref{eq:eta-cmb})
the primordial D/H abundance is predicted to be\cite{cfo1,cyburt}:
\begin{equation}
({\rm D/H})_p = 2.55^{+0.21}_{-0.20} \times 10^{-5}
\label{dpred}
\end{equation}
As one can see, this value is in very good agreement with the observational
value.

\subsection{\he4}

We observe \he4 in clouds of ionized hydrogen (HII regions), the most
metal-poor of which are in dwarf galaxies. There is
now a large body of data on \he4 and CNO in these systems\cite{iz}. 
Of the modern \he4 determinations, the work of Pagel et
al.\cite{pagel} established the analysis techniques that were soon to
follow\cite{follow}.  Their value of $Y_p$ $=$ 0.228 $\pm$ 0.005 was
significantly lower than that of a
sample of 45 low metallicity HII regions, observed and
analyzed in a uniform manner\cite{iz}, with a derived value of $Y_p$ $=$ 0.244
$\pm$ 0.002.  An analysis based on the
combined available data as well as unpublished data yielded an
intermediate value of 0.238 $\pm$ 0.002 with an estimated systematic
uncertainty of 0.005~\cite{osk}.  An extended data set 
including 89 HII regions obtained $Y_p$ $=$ 0.2429
$\pm$ 0.0009~\cite{iz2}.  However, the recommended value is based on
the much smaller subset of 7 HII regions, finding $Y_p$ $=$ 0.2421
$\pm$ 0.0021.

\he4 abundance determinations depend on a number of physical 
parameters associated with the HII region in addition to the overall 
intensity of the He emission line.  These include, the temperature,
electron density, optical depth and degree of underlying absorption.
A self-consistent analysis may use
multiple  \he4 emission lines to 
determine the He abundance, the electron density and the optical
depth. In~\cite{iz}, five He lines were used, underlying He absorption was assumed to be negligible 
and used temperatures based on OIII data.

\begin{figure}
\begin{center}
 \includegraphics[height=.23\textheight, angle=0]{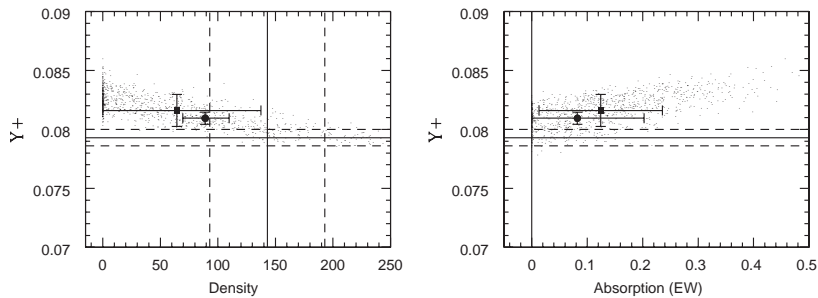}
\end{center}
\caption{{Results of modeling of 6 He~I line observations of NGC 346\protect\cite{peim}.
The solid lines show the original derived values and the dashed lines
show the  1 $\sigma$ errors on those values.
The solid circles (with error bars) show the results of the $\chi ^2$
minimization solution (with calculated  errors)\protect\cite{os2}.
The small points show the results of Monte Carlo realizations
of the original input spectrum.
The solid squares (with error bars) show the means and dispersions
of the output values for the $\chi ^2$ minimization solutions of
the Monte Carlo realizations.}}
\label{degen}
\end{figure}

The question of systematic uncertainties was addressed in some detail
in~\cite{OSk}. It was shown that there exist severe degeneracies
inherent in the self-consistent method, particularly when the effects
of underlying absorption are taken into account.  
The results of a Monte-Carlo reanalysis\cite{os2} of NCG 346\cite{peim} is shown in 
Fig. \ref{degen}. In the left panel, solutions for the \he4 abundance 
and electron density are shown (symbols are described in the caption). 
In the right panel, a similar plot with the \he4 abundance and 
the equivalent width for underlying absorption is shown.
As one can see,  solutions with
no absorption and high density are often indistinguishable (i.e., in a
statistical sense they are equally well represented by the data) from
solutions with underlying absorption and a lower density.  In the
latter case, the He abundance is systematically higher.  These
degeneracies are markedly apparent when the data is analyzed using
Monte-Carlo methods which generate statistically viable
representations of the observations as shown in Fig. \ref{degen}. When this is done, not only are the
He abundances found to be higher, but the uncertainties are also found
to be significantly larger than in a direct self-consistent approach.

\begin{figure}
\begin{center}
 \includegraphics[height=.5\textheight, angle=90]{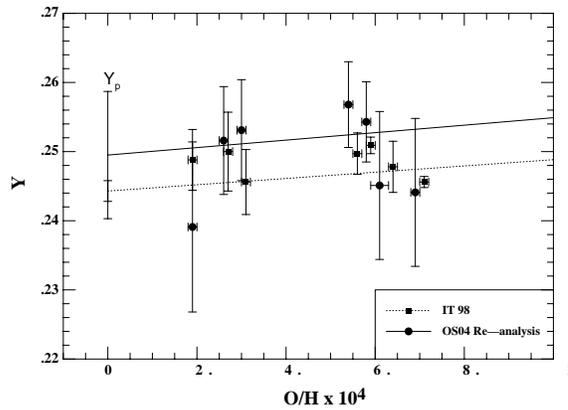}
\end{center}
\caption{{A comparison of the results for the best targets\protect\cite{iz} and a re-analysis
of the spectra for those targets\protect\cite{os2}. }}
\label{He4}
\end{figure}

Recently a careful study of the systematic uncertainties in \he4,
particularly the role of underlying absorption has been
performed using a subset of the highest quality from the data of Izotov and Thuan\cite{iz}.
All of the physical parameters listed above including the \he4 abundance were
determined self-consistently with Monte Carlo methods\cite{OSk}. 
Note that the \he4 abundances are systematically higher, and the 
uncertainties are several times larger than quoted in~\cite{iz}. 
In fact this study has shown that the determined value of 
$Y_p$ is highly sensitive to the method of analysis used.
The result is shown in 
Fig. \ref{He4} together with a comparison of the previous result.
The extrapolated \he4 abundance was determined to be 
$Y_p = 0.2495 \pm 0.0092$.  The value of $\eta$ corresponding 
to this abundance is $\eta_{10} = 6.9^{+11.8}_{-4.0}$
and clearly overlaps with $\eta_{CMB}$. 
Conservatively, it would be difficult at this time to exclude any value of
$Y_p$ inside the range 0.232 -- 0.258.

At the WMAP value for $\eta$, the \he4 abundance is predicted to be\cite{cfo1,cyburt}:
\begin{equation}
\label{eq:Yp}
Y_p = 0.2485 \pm 0.0005
\end{equation}
This value is considerably higher than any prior determination of the 
primordial \he4 abundance,
it is in excellent agreement with the most recent analysis of the \he4 abundance\cite{os2}. 
Note also that the large uncertainty ascribed to this value
indicates that the while \he4 is certainly consistent with the WMAP 
determination of the baryon density, it does not provide for
a highly discriminatory test of the theory at this time.

\subsection{\li7/H}

The systems best suited for Li observations are metal-poor halo stars
in our Galaxy. Observations
have long shown\cite{sp} that Li does not vary
significantly in Pop II stars with metallicities $\la1/30$ of solar
--- the ``Spite plateau''. Recent precision data
suggest a small but significant correlation between Li and Fe~\cite{rnb} 
which can be understood as the result of Li production
from Galactic cosmic rays\cite{van}. Extrapolating to
zero metallicity one arrives at a primordial value\cite{rbofn}
${\rm Li/H}|_p =
 (1.23^{+0.34}_{-0.16}) \times 10^{-10}$.
 
 \begin{figure}[h] 
\begin{center}
\includegraphics[height=.35\textheight]{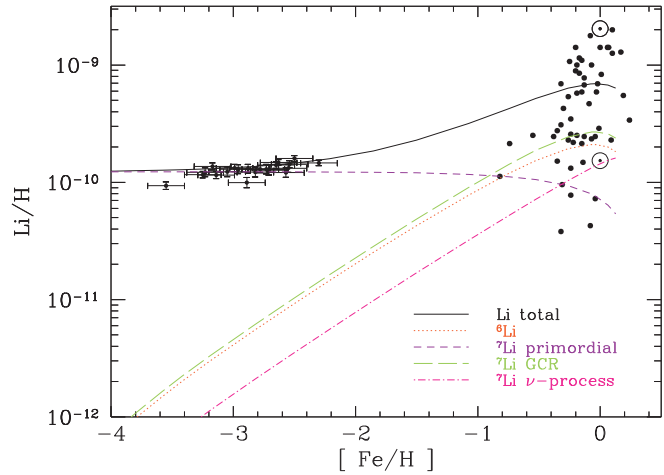}
\end{center}
\caption{Contributions to the total predicted lithium abundance from
the adopted GCE model of~\protect\cite{fo99}, compared with low
metallicity  stars  and a sample of high
metallicity stars. The solid curve is the sum of all  components.}
\label{li2fig}
\end{figure}

Figure \ref{li2fig} shows the different Li
components for a model with (\li7/H)$_p = 1.23 \times 10^{-10}$. 
The linear slope produced by the model is independent of the input
primordial value. The model of ref.~\cite{fo99} 
includes in addition to primordial \li7, lithium produced in
galactic cosmic ray nucleosynthesis (primarily $\alpha + \alpha$ fusion),
and \li7 produced by the $\nu$-process during type II supernovae. As one
can see, these processes are not sufficient to  reproduce the population
I abundance of \li7, and additional production sources are needed. 

Recent data\cite{bon1} with temperatures
based on H$\alpha$ lines (considered to give systematically high
temperatures) yields \li7/H = $(2.19 \pm 0.28) \times
10^{-10}$. These results are based on a globular cluster sample (NGC 6397).  
This result is consistent with previous Li measurements of the same cluster
which gave  \li7/H = $(1.91 \pm 0.44) \times
10^{-10}$~\cite{pm} and  \li7/H = $(1.69 \pm 0.27) \times
10^{-10}$~\cite{thev}.  A related study (also of globular
cluster stars) gives \li7/H = $(2.29 \pm 0.94) \times10^{-10}$~\cite{bon2}.

 The \li7 abundance based on the WMAP baryon density is predicted to be\cite{cfo1,cyburt}:
\begin{equation}
{\rm \li7/H} = 4.26^{+0.73}_{-0.60} \times 10^{-10}
\label{li7}
\end{equation}
This value is in clear contradiction with most estimates of the 
primordial Li abundance. 
It is a factor of $\sim 3$ higher than the value observed in most halo stars,
and  just about 0.2
dex over the globular cluster value.

\subsection{Concordance}

\begin{figure}
 \includegraphics[height=.27\textheight, angle=0]{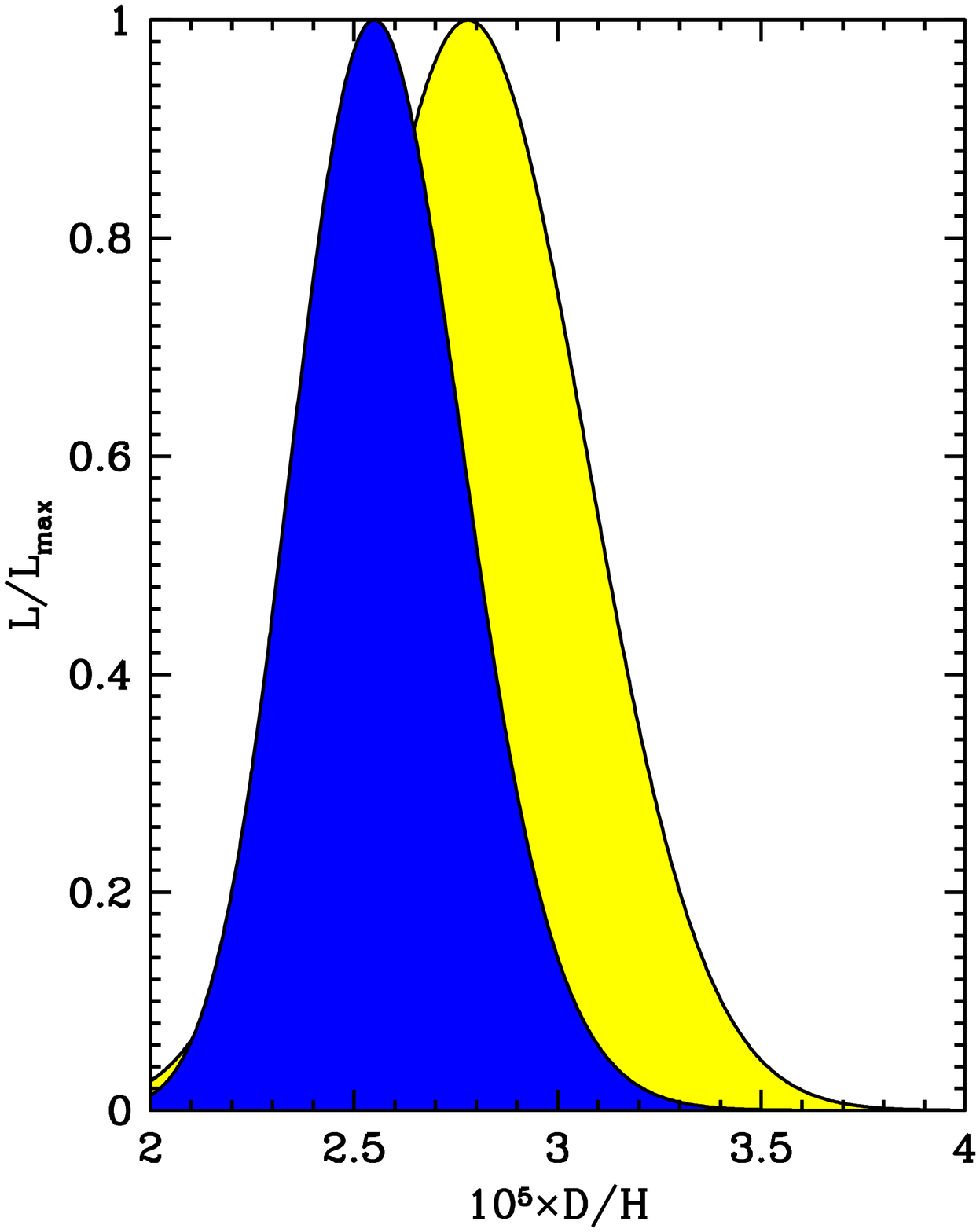}
  \includegraphics[height=.27\textheight, angle=0]{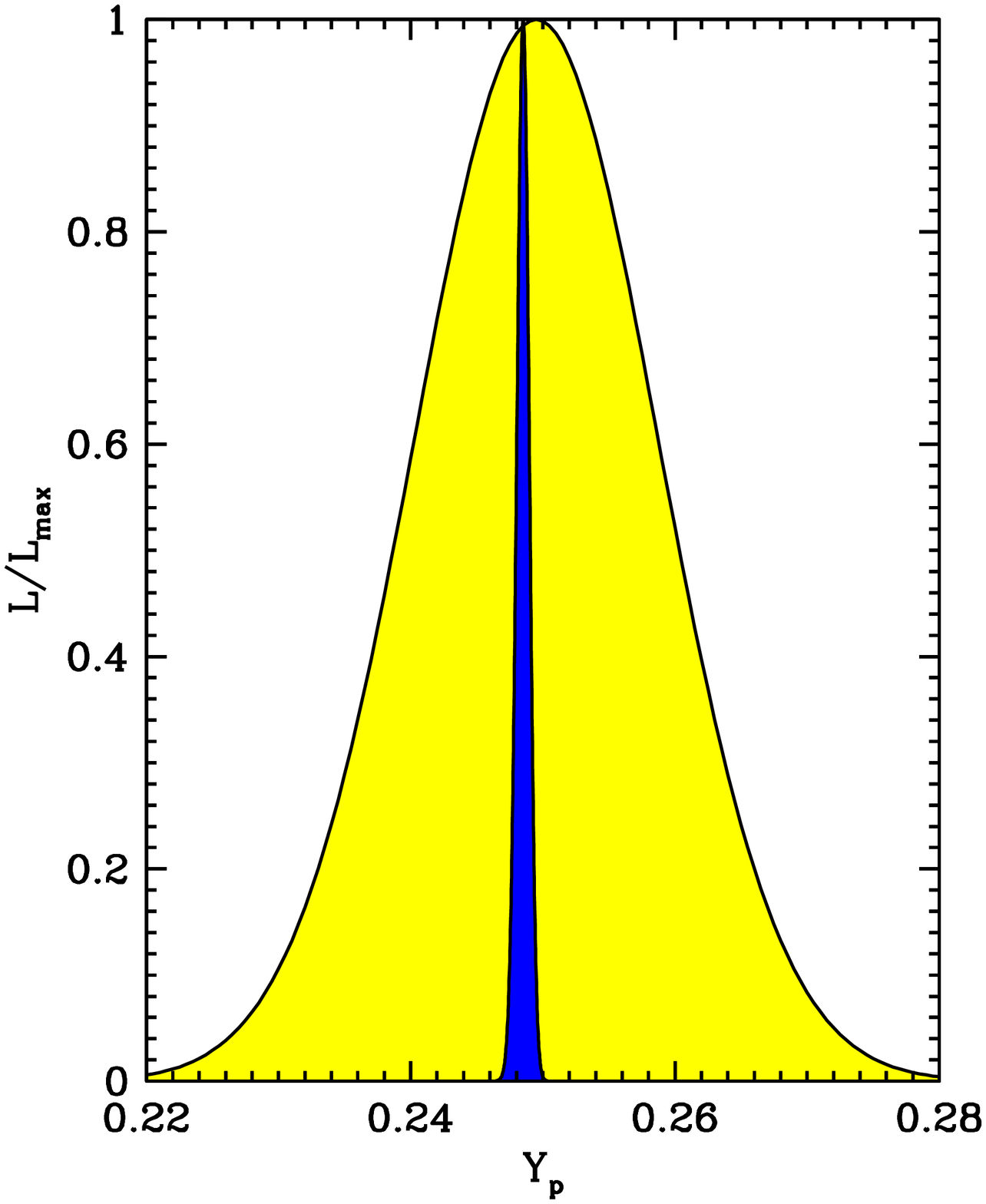}
   \includegraphics[height=.27\textheight, angle=0]{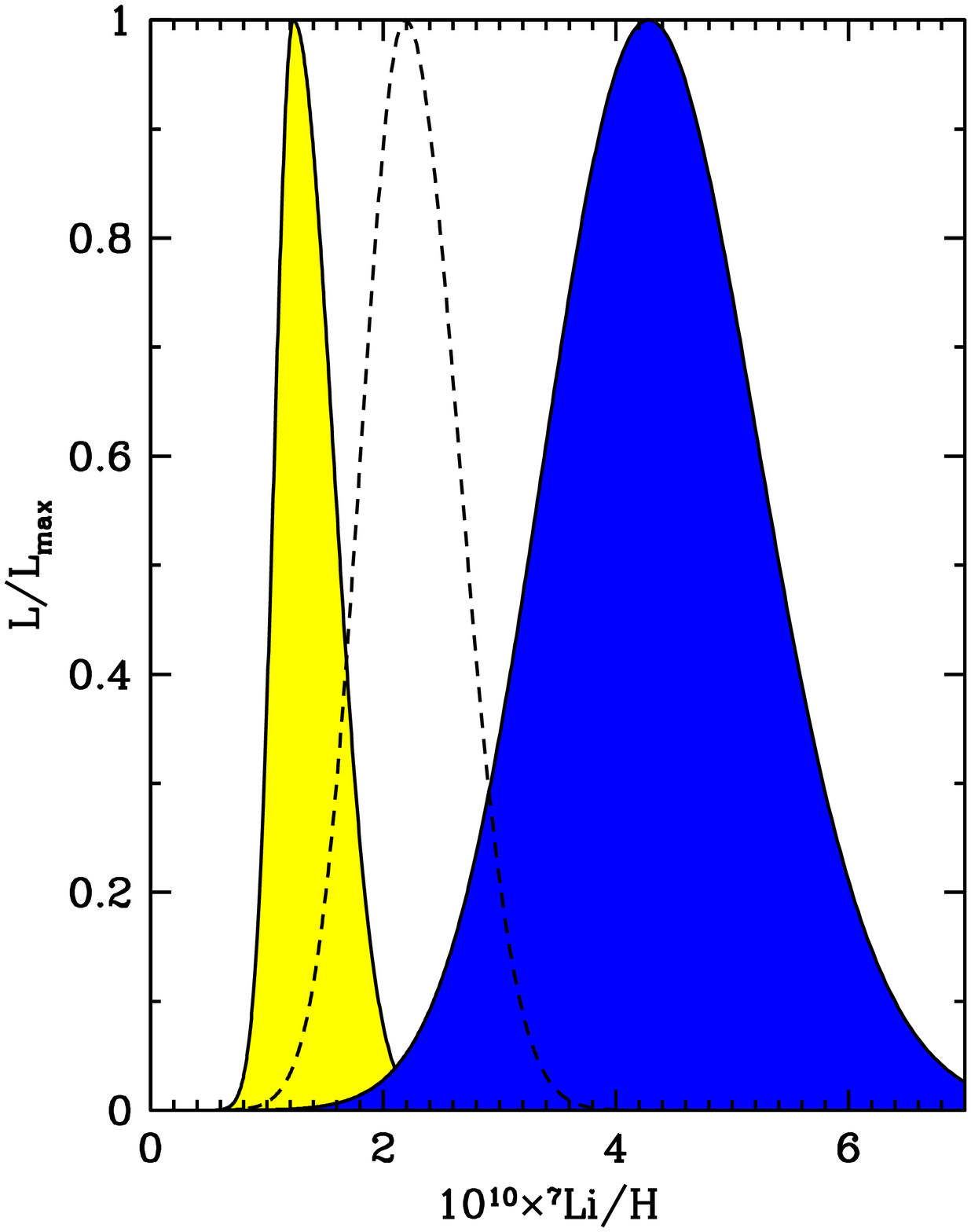}
\caption{
\label{fig:abs-MAP}
Primordial light element abundances as predicted by BBN and
WMAP (dark shaded regions)\protect\cite{cfos}.  
Different observational assessments of primordial abundances 
are plotted as follows:
{\bf (a)} the light shaded
region shows ${\rm D/H} = (2.78 \pm 0.29) \times 10^{-5}$;
{\bf (b)} the light shaded region shows $Y_p = 0.249 \pm 0.009$;
{\bf (c)} the light shaded region shows
\li7/H = $1.23^{+0.34}_{-0.16} \times 10^{-10}$,
while the dashed curve shows  
\li7/H = $(2.19\pm 0.28) \times 10^{-10}$.
}
\end{figure}

In Fig. \ref{fig:abs-MAP}, we show the direct comparison between
the BBN predicted abundances given in eqs. (\ref{dpred}), (\ref{eq:Yp}), and (\ref{li7}), using the 
WMAP value of $\eta_{10} = 6.25 \pm 0.25$ with the observations\cite{cfos}.
As one can see, there is very good agreement between theory and observation for 
both D/H and \he4.  Of course, in the case of \he4, concordance is almost guaranteed
by the large errors associated to the observed abundance. 
In contrast, as was just noted above, there is a marked discrepancy in the 
case of \li7.

 The quoted value for the \li7 abundance assumes that the Li abundance in the stellar
sample reflects the initial abundance at the birth of the star.
However, an important source of systematic uncertainty comes from the
possible depletion of Li over the $\ga 10$ Gyr  age of
the Pop II stars. The atmospheric Li abundance will suffer depletion
if the outer layers of the stars have been transported deep enough
into the interior, and/or mixed with material from the hot interior;
this may occur due to convection, rotational mixing, or diffusion.
Standard stellar evolution models
predict Li depletion factors which are very small
($<$0.05~dex) in very metal-poor turnoff stars\cite{ddk}. 
However,  there is no reason to 
believe that such simple models incorporate all effects which lead to
depletion such as rotationally-induced mixing and/or diffusion.
Current estimates for possible depletion factors are in the range
$\sim$~0.2--0.4~dex~\cite{dep}. 
As noted above, this data sample\cite{rnb}
shows a negligible intrinsic spread in Li leading to the conclusion
that depletion in these stars is as low as 0.1~dex.

Another important source for potential systematic uncertainty stems
from the fact that the Li abundance is not directly
observed but rather, inferred from an absorption line strength and
a model stellar atmosphere. Its determination
depends on a set of physical parameters and a model-dependent analysis
of a stellar spectrum.  Among these parameters, are the metallicity
characterized by the iron abundance (though this is a small effect),
the surface gravity which for hot stars can lead to an underestimate
of up to 0.09 dex if $\log g$ is overestimated by 0.5, though this effect
is negligible in cooler stars.  Typical uncertainties in $\log g$ are
$\pm 0.1 - 0.3$.  The most important source for error is the
surface temperature.  Effective-temperature calibrations for stellar
atmospheres can differ by up to 150--200~K, with higher temperatures
resulting in estimated Li abundances which are higher by $\sim
0.08$~dex per 100~K.  Thus accounting for a difference of 0.5 dex
between BBN and the observations, would require a serious offset of
the stellar parameters. While there has been a recent analysis\cite{mr}
which does support higher temperatures, the consequences of the higher
temperatures on the inferred abundances of related elements such as Be, B, and O
observed in the same stars is somewhat negative\cite{fov}.

Finally a potential source for systematic uncertainty
lies in the BBN calculation of the \li7 abundance.  As one can
see from Fig. \ref{cfo1}, the predictions for \li7 carry the largest uncertainty
of the 4 light elements which stem from uncertainties in the nuclear rates.
The effect of changing the yields of certain BBN reactions was
recently considered by Coc et al.\cite{coc}.  In particular, they
concentrated on the set of cross sections which affect \li7 and are
poorly determined both experimentally and theoretically.  In many
cases however, the required change in cross section far exceeded any
reasonable uncertainty.  Nevertheless, it may be possible that certain
cross sections have been poorly determined.  In~\cite{coc}, it was
found for example, that an increase of either the $\li7(d,n)2\he4$ or
$\be7(d,p)2\he4$ reactions by a factor of 100 would reduce the \li7
abundance by a factor of about 3.

The  possibility
of systematic errors in the
$\he3(\alpha,\gamma)\be7$ reaction, which is the only important \li7
production channel in BBN, was considered in detail in~\cite{cfo4}.  
The absolute value of the cross section
for this key reaction is known relatively poorly both experimentally
and theoretically.  However, the agreement between the standard solar model and
solar neutrino data thus provides additional constraints on variations
in this cross section.  Using the standard solar model of
Bahcall\cite{bah}, and recent solar neutrino data\cite{sno}, one can exclude systematic
variations of the magnitude needed to resolve the BBN \li7
problem at the $\ga 95\%$ CL~\cite{cfo4}.  Thus the ``nuclear fix'' to the
\li7 BBN problem is unlikely.

Finally, we turn to \he3.  Here, the only observations available are
in the solar system and (high-metallicity) HII regions in our Galaxy\cite{bbrw}. 
This makes inference of the primordial abundance
difficult, a problem compounded by the fact that stellar
nucleosynthesis models for \he3 are in conflict with observations\cite{osst}. 
Consequently, it is not appropriate to use \he3 as a
cosmological probe\cite{brb}; instead, one might hope to turn the problem around
and constrain stellar astrophysics using the predicted primordial \he3
abundance\cite{vofc}.  For completeness, we note that the 
 \he3 abundance is predicted to be:
\begin{equation}
{\rm \he3/H} = 9.28^{+0.55}_{-0.54} \times 10^{-6}
\end{equation}
at the WMAP value of $\eta$.

\section{Constraints on Decaying Particles and Gravitino Dark Matter from BBN}

As an example of constraints on particle properties from BBN, I will
concentrate here on life-time and abundance limits on decaying 
particles as it ties in well with the previous discussion on supersymmetric dark matter.
There are of course many other constraints on particle properties
which can be derived from BBN, most notably the limit on the number of
relativistic degrees of freedom.  For a recent update on
these limits, see~\cite{cfos}.

Because there is good overall agreement between the theoretical predictions
of the light element abundances and their observational determination,
any departure from the standard model (or either particle physics, cosmology, or BBN)
generally leads to serious inconsistencies among the element abundances.

Gravitinos have long been known to be potentially problematic in cosmology\cite{grprob}.
If gravitinos are present with equilibrium
number  densities, we can write  their energy density as
\begin{equation}
\rho_{3/2} = m_{3/2} n_{3/2} = m_{3/2} \left( 3\zeta(3) \over \pi^2\right)
T_{3/2}^2
\end{equation}
where today one expects that the gravitino temperature $T_{3/2}$ is reduced
relative to the photon temperature due to the annihilations of particles
dating back to the Planck time\cite{oss}.  Typically one can expect the gravitino abundance
$Y_{3/2} \equiv n_{3/2}/n_\gamma \sim (T_{3/2}/T_\gamma)^3 \sim 10^{-2}$. 
Then for $\Omega_{3/2} h^2 \la
1$, we obtain the limit that $m_{3/2} \la 1$ keV.

Of course, the above mass limit assumes a stable gravitino, the problem
persists however, even if the gravitino decays, since its gravitational decay
rate is very slow.  Gravitinos decay when their decay rate, $\Gamma_{3/2}
\simeq m_{3/2}^3/M_P^2$,  becomes comparable to the expansion rate of the
Universe (which becomes dominated by the mass density of gravitinos), $H
\simeq m_{3/2}^{1/2} T_{3/2}^{3/2}/M_P$.  Thus decays occur at $T_d \simeq
m_{3/2}^{5/3}/M_P^{2/3}$. After the decay, the Universe is ``reheated" to 
a temperature 
\begin{equation}
T_R \simeq \rho(T_d)^{1/4} \simeq m_{3/2}^{3/2}/M_P^{1/2}
\end{equation}
The Universe must reheat
sufficiently so that big bang nucleosynthesis occurs in a standard radiation
dominated Universe. For $T_R \ga 1$ MeV, we must require $m_{3/2} \ga 20$ TeV.
This large value threatens the solution of the hierarchy problem.

Inflation could alleviate the gravitino problem by diluting
the gravitino abundance to safe levels\cite{grinf}. 
If gravitinos satisfy the noninflationary bounds,
then their reproduction after inflation is never a problem.
For gravitinos with mass of order 100 GeV,
dilution without over-regeneration will also solve the problem,
but there are several
factors one must contend with in order to be cosmologically safe.
Gravitino decay products can also upset
the successful predictions of Big Bang nucleosynthesis,
 and decays into LSPs (if R-parity is conserved) can also yield
too large a mass density in the now-decoupled LSPs\cite{EHNOS}.
 For unstable
gravitinos, the most restrictive bound on their number density comes form the
photo-destruction of the light elements produced during nucleosynthesis\cite{decays,cefo,kawa}.

Here, we will consider electromagnetic decays, meaning that the decays inject
electromagnetic radiation into the early universe. If the decaying
particle is abundant enough or massive enough, the injection of electromagnetic
radiation can photo-erode the light elements created during primordial
nucleosynthesis.  The theories we have in mind are generally supersymmetric, in
which the gravitino and neutralino are the next-to-lightest and
lightest supersymmetric particles, respectively (or vice versa), but the constraints
hold for any decay producing electromagnetic radiation.  We thus
constrain the abundance of such a particle given its mean lifetime
$\tau_X$.  The abundance is constrained through the parameter
$\zeta_X\equiv m_X n_X/n_\gamma$.

\begin{figure}
\vskip 0.5in
\vspace*{-0.75in}
\begin{center}
\epsfig{file=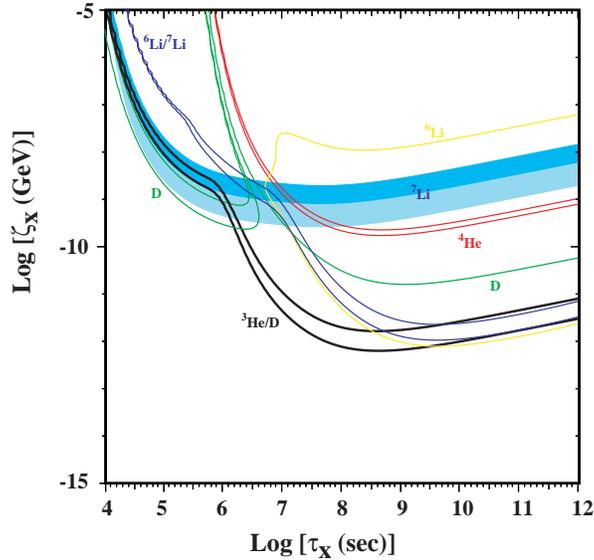,height=3in}
\hfill
\end{center}
\caption{
{\it 
The constraints imposed by the astrophysical observations of $^4$He (red 
lines), $D/H$ (green lines), $^6$Li (yellow line), $^6$Li/$^7$Li (blue 
lines), $^7$Li (blue band) and $^3$He (black lines).}}
\label{fig:tauzetacleft}
\end{figure}

The BBN limits in the $(\zeta_X, \tau_X)$ plane is shown in Fig. \ref{fig:tauzetacleft}\cite{eov}.
The constraint placed by the \he4 abundance comes from its lower
limit, as this scenario destroys \he4.  Shown are the limits assuming 
$Y_{min} = 0.232$ and 0.227~\cite{cefo,cfos,eov}.  The area above these curves are excluded.
The deuterium lines correspond to the contours
$(1.3 \; {\rm or} \; 2.2) \times 10^{-5} \; < \; {{\rm D}/ {\rm H}} \;   < \; 5.3 \times 10^{-5}$.
The first of the lower bounds is the higher line to the left of the cleft, 
and represents the very conservative lower limit on 
D/H assumed in~\cite{cefo}. The range 1.3 -- 5.3 $\times 10^{-5}$
effectively brackets all recent observations of D/H in quasar absorption 
systems as discussed above. The second of the lower bounds is the lower line on the left 
side and represents  the 
2-$\sigma$ lower limit in the best set of D/H observations. The upper bound 
is the line to the right of 
the cleft. 
{\it A priori}, there is also a narrow strip 
at larger $\zeta_X$ and $\tau_X$ where the D/H ratio also falls within
the acceptable range  but this is excluded by the 
observed $^4$He abundance.

The constraint imposed by the $^6$Li abundance is shown~\cite{cefo} as a solid 
yellow line in Fig.~\ref{fig:tauzetacleft}. 
Also shown, as solid blue lines, are two contours 
representing upper limits on the $^6$Li/$^7$Li ratio:
$ {^6{\rm Li} / ^7{\rm Li}} \; < \; 0.07 \; {\rm or} \; 0.15$. 
The lower number was used in~\cite{cefo} and
represented the upper limit available at the time, which was essentially
based on multiple observations of a single star. The most recent data\cite{Asplund2} 
includes observations of several stars. The Li
isotope ratio for most metal-poor stars in the sample is as high as 0.15,
and we display that upper limit here\cite{eov}.  The main effect of this
constraint is to disallow a region in the near-vertical cleft between the
upper and lower limits on D/H, as seen in Fig.~\ref{fig:tauzetacleft}.

The blue shaded band in Fig. \ref {fig:tauzetacleft} corresponds to a \li7 abundance of 
$ 0.9 \times 10^{-10} \; < \; {^7{\rm Li} / {\rm H}} \; < \; (2 \; {\rm 
or} \; 3) \times 10^{-10}, $
with the $^7$Li abundance decreasing as $\zeta_X$ increases and the 
intensity of the shading changing at the intermediate value. In~\cite{cefo},
only the lower bound was used due the existing discrepancy between
the primordial and observationally determined values. It is 
apparent that 
$^7$Li abundances in the lower part of the range  are 
possible only high in the Deuterium cleft, and even then only if one 
uses the recent and more relaxed limit on the $^6$Li/$^7$Li ratio. 
Values of the $^7$Li abundance in the upper part of the range 
are possible, however, even if one uses the more stringent 
constraint on $^6$Li/$^7$Li. In this case, the allowed region of parameter 
space would also extend to lower $\tau_X$, if one could tolerate values of 
D/H between 1.3 and $2.2 \times 10^{-5}$.

Finally, we show the impact of the \he3 constraint\cite{kawa,eov}.
Since Deuterium is more fragile than $^3$He, whose 
abundance is thought to have remained roughly constant since primordial 
nucleosynthesis when comparing the BBN value to it proto-solar abundance,
one would expect, in principle,  the $^3$He/D ratio to have been 
increased by stellar processing. 
Since D is totally destroyed in stars, the ratio of \he3/D can only 
increase in time or remain constant
if \he3 is also completely destroyed in stars. 
The present or proto-solar value of 
$^3$He/D can therefore be used to set an upper limit on the primordial 
value. Fig.~\ref{fig:tauzetacleft} displays the upper limits 
$ {^3{\rm He} / D} \; < \; 1 \; {\rm or} \; 2$
as solid black lines. Above these contours, the value of $^3$He/D
increases very rapidly, and points high in the Deuterium cleft of 
Fig.~\ref{fig:tauzetacleft} have absurdly high values of $^3$He/D, 
exceeding the limit  by an order of magnitude or more. 

The previous upper limit on $\eta_X$~\cite{cefo} corresponded to the
constraint $m_X n_X / n_\gamma < 5.0 \times 10^{-12}$~GeV for $\tau_X =
10^8$~s. The weaker (stronger) version of the $^3$He constraint adopted 
corresponds\cite{eov} to
\begin{equation}
m_X {n_X \over n_\gamma} \; < \; 2.0 (0.8) \times 10^{-12}~{\rm GeV}
\label{new8}
\end{equation}
for $\tau_X = 10^8$~s.

Returning to the case of a decaying gravitino, recall that thermal reactions are 
estimated to produce an abundance of gravitinos given by~\cite{EKN,cefo}:
\begin{equation}
{n_{m_{3/2}} \over n_\gamma} \; = \; (0.7 - 2.7) \times 10^{-11} \times 
\left( {T_R \over 10^{10}~{\rm GeV}} \right).
\label{Y32}
\end{equation}
Assuming that $m_{3/2} = 100$~GeV and $\tau_X = 10^8$~s, and imposing the 
constraints (\ref{new8}), we now find
\begin{equation}
T_R \; < \; (0.8 - 2.8) \times 10^7~{\rm GeV}, \; ((0.3 - 1.1) \times 
10^7~{\rm GeV})
\label{TR}
\end{equation}
for the weaker (stronger) version of the $^3$He constraint.

\begin{figure}
\vskip 0.5in
\vspace*{-0.75in}
\begin{minipage}{8in}
\epsfig{file=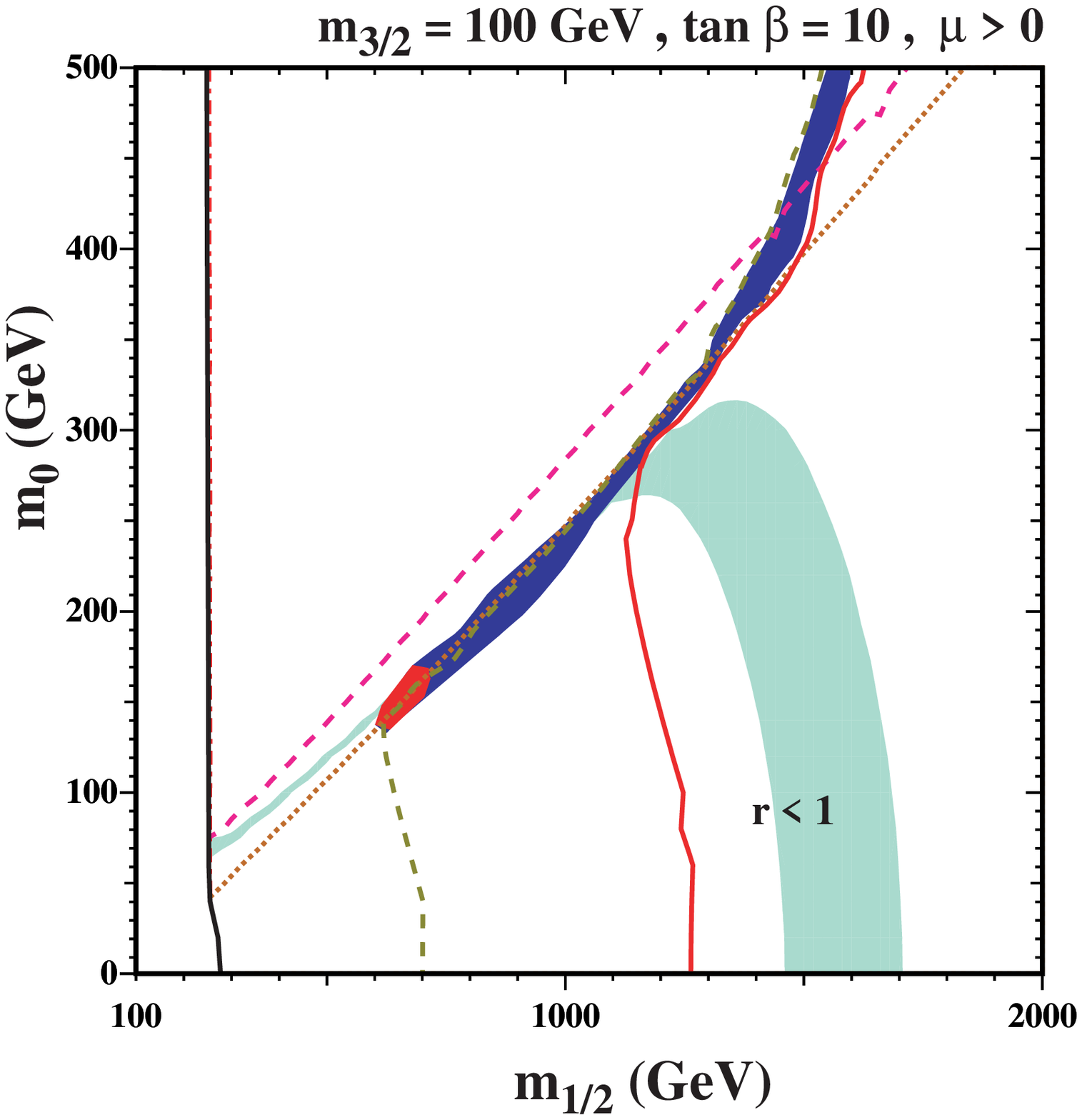,height=2.4in}
\hspace*{-0.17in}
\epsfig{file=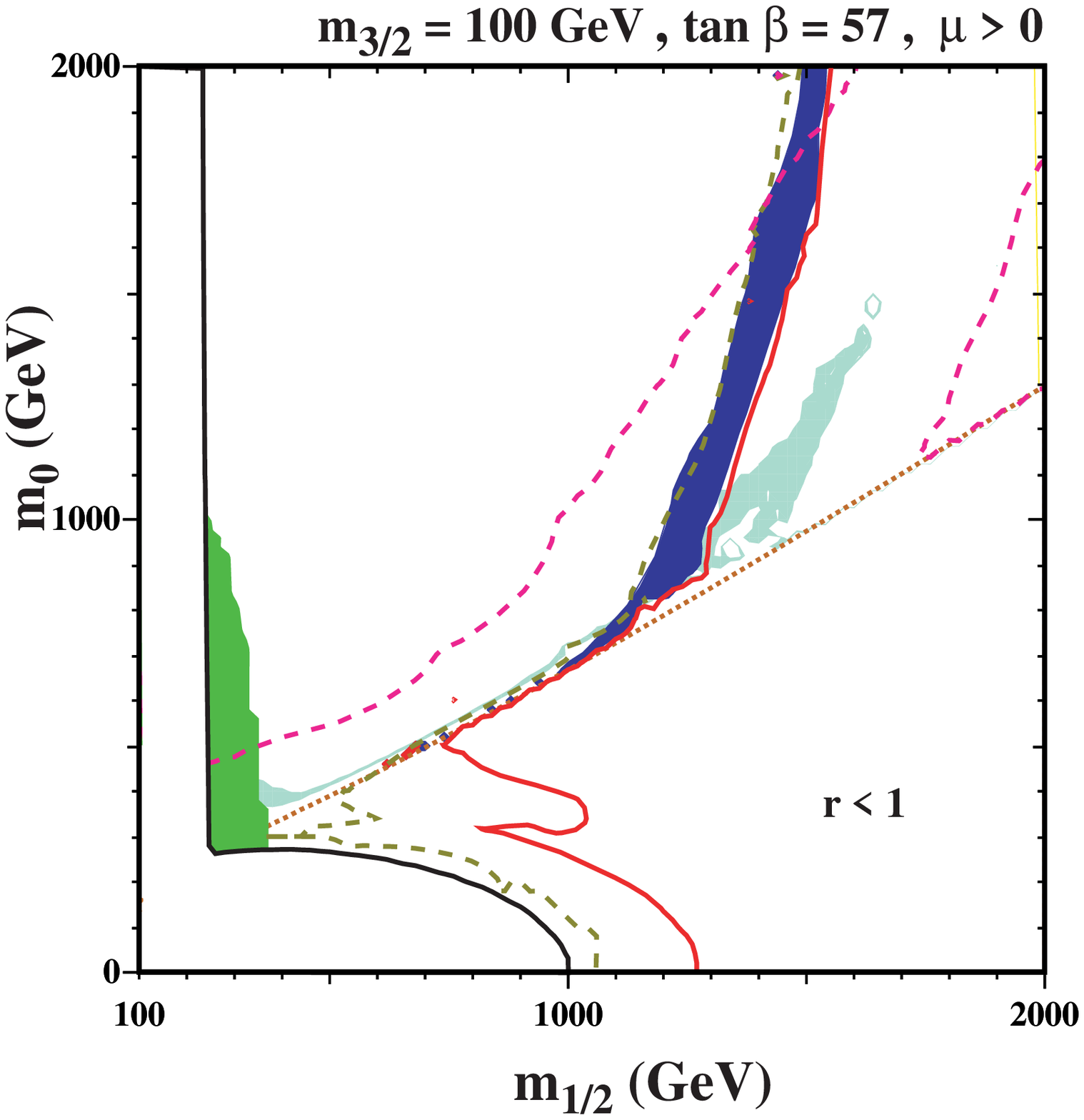,height=2.4in}
\hfill
\end{minipage}
\caption{
{\it 
The $(m_{1/2}, m_0)$ planes for $\mu > 0$, $m_{3/2} = 100$~GeV and (a)  $\tan
\beta = 10$  (b)  $\tan \beta = 57$. We restrict our attention 
to the regions between the solid black lines, where the gravitino is the 
LSP and the NSP lifetime exceeds $10^4$~s. In each panel, 
the near-vertical dashed black (dash-dotted red) line is the constraint
$m_{\chi^\pm} > 104$~GeV ($m_h > 114$~GeV), the upper (purple) dashed line 
is the
constraint $\Omega_{3/2} h^2 < 0.129$, and the light green shaded region
is that where the NSP would have had $0.094 \le \Omega h^2 \le
0.129$ if it had not decayed. The solid red (dashed grey-green) line is 
the region now (previously) allowed by the light-element abundances: $r < 
1$ as
described in the text. The red (blue) shaded region is that where the
\li7 abundance could have been improved by NSP decays, but which is now
excluded by the \he3 ({\rm D}) constraint.}}
\label{fig:CMSSM10} 
\end{figure}

Finally, we consider the possibility that gravitinos are stable and the LSP\cite{morefeng,eoss5}.
In this case, in the CMSSM, the next lightest supersymmetric particle (NSP) is either 
the neutralino or the stau. 
In Fig.~\ref{fig:CMSSM10}, we fix the
ratio of supersymmetric Higgs vacuum expectation values $\tan \beta = 10$ (left panel),
and $\tan \beta = 57$ (right panel), and assume
$m_{3/2} = 100$~GeV. In each panel of Fig.~\ref{fig:CMSSM10}, we
display accelerator, astrophysical and cosmological constraints in the
corresponding $(m_{1/2}, m_0)$ planes as discussed above for the 
CMSSM, concentrating on the regions to the
right of the near-vertical black lines, where the gravitino is the LSP.
The NSP is the $\tau$ lepton below the (red) dotted line. 

Below and to the right of the upper (purple) dashed lines, the density of
relic gravitinos produced in the decays of other supersymmetric particles
is always below the WMAP upper limit: $\Omega_{3/2} h^2 \le 0.129$. 
The code used in~\cite{cefo}, when combined with the observational
constraints used in~\cite{cefo}, yielded the astrophysical constraint
represented by the dashed grey-green lines in both panels of
Fig.~\ref{fig:CMSSM10} and did not include the constraint due to \he3/D.
These constraints on the CMSSM parameter plane were computed in~\cite{eoss5}. 
For each point in the $(m_{1/2},m_0)$, the relic density of
either $\chi$ or ${\tilde \tau}$ is computed and $\zeta_X$ is determined
using $\Omega_X h^2 = 3.9 \times 10^7$ GeV $\zeta_X$.  When $X = {\tilde
\tau}$, $\zeta_X$ is reduced by a factor of 0.3, as only 30\% of stau
decays result in electromagnetic showers which affect the element
abundances at these lifetimes. In addition, at each point, the lifetime of
the NSP is computed. Then for each $\tau_X$, the limit on $\zeta_X$ is
found from the results shown in Fig. \ref{fig:tauzetacleft}. The region to
the right of this curve where $r = \zeta_X/\zeta_X^{limit} < 1$ is
allowed.

The astrophysical constraints obtained with the newer abundance limits\cite{eov}
yields the solid red lines in Fig.~\ref{fig:CMSSM10}. The
examples where $\tau_X$ and $\zeta_X$ for the NSP decays fall within the
ranges shown by the blue band of Fig. \ref{fig:tauzetacleft}, and hence are suitable for
modifying the $^7$Li abundance\cite{jed,feng}, are shown as red and blue shaded regions
in each panel of Fig.~\ref{fig:CMSSM10}. 
 If we had been able to allow a Deuterium abundance as low
as D/H $\sim (1-2) \times 10^{-5}$, the blue shaded region would have been
able to resolve the Li discrepancy in the context of the CMSSM with
gravitino dark matter. The blue region that we now regard as excluded by
the lower limit on D/H, which is stronger than that used in~\cite{cefo},
extends to large $m_{1/2}$. The red shaded region, which is consistent
even with this limit on D/H, but yields very large $^3$He/D.
Fig.~\ref{fig:CMSSM10} show as solid red lines the additional
restrictions these constraints impose on the $(m_{1/2}, m_0)$ planes\cite{eov}.

\section{The variation of fundamental constants}

There has been considerable interest of late in the possible 
variation of the fundamental constants.
The construction of theories with variable ``constants'' is straightforward. 
Consider for example a gravitational Lagrangian which contains the term
\begin{equation}
{\mathcal L} \sim \phi R,
\end{equation}
where $\phi$ is some scalar field and $R$ is the Einstein curvature scalar.
The gravitational constant is determined if the dynamics of the theory 
fix the expectation value of the scalar field so that
\begin{equation}
G_N = {1 \over 16 \pi \langle \phi \rangle}.
\end{equation}
Similarly a coupling in the Lagrangian of a scalar to the Maxwell term $F^2$,
fixes the fine-structure constant
\begin{equation}
{\mathcal L} \sim \phi F^2, \qquad \alpha = {1 \over 16 \pi \langle \phi \rangle}.
\end{equation}
Indeed, gravitational theories of the Jordan-Brans-Dicke type
do contain the possibility for a  time-varying gravitational constant.
However, these theories can always be re-expressed such that
$G_N$ is fixed and other mass scales in the theory become time dependent (i.e.,
dependent on the scalar field).  For example, the JBD action can be written as
\begin{equation}
S = \int d^4x \sqrt{g} \left[ \phi R - {\omega
\over \phi} 
\partial_\mu \phi \partial^\mu \phi + \Lambda
+{\mathcal L}_m(\psi_{matter}, g_{\mu\nu}) \right],
\end{equation}
where $\omega$ is a number which characterizes the degree of departure from
general relativity (GR is recovered as $\omega \to \infty$), $\Lambda$ is the 
cosmological constant,
and the matter action for electromagnetism and a single massive fermion can 
be written as
\begin{equation}
{\mathcal L}_m = -{1 \over 4 e^2} F^2 - 
{\overline \Psi} \not \!\!D \Psi 
- m {\overline
\Psi} \Psi.
\end{equation}
Written this way, if the scalar field $\phi$ evolves, then $G_N$ does as well.
In another conformal frame, the JBD action can be rewritten as
\begin{equation}
S = \int d^4x \sqrt{\overline g} \left[
{\overline R} - (\omega +
{3\over 2}) 
{(\partial_\mu \phi)^2 \over \phi^2}
 - {{\overline
\Psi} \not \!\!D \Psi \over \phi^{3/2}} 
- {m {\overline
\Psi \Psi} \over \phi^2}- {1 \over 4 e^2} F^2  +
 {\Lambda \over \phi^2} \right].
\end{equation}
In this frame, Newton's constant {\em is} constant, but the fermion 
mass (after $\Psi$ is rescaled) varies as $\phi^{-1/2}$ and the cosmological
constant varies as $1/\phi^2$.  The physics described by either of these two 
actions is identical and the two frames can not be distinguished as
the measurable dimensionless quantity $Gm^2 \propto \phi^{-1}$ in both frames.  
While the fine-structure constant remains constant
in this construction, it is straight forward to consider theories
where it is not.  In what follows, I will restrict attention to 
variations in the fine-structure constant.

In any unified theory in which
the gauge fields have a common origin, variations in the fine structure
constant will be accompanied by similar variations in the other gauge
couplings\cite{co} (see also,~\cite{ds}). 
 In other words, variations of the gauge coupling at the
unified scale will induce variations in all of the gauge couplings at the
low energy scale.

It is easy to see that the running of the strong coupling constant has
dramatic consequences for the low energy hadronic parameters, including the
masses of nucleons\cite{co}.
Indeed the masses are determined by the QCD scale, $\Lambda$, which is
related to the ultraviolet scale, $M_{UV}$, by dimensional transmutation:
\begin{equation}
\alpha_s(M_{UV}^2) \equiv {g_s^2(M_{UV}^2) \over 4 \pi} =
{4 \pi \over b_3\ln (M_{UV}^2/\Lambda^2)},
\end{equation}
where $b_3$ is a usual renormalization group coefficient that depends on
the number
of massless degrees of freedom, running in the loop.
Clearly,  changes in $g_s$ will induce (exponentially) large
changes in $\Lambda$:
\begin{equation}
{\Delta \Lambda \over \Lambda} = {2 \pi \over 9 \alpha_s(M_{UV})}
{\Delta \alpha_s(M_{UV})\over \alpha_s(M_{UV})} \gg
{\Delta \alpha_s(M_{UV})\over \alpha_s(M_{UV})},
\label{deltaLambda}
\end{equation}
where for illustrative purposes we took the beta function of QCD with
three
fermions. On the other hand, the electromagnetic coupling $\alpha$ never
experiences
significant running from $M_{UV}$ to $\Lambda$ and thus
$\Delta \Lambda / \Lambda\gg\Delta \alpha/\alpha $. A more elaborate
treatment
of the renormalization group equations
above $M_Z$~\cite{lss} leads to the result that is in
perfect agreement with~\cite{co}:
\begin{equation}
{\Delta \Lambda \over \Lambda}\simeq 30 {\Delta \alpha \over \alpha}.
\label{enhance}
\end{equation}

In addition, we expect that not only the gauge couplings will vary,
but all Yukawa couplings are expected to vary as well.  In~\cite{co}, the
string motivated dependence was found to be
\begin{equation}
{\Delta h \over h} = {\Delta \alpha_U \over \alpha_U}
\end{equation}
where $\alpha_U $ is the gauge coupling at the unification scale and $h$
is the Yukawa coupling at the same scale. However in theories in which
the electroweak scale is derived by dimensional transmutation, changes in
the Yukawa couplings (particularly the top Yukawa) leads to exponentially
large changes in the Higgs vev.
In such theories, the Higgs expectation value
corresponds to the renormalization point and
is given qualitatively by
\begin{equation}
v\sim M_P \exp (- 2 \pi c / \alpha_t)
\label{rad1}
\end{equation}
where $c$ is a constant of order 1, and $\alpha_t = h_t^2/4\pi$.
Thus small changes in $h_t$ will induce large changes in $v$.
For $c \sim h_t \sim 1$,
\begin{equation}
{\Delta v \over v} \sim 80 {\Delta \alpha_U \over \alpha_U}
\label{enhance2}
\end{equation}
This dependence gets translated into
a variation in all low energy particle masses.  In short, once we
allow $\alpha$ to vary, virtually all masses and couplings are expected
to vary as well, typically much more strongly than the variation induced
by the Coulomb interaction alone.
 Unfortunately, it is very hard to make a
quantitative prediction for $\Delta v/v$ simply because we do not know
exactly how the dimensional transmutation happens in the Higgs sector,
and the answer will  depend, for example, on such things as the dilaton
dependence of the  supersymmetry breaking parameters. This uncertainty is
characterized in Eq. (\ref{rad1}) by the parameter $c$.
For the purpose of the present discussion
it is reasonable to assume that $\Delta v/v$ is comparable but not
exactly equal to $\Delta \Lambda/\Lambda$. That is, although they are both
$O(10-100) \Delta \alpha/\alpha $, their difference $|\Delta
\Lambda/\Lambda - \Delta v/v|$ is of the same order of magnitude which we
will take as 
$ \sim 50 \Delta \alpha/\alpha$.
 
Much of the recent excitement over the possibility of a time variation in the
fine structure constant stems from a series of recent observations of quasar absorption
systems and a detailed fit of positions of the absorption lines for several heavy elements
using the ``many-multiplet'' method\cite{webb,murphy3}.
A related though less sensitive method for testing the variability of $\alpha$, is the 
alkali doublet method, which neatly describes the physics involved.

Absorption clouds are prevalent along the lines of sight towards distant, high redshift
quasars.  As such, the quasar acts as a bright source, and the
absorption features seen in these clouds reflect their chemical abundances.
Consider an absorption feature in a doublet system involving 
for example, $S_{1/2} \to P_{3/2}$ and    $S_{1/2} \to P_{1/2}$ transitions. While
the overall wavelength position of the doublet is a measure of the redshift of the absorption
cloud, the separation of the two lines is a measure of the fine structure constant.
This is easily seen by recalling the energy splitting due to the spin-orbit coupling,
\begin{equation}
\Delta E \sim {e^2 \over m^2 r^3} S \cdot L \sim m e^8, \qquad 
{\Delta E \over E} \sim e^4 \sim \alpha^2.
\end{equation}
Since the line splitting $\Delta \lambda /\lambda \sim \Delta E / E$, the relative change in the
line splitting is directly proportional to $\Delta \alpha / \alpha$.
The many multiplet method compares transitions from different multiplets and different atoms
and utilizes the effects of relativistic corrections on the spectra.
The alkali doublet method\cite{doublet} has been applied to quasar absorption spectra,
but the sensitivity of the method only limits the variation in 
$\alpha$ within an of order $10^{-5}$.  Similarly, at present,
considerations based on OIII emission line systems\cite{bahcall} are also only able to set limits on the 
variation of $\alpha$ at the level of $10^{-4}$.  

In contrast, the many multiplet method based on the relativistic corrections to atomic transitions using several
transition lines from several elemental species allows for sensitivities which approach
the level of $10^{-6}$~\cite{webb,murphy3,chand}.
This method compares the line shifts of elements which are particularly
sensitive to changes in $\alpha$ with those that are not. At relatively low redshift ($z < 1.8$), 
the method relies on the comparison of Fe lines to Mg lines. At higher redshift, the comparison
is mainly between Fe and Si.  At all redshifts, other elemental transitions are also included
in the analysis.
Indeed, when this method is applied to a set of Keck/Hires data, a statistically
significant trend for a variation in $\alpha$ was reported: 
$\Delta \alpha / \alpha = (-0.54 \pm 0.12) \times 10^{-5}$ over a redshift 
range $0.5 \la z \la 3.0$.  The minus sign indicates a smaller value of 
$\alpha$ in the past.  

More recent observations taken at VLT/UVES using the many multiplet method have not been able to 
duplicate the previous result\cite{chand,quast}.  
The use of Fe lines in~\cite{quast} on a single absorber found
$\Delta \alpha / \alpha = (0.01 \pm 0.17) \times 10^{-5}$.  However, since the previous
result relied on a statistical average of over 100 absorbers, it is not clear that 
these two results are in contradiction.  In~\cite{chand}, the use of Mg and Fe lines
in a set of 23 high signal-to-noise systems yielded the result
$\Delta \alpha / \alpha = (-0.06 \pm 0.06) \times 10^{-5}$
and therefore represents a more significant disagreement and can be used to 
set very stringent limits on the possible variation in $\alpha$.

There exist various sensitive experimental checks that
constrain the variation of coupling constants (see e.g.,~\cite{Sister}).
Limits can be derived from cosmology (from both big bang nucleosynthesis
and the microwave background), the Oklo reactor, long-lived isotopes found
in meteoritic samples, and atomic clock measurements.

The most far-reaching limit (in time) on the variation of $\alpha$ comes from BBN.
The limit is primarily due to the limit on \he4.
Changes in the fine structure constant affect directly the neutron-proton mass
difference which can be expressed as $\Delta m_N \sim a \alpha \Lambda_{QCD} + bv$,
where $\Lambda_{QCD} \sim {\mathcal O}(100)$ MeV is the mass scale associated with
strong interactions, $v\sim {\mathcal O}(100)$ GeV  determines the weak scale,
and $a$ and $b$ are
numbers which fix the final contribution to $\Delta m_N$ to be 
$-0.8$ MeV and 2.1 MeV, respectively.
From the previous discussion on BBN, changes in $\alpha$ directly induce changes in $\Delta m_N$,
which affects the neutron to proton ratio.
The relatively good agreement between theory and observation,
$|\Delta Y / Y| \la 3.5 \%$ allows one to set a limit $|\Delta \alpha / \alpha| \la 0.06$ 
($\Delta Y / Y$ scales with $\Delta \alpha / \alpha$)~\cite{bbn,co,cfos}. Since this 
limit is applied over the age of the Universe,
we obtain a limit on the rate of change
$|\dot \alpha / \alpha| \la 4 \times 10^{-12}$ yr$^{-1}$ over the last 13 Gyr.
When coupled variations of the couplings are considered,
the above bound is improved
by about 2 orders of magnitude to ${\Delta \alpha /
\alpha} \la 10^{-4}$ as confirmed in a numerical
calculation\cite{bbn2}.

One can also derive cosmological bounds based on the
microwave background. Changes in the fine-structure constant lead directly to
changes in the hydrogen binding energy, $E_b$.  As the Universe expands, its radiation
cools to a temperature, $T_{dec}$, at which protons and electrons can combine to form neutral 
hydrogen atoms, allowing the photons to decouple and free stream.
Measurements of the microwave background
can determine this temperature to reasonably high accuracy 
(a few percent)\cite{wmap}.  At decoupling $\eta^{-1}\exp(-E_b/T_{dec}) \sim 1$.
Thus, changes in $\alpha$ of at most a few percent can be tolerated over the time 
scale associated
with decoupling (a redshift of $z \sim 1100$)\cite{cmb}.

 Interesting constraints on the variation of $\alpha$ can be obtained from
the Oklo phenomenon concerning the operation of a natural reactor in a rich 
uranium deposit in Gabon approximately two billion years ago. The observed 
isotopic abundance distribution at Oklo can be related to the cross section for
neutron capture on $^{149}$Sm~\cite{Oklo}. This cross section depends
sensitively on the neutron resonance energy $E_r$ for radiative capture
by $^{149}$Sm into an excited state of $^{150}$Sm. The observed isotopic
ratios only allow a small shift of $|\Delta E_r|\la E_r$ from the
present value of $E_r=0.0973$ eV. This then constrains the possible
variations in the energy difference between the excited state of $^{150}$Sm
and the ground state of $^{149}$Sm over the last two billion years.
A contribution to this energy difference comes from the Coulomb energy
$E_C=(3/5)(e^2/r_0)Z^2/A^{1/3}$ ($r_0=1.2$ fm) for a nucleus with $Z$ protons
and $(A-Z)$ neutrons. This contribution clearly scales with $\alpha$ and is
$E_C({^{150}{\rm Sm}})-E_C({^{149}{\rm Sm}})=1.16(\alpha/\alpha_0)$ MeV, where
$\alpha_0$ is the present value of $\alpha$. Considering the time
variation of $\alpha$ alone, $|\Delta E_r|\sim 1.16|\Delta\alpha/\alpha|$ MeV
and a limit $|\Delta\alpha/\alpha|\la 10^{-7}$ can
be obtained\cite{Oklo}. However, if all fundamental couplings are allowed
to vary interdependently, a much more stringent limit 
$|\Delta\alpha/\alpha|<(1-5)\times 10^{-10}$
may be obtained\cite{opqccv}.

Bounds on the variation of the fundamental couplings can also be
obtained from our knowledge of the lifetimes of certain long-lived nuclei.
In particular, it is possible to use precise meteoritic
data to constrain nuclear decay rates back to the time of solar system
formation (about 4.6 Gyr ago). Thus, we can derive a constraint on
possible variations at a redshift $z \simeq 0.45$ bordering the
range ($z = 0.5$--3.0) over which such variations are claimed to be observed.
The pioneering study on the effect of variations of fundamental
constants on radioactive decay rates was performed by Peebles
and Dicke and  by Dyson\cite{PD}. The $\beta$-decay rate, 
$\lambda_\beta$, depends on some power $n$ of the energy $Q_\beta$ released 
during the decay, $\lambda_\beta\propto Q_\beta^n$. A contribution to $Q_\beta$ 
again comes from the Coulomb energy $E_C\propto\alpha$.
Isotopes with the lowest $Q_\beta$ are 
typically most sensitive to changes in $\alpha$ as
$\Delta\lambda_\beta/\lambda_\beta=n(\Delta Q_\beta/Q_\beta)$ is large
for small $Q_{\beta}$. The isotope with the smallest $Q_{\beta}$ 
($2.66\pm0.02$~keV) is $^{187}$Re, which decays into $^{187}$Os. If some
radioactive $^{187}$Re was incorporated into a meteorite formed in the early
solar system, the present abundance of $^{187}$Os in the meteorite is
$({^{187}{\rm Os}})_0=({^{187}{\rm Os}})_i+({^{187}{\rm Re}})_0
[\exp(\lambda_{187}t_a)-1]$, where the subscripts 
``$i$'' and ``0'' denote the initial and present abundances, respectively,
$\lambda_{187}$ is $\lambda_\beta$ for $^{187}$Re,
and $t_a$ is the age of the meteorite.
The above correlation between the present meteoritic abundances of $^{187}$Os
(daughter) and $^{187}$Re (parent) can be generalized to other daughter-parent 
pairs. All these correlations can be used to derive the product of the
relevant decay rate and the meteoritic age. Using the decay rates of
$^{238}$U and $^{235}$U from laboratory measurements, the correlations for the 
$^{206}$Pb-$^{238}$U and $^{207}$Pb-$^{235}$U pairs give a precise age of 
$t_a=4.558$ Gyr for angrite meteorites\cite{lugm}. This determination of $t_a$ 
has the advantage that the decay rates of $^{238}$U and $^{235}$U, 
and hence $t_a$, are rather
insensitive to the variation of fundamental couplings\cite{PD}.
The above age for angrite meteorites allows for a precise determination of
$\lambda_{187}$ from the correlation for the $^{187}$Os-$^{187}$Re pair in iron 
meteorites formed within 5 Myr of the angrite meteorites\cite{smo}. 
Comparing this value of $\lambda_{187}$, which covers the decay over the past 4.6 Gyr, 
with the present value from a laboratory measurement\cite{lind} limits 
the possible variation of $\alpha$ to 
$\Delta\alpha/\alpha=(8\pm 8)\times 10^{-7}$~\cite{opqmvcc}.
Once again, if all fundamental couplings are allowed
to vary interdependently, a more stringent limit 
$\Delta\alpha/\alpha=(2.7\pm 2.7)\times 10^{-8}$
may be obtained.  

Finally, there are a number of present-day laboratory limits on the
variability of the fine-structure constant using two kinds of atomic clocks:
one based on hyperfine transitions involving changes only in the total 
spin of electrons and the nucleus and the other based on electronic transitions 
involving changes in the spatial wavefunction of electrons. 
The electronic transition frequency $\nu_{el}$ depends on a relativistic
correction $F_{rel}(\alpha)$, which is a function of $\alpha$ and is
different for different atoms. Relative to $\nu_{el}$, the hyperfine 
transition frequency $\nu_{hf}$ has an extra dependence on 
$(\mu_{nucl}/\mu_B)\alpha^2$, where
$\mu_{nucl}$ is the magnetic moment of the relevant nucleus and $\mu_B$ is
the atomic Bohr magneton. For atoms A and B,
$\nu_{hf,A}/\nu_{hf,B}\propto(\mu_{nucl,A}/\mu_{nucl,B})
F_{rel,A}(\alpha)/F_{rel,B}(\alpha)$ and
$\nu_{hf,A}/\nu_{el,B}\propto
(\mu_{nucl}/\mu_B)\alpha^2F_{rel,A}(\alpha)/F_{rel,B}(\alpha)$.
If only the variation of $\alpha$ is considered,
this can be tested by comparing $\nu_{hf,A}$ with $\nu_{hf,B}$
or $\nu_{hf,A}$ with $\nu_{el,B}$ over a period of time. 
Three recent experiments have led to marked improvement in 
the limit on the variation of $\alpha$:
$\Delta \alpha / \alpha < 6 \times 10^{-15}$ from
comparing hyperfine transitions in $^{87}$Rb and $^{133}$Cs
over a period of about 4 years\cite{marion},
$\Delta \alpha / \alpha < 4 \times 10^{-15}$ from
comparing an electric quadrupole transition in $^{199}$Hg$^+$ to
the ground-state hyperfine transition in $^{133}$Cs over a 3 year period\cite{bize},
and $\Delta \alpha / \alpha = (1.1 \pm 2.3) \times 10^{-15}$ from
comparing the 1S-2S transition in atomic hydrogen to  
the hyperfine transition in $^{133}$Cs
over a 4 year period\cite{fischer}. If both $\alpha$ and $\mu_{nucl}/\mu_B$
are allowed to vary, then constraints on these two distinct variations
can be obtained by combining the latter two experiments, which give
$\Delta \alpha / \alpha = (-0.9 \pm 4.2)  \times 10^{-15}$
or $\dot \alpha / \alpha \la 10^{-15} $ yr$^{-1}$~\cite{fischer}.

\begin{figure}
\begin{center}
\epsfig{file=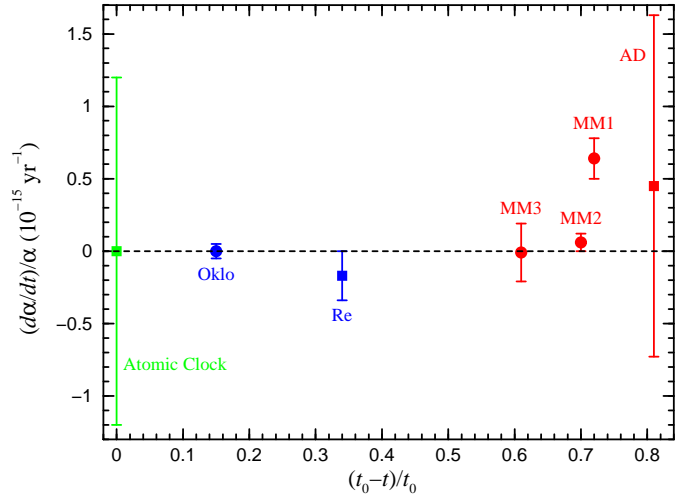, height=4.0in, angle=270}
\end{center}
\caption{{Constraints on the rate of variation $(d\alpha/dt)/\alpha$
as a function of the fractional ``look-back'' time
$(t_0-t)/t_0$, where $t_0\approx 13$ Gyr is the present age
of the Universe. The results shown are taken from data
on atomic clocks \protect\cite{bize}, considerations of
the Oklo phenomenon \protect\cite{Oklo}, meteoritic
data on $^{187}$Re decay \protect\cite{smo,opqmvcc}, 
and many-multiplet (MM1\protect\cite{murphy3},
MM2 \protect\cite{chand},
MM3 \protect\cite{quast}) and alkali-doublet
(AD \protect\cite{doublet}) analyses of quasar absorption
spectra. For convenience, the results for MM1, MM2, MM3,
and AD are shown at the mean redshift for the data used and then converted to 
$(t_0 - t)/t_0$.
Note that the result for MM1 actually covers
$(t_0-t)/t_0=0.37$--0.84. Except for this result, all
others are consistent with no time variation of $\alpha$.
}}
\label{alpha}
\end{figure}

A summary of the constraints on $\alpha$ is found in Fig. \ref{alpha},
taken from ref. \cite{oq}.

The result found in~\cite{chand} and in the statistically dominant subsample of 
74 out of the 128 low redshift absorbers used in~\cite{murphy3} are sensitive to
the assumed isotopic abundance ratio of Mg.  In both analyses, a solar ratio of
$^{24}$Mg:$^{25}$Mg:$^{26}$Mg = 79:10:11 was adopted. However, the resulting 
shift in $\alpha$ is very sensitive to this ratio. Furthermore,
it is commonly assumed that the heavy Mg isotopes are absent
in low metallicity environments characteristic of quasar absorption systems.
Indeed, had the analyses assumed
only pure $^{24}$Mg is present in the quasar absorbers, a much more significant
result would have been obtained.  The Keck/Hires data\cite{murphy3} would have yielded
$\Delta \alpha / \alpha = (-0.98 \pm 0.13) \times 10^{-5}$ for the low redshift subsample
and $\Delta \alpha / \alpha = (-0.36 \pm 0.06) \times 10^{-5}$ for the VLT/UVES
data\cite{chand}.  

The sensitivity to the Mg isotopic ratio has led to a new interpretation of
the many multiplet results\cite{amo}.  The apparent variation in $\alpha$ 
in the Fe-Mg systems
can be explained by the early nucleosynthesis of $^{25,26}$Mg.
The heavy Mg isotopes are efficiently produced in 
intermediate mass stars, particular in stars with masses 4-6 times the mass of the sun,
when He and H are burning in shells outside the C and O core.
There may even be evidence for enhanced populations of intermediate
mass stars at very low metallicity.

Recall the dispersion seen in D/H observations in quasar absorption systems
as seen in  Fig. \ref{D}.
Is there a real dispersion in D/H in these high redshift systems?  
The data may show an inverse 
correlation of D/H abundance  with Si~\cite{omeara,pet}. This
may be an artifact of poorly determined Si abundances, or
(as yet unknown) systematics affecting the
D/H determination in high-column density (damped Lyman-$\alpha$, 
hereafter DLA) 
or low-column density 
(Lyman limit systems) absorbers.
On the other hand, if the correlation is real it would indicate that 
chemical evolution processes have occurred in these systems
and that some processing of  D/H must have
occurred even at high redshift.

\begin{figure}[ht]
\begin{center}
 \includegraphics[height=.5\textheight, angle=270]{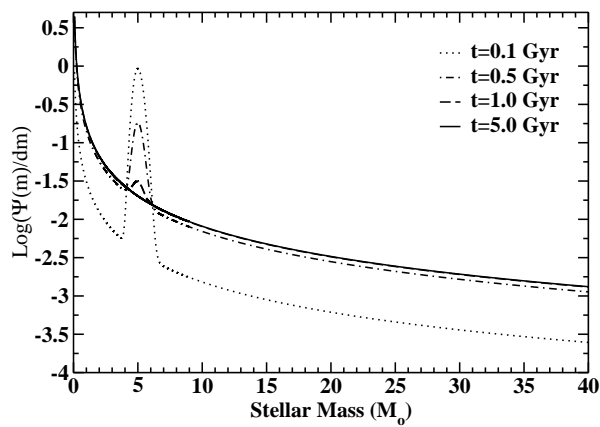}
\end{center}
\caption{{An IMF with an early enhancement of intermediate mass stars\protect\cite{amo2}.}}
\label{IMF}
\end{figure}

It is interesting to speculate\cite{cfosv} that the possible high
redshift destruction of D/H is real
and related to the chemical evolutionary history of 
high red shift systems. For example, these observations could be signatures
of an early population of intermediate-mass stars characterized by an
initial mass function different from that of the solar neighborhood.
An example of such an IMF is shown in Fig. \ref{IMF}~\cite{amo2}.

There are a number of immediate consequences of an IMF of the type shown in
Fig. \ref{IMF}.  In addition to the destruction of D/H at low metallicity,
one expects observable C and N enhancements in high redshift absorption systems.
In addition, one also expects an enhancement of the heavy Mg isotopes, $^{25,26}$Mg,
which may account\cite{amo,amo2} for the apparent variation of the fine-structure constant 
in quasar absorption systems. 
 In this sense,  the many multiplet method can be used to 
trace the chemical history of primitive absorption clouds\cite{amo}.  
This hypothesis will be tested by future observations
and examinations of correlation among other heavy elements 
produced in intermediate mass stars\cite{iso}.

\section*{Acknowledgments}
  I would like to thank T. Ashenfelter, M. Cass\'{e}, R. Cyburt, J. Ellis, 
  T. Falk, B. Fields, K. Kainulainen, G. Mathews, M. Pospelov, Y. Qian, Y. Santoso, 
  J. Silk, E. Skillman, V. Spanos, M. Srednicki, and 
  E. Vangioni for recent (and enjoyable) collaborations on BBN.
  This work was partially supported by DOE grant
DE-FG02-94ER-40823.

\end{document}